\documentclass{emulateapj}
\usepackage{graphicx}
\usepackage[normalem]{ulem}
\usepackage{natbib}
\shorttitle{SN2013ge}
\shortauthors{M. R. Drout et al.}

\begin{document}

\title{The Double-Peaked SN\,2013\lowercase{ge}: a Type I\lowercase{b/c} SN with an Asymmetric Mass Ejection or an Extended Progenitor Envelope}

\author{M. R. Drout\altaffilmark{1}, D. Milisavljevic\altaffilmark{1}, J. Parrent\altaffilmark{1}, R. Margutti\altaffilmark{1}, A. Kamble\altaffilmark{1}, A. M. Soderberg\altaffilmark{1}, P. Challis\altaffilmark{1}, R. Chornock\altaffilmark{2}, W. Fong\altaffilmark{3,4}, S. Frank\altaffilmark{5}, N. Gehrels\altaffilmark{6}, M. L. Graham\altaffilmark{7}, E. Hsiao\altaffilmark{8,9}, K. Itagaki\altaffilmark{10},M. Kasliwal\altaffilmark{11}, R. P. Kirshner\altaffilmark{1}, D. Macomb\altaffilmark{12}, G. H. Marion\altaffilmark{13}, J. Norris\altaffilmark{12}, M. M. Phillips\altaffilmark{9}}

\altaffiltext{1}{Harvard-Smithsonian Center for Astrophysics, 60 Garden Street, Cambridge, MA 02138}
\altaffiltext{2}{Astrophysical Institute, Department of Physics and Astronomy, 251B Clippinger Lab, Ohio University, Athens, OH 45701, USA}
\altaffiltext{3}{Steward Observatory, University of Arizona, 933 North Cherry Avenue, Tucson, AZ 85721, USA}
\altaffiltext{4}{Einstein Fellow}
\altaffiltext{5}{Department of Astronomy, The Ohio State University, 140 West 18th Avenue, Columbus, OH 43210, USA}
\altaffiltext{6}{NASA Goddard Space Flight Center, Greenbelt, MD 20771, USA}
\altaffiltext{7}{Department of Astronomy, University of California, Berkeley, CA 94720-3411, USA}
\altaffiltext{8}{Department of Physics and Astronomy, Aarhus University, Ny Munkegade 120, 8000 Aarhus C, Denmark}
\altaffiltext{9}{Carnegie Observatories, Las Campanas Observatory, Colina El Pino, Casilla 601, Chile}
\altaffiltext{10}{Itagaki Astronomical Observatory, Teppo-cho, Yamagata, Yamagata 990-2492, Japan}
\altaffiltext{11}{Observatories of the Carnegie Institution for Science, 813 Santa Barbara Street, Pasadena CA 91101, USA}
\altaffiltext{12}{Boise State University, Dept. Of Physics, 1910 Univ. Drive Boise, Boise ID 83725 USA}
\altaffiltext{13}{Department of Astronomy, University of Texas at Austin, Austin, TX 78712, USA}

\begin{abstract}
We present extensive multiwavelength (radio to X-ray) observations of the Type Ib/c SN\,2013ge from $-$13 to $+$457 days relative to maximum light, including a series of optical spectra and \emph{Swift} UV-optical photometry beginning 2 $-$ 4 days post-explosion. This data set makes SN\,2013ge one of the best observed normal Type Ib/c SN at early times---when the light curve is particularly sensitive to the progenitor configuration and mixing of radioactive elements---and reveals two distinct light curve components in the UV bands. The first component rises over 4 $-$ 5 days and is visible for the first week post-explosion. Spectra of the first component have blue continua and show a plethora of moderately high-velocity ($\sim$15,000 km s$^{-1}$) but narrow ($\sim$3500 km s$^{-1}$) spectroscopic features, indicating that the line-forming region is restricted. The explosion parameters estimated for the bulk explosion (M$_{\rm{ej}}$ $\sim$ 2 $-$ 3 M$_\odot$; E$_{\rm{K}}$ $\sim$ 1 $-$ 2 $\times$10$^{51}$ erg) are standard for Type Ib/c SN, and there is evidence for \emph{weak} He features at early times---in an object which would have otherwise been classified as Type Ic. In addition, SN\,2013ge exploded in a low metallicity environment ($\sim$0.5 Z$_\odot$) and we have obtained some of the deepest radio and X-ray limits for a Type Ib/c SN to date, which constrain the progenitor mass-loss rate to be $\dot{M}$ $<$ 4 $\times$ 10$^{-6}$ M$_\odot$ yr$^{-1}$.  We are left with two distinct progenitor scenarios for SN\,2013ge, depending on our interpretation of the early emission.  If the first component is cooling envelope emission, then the progenitor of SN\,2013ge either possessed an extended ($\gtrsim$ 30 R$_\odot$) envelope or ejected a portion of its envelope in the final $\lesssim$1 year before core-collapse. Alternatively, if the first component is due to outwardly mixed $^{56}$Ni, then our observations are consistent with the asymmetric ejection of a distinct clump of nickel-rich material at high velocities. Current models for the collision of a SN shock with a binary companion cannot reproduce both the timescale and luminosity of the early emission in SN\,2013ge. Finally, the spectra of the first component of SN\,2013ge are similar to those of the rapidly-declining SN\,2002bj.
\end{abstract}

\email{mdrout@cfa.harvard.edu}
\keywords{supernovae:general; supernova:individual(SN2013ge)}

\section{Introduction}\label{Sec:Intro}

Type Ib/c supernovae (SN) are an observational subclass of stellar explosions. They are identified mainly by a lack of either strong hydrogen or strong silicon features in their optical spectra (see \citealt{Wheeler1995}, \citealt{Filippenko1997} for a review of SN classifications).  This class can be further divided into Type Ib SN, which show conspicuous lines of helium in their spectra, and Type Ic SN, which do not. These events are physically understood to be the core-collapse of massive stars that were stripped of their hydrogen envelopes.  Main progenitor channels include isolated Wolf Rayet (WR) stars with massive winds \citep{Begelman1986,Woosley1995} and lower mass helium stars stripped by binary companions \citep{Wheeler1985,Podsiadlowski1992,Yoon2010}.  
 
Two of the main power sources that contribute to the rising phase of a SN light curve are the radioactive decay of $^{56}$Ni synthesized in the explosion and the cooling envelope emission produced when the ejecta radiates away energy deposited by the SN shock \citep[e.g.][]{Piro2013}.  In stripped-envelope SN, $^{56}$Ni powers a majority of the light curve, while cooling envelope emission is only predicted to be visible for a few days post-explosion.  As a result, early observations of Type Ib/c SN provide a particularly sensitive probe of both the structure of the progenitor star prior to explosion \citep{Nakar2010,Rabinak2011,Nakar2014} and the degree to which radioactive materials are mixed into the outer ejecta \citep{Dessart2012,Piro2013}.  

Constraints on the structure of the progenitor star from cooling envelope emission are valuable as the final radii of putative Type Ib/c progenitors are predicted to vary by an order of magnitude or more depending on their initial conditions (mass, metallicity) and evolutionary history (single versus binary) \citep{Yoon2010}.  While no cooling envelope emission has been observed for a normal (not broad-lined) Type Ic SN to date, non-detections have been used to place constraints on the progenitor radii in several objects (e.g. PTF10vgv; \citealt{Corsi2012}).  In addition, when interpreted as shock breakout emission, the early X-ray/UV peak observed from the Type Ib SN\,2008D constrains its progenitor radius to be $\lesssim$ 12 R$_\odot$ (\citealt{Soderberg2008}; but see also e.g.\ \citealt{Mazzali2008b}, \citealt{Bersten2013} for alternative interpretations of this emission). 

In addition, recent observations have highlighted gaps in our understanding of the final state of the progenitors for some SN.  For example, a handful of SN with double-peaked optical light curves have been discovered.  When interpreted as cooling envelope emission, the first peak requires that the progenitor star possessed a low-mass extended envelope, which differs from the standard hydrostatic models of stellar structure \citep{Nakar2014,Bersten2012,Nakar2015}.  While a majority of these events are of Type IIb (e.g. SN\,1993J \citealt{Wheeler1993}; SN\,2011dh \citealt{Arcavi2011}; SN\,2013df \citealt{VanDyk2014}), a similar morphology has also been observed in the Type Ibn iPTF\,13beo \citep{Gorbikov2014} and the Ic-BL SN\,2006aj associated with Gamma-Ray Burst (GRB) 060218 \citep{Campana2006,Nakar2015}.  Further, X-ray observations point to a subset of long GRB progenitors which either underwent enhanced mass-loss in the final years before explosion, or possess low-mass, extended, progenitor envelopes \citep{Margutti2015}.

Constraints on the mixing of radioactive material from rising light curves can also provide insight into the progenitor structure and explosion mechanism for various subclasses of Type Ib/c SN.  Mixing in core-collapse SN can be accomplished by a number of mechanisms including a large-scale asymmetry of the explosion (e.g. a ``jet''-like explosion), a large-scale asymmetry of the SN shock produced by the neutrino or magneto-rotational mechanism \citep[e.g.][]{Scheck2006,Marek2009,Maeda2002,Burrows2007}, and smaller-scale Rayleigh-Taylor and Kelvin-Helmholtz instabilities at the shock front (\citealt{Kifonidis2006,Joggerst2009,Hammer2010}).  The observed signature of such mixing will vary depending on its origin, ranging from small modifications to the timescale and colors on the rise for well-mixed shallow deposits of $^{56}$Ni \citep{Dessart2012} to double-peaked light curves for asymmetric ejections of material. Models of the latter type have been investigated as a possible source for the double-peaked light curves observed in the Type Ib SN\,2008D \citep{Bersten2013} and SN\,2005bf (\citealt{Folatelli2006}; but see also \citealt{Maeda2007}).  

\begin{figure}[!ht]
\begin{center}
\includegraphics[width=\columnwidth]{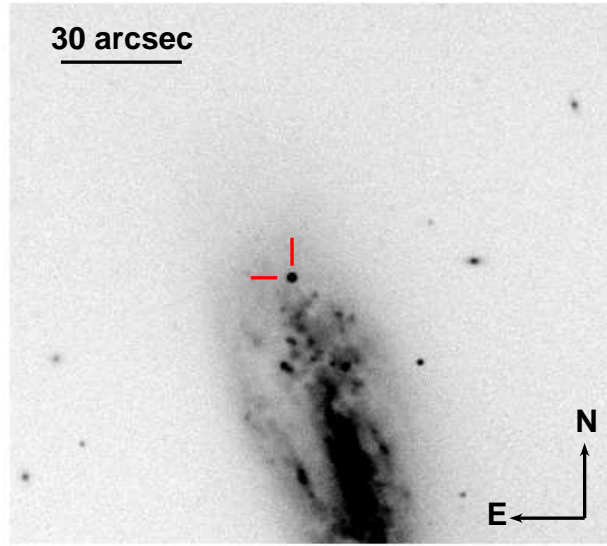}
\caption{r-band image from MMTCam, obtained on 2014 Apr.\ 1, showing the location of SN\,2013ge (red cross hairs) on the outskirts of NGC\,3287.   \label{fig:finder}}
\end{center}
\end{figure}  

Understanding the mixing of radioactive material is especially vital for constraining what distinguishes whether a given progenitor will explode as a Type Ib or Ic SN.   The production of \ion{He}{1} lines in SN spectra requires non-thermal excitation of the helium atoms, likely from the $\gamma$-rays produced by the radioactive decay of $^{56}$Ni \citep[e.g.][]{Lucy1991,Dessart2012,Hachinger2012}.  Thus, in order to produce a Type Ib SN, mixing may be required, while the observation of a Type Ic SN may not necessarily imply that its progenitor was He poor.  Early observations that constrain mixing in a normal Type Ic SN can therefore help to distinguish whether or not these events have an intrinsically lower helium abundance.
  
Thus, particularly when they are coupled with other multiwavelength observations, early light curves offer us insight into topics such as the evolutionary path of the progenitor, the explosion mechanism, and the properties that dictate whether a given star will explode as a Type Ib or Ic SN.   In this paper we present detailed observations of the Type Ib/c SN\,2013ge, which span radio to X-ray and include spectroscopy and UV-optical photometry beginning 2$-$4 days after the epoch of first light.  These early observations show behavior which has not been observed in any Type Ib/c SN to date: a distinct light curve component visible in the blue bands for the first week after explosion, which---on its rising portion---shows spectral features with moderately high expansion velocities but \emph{narrow} line widths.

In Section~\ref{Sec:Obs} we describe the observations obtained for SN\,2013ge.  In Sections~\ref{Sec:Photom} \&~\ref{sec:spec} we describe the photometric and spectroscopic properties of SN\,2013ge, respectively, while in Section~\ref{Sec:Enviro} we examine the properties of the circumstellar medium (CSM) surrounding the progenitor star. In Section~\ref{Sec:Discuss} we discuss the consequences of various observed properties on our understanding of the progenitor of SN\,2013ge.

\begin{figure*}[!ht]
\begin{center}
\includegraphics[width=\textwidth]{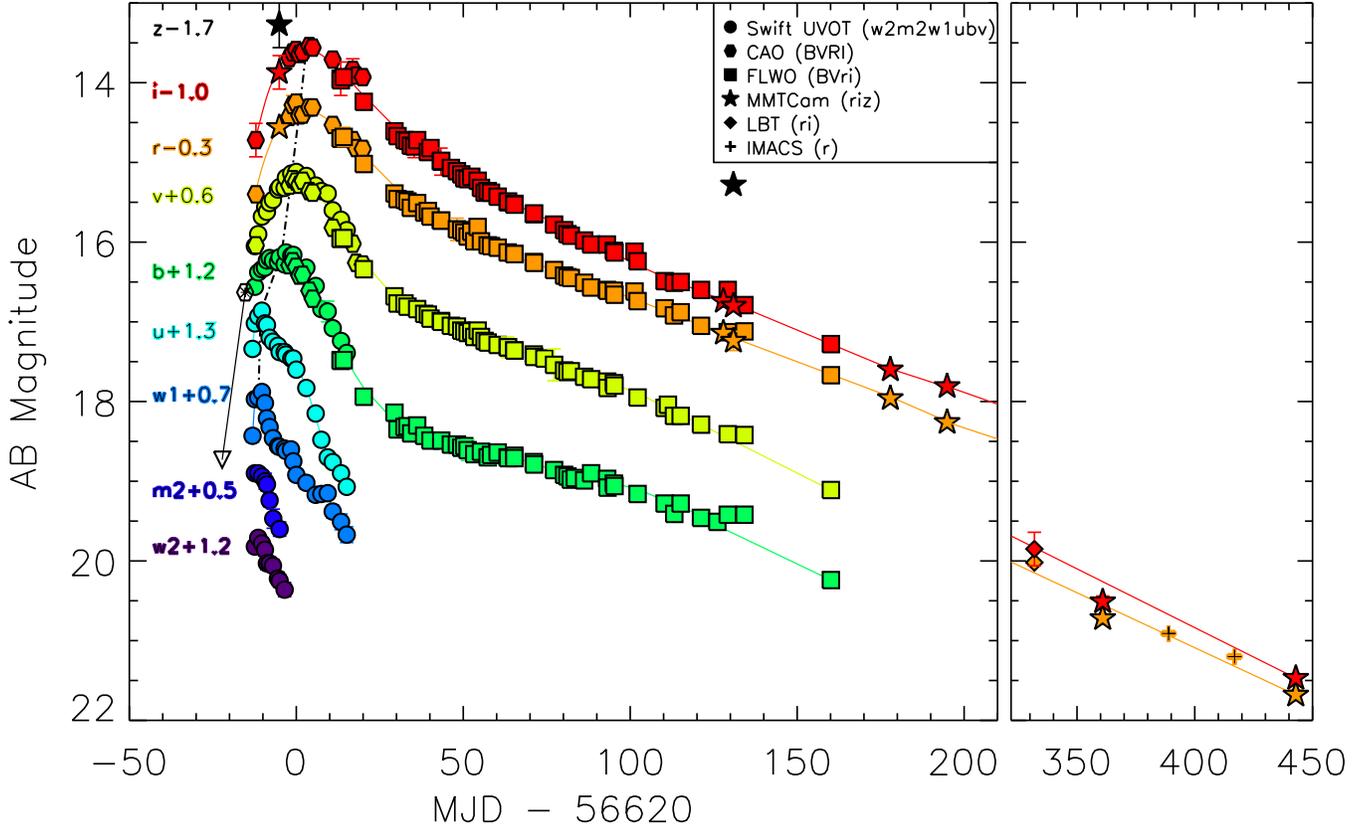}
\vspace{0.02in}
\caption{Multi-band photometry for SN\,2013ge. Symbol shape indicates the source of the photometry and color indicates the observed band, as labeled.  The unfiltered discovery photometry and pre-explosion limit are shown as a hexagon with an asterisk and an open downward facing triangle, respectively.  See Section~\ref{sec:LCcomb} for further information on the procedures applied to place data on this plot.  The dashed-dotted line traces the epoch of maximum light for each observed band. \label{Fig:OpticalPhotom}}
\end{center}
\end{figure*}

\section{Observations}\label{Sec:Obs}

\subsection{UV and Optical Photometry}

We obtained UV and optical photometric observations of SN\,2013ge from a wide variety of instruments, spanning 466 days. In the sections below we describe the data acquisition, reduction, and calibration for each instrument and in Section~\ref{sec:LCcomb} we discuss the consistency of the combined light curve.  The location of the transient on the outskirts of NGC\,3287 is shown in Figure~\ref{fig:finder}.

\subsubsection{Discovery Photometry and Pre-Explosion Limit}

SN\,2013ge was discovered by Koichi Itagaki on 2013 Nov.\ 8.8 (all times UT) using the 0.5-m reflector at the Takanezawa station, Tochigi-ken, and was undetected prior to discovery on 2013 Nov.\ 1.7 with the same instrument (CBAT 3601).  We have reanalyzed these unfiltered images for this manuscript.  We performed point spread function (PSF) photometry on the SN and nearby field stars in the discovery image using standard packages in IRAF\footnote{IRAF is distributed by NOAO, which is operated by the Association for Research in Astronomy, Inc.\, under cooperative agreement with the NSF.}, and absolute calibration was performed using Bessell R-band magnitudes of nearby field stars. No formal color correction to the Bessell system was performed.  This same procedure was also carried out on fake sources injected into the pre-discovery image. This yields an unfiltered discovery magnitude of 16.9 $\pm$ 0.1 mag and a pre-explosion limit of 19.0 mag. 

\subsubsection{\emph{Swift} UVOT Photometry}

We observed SN\,2013ge with the UV Optical Telescope onboard \emph{Swift} (UVOT, \citealt{Gehrels2004}, \citealt{Roming05}) from 2013 Nov.\ 11 to 2013 Dec.\ 9 ($-$11 to $+$17 days).  The \emph{Swift}-UVOT photometric data were extracted following the prescriptions of \citet{Brown09}.  We used a variable aperture with radius 3'' -- 5'' to maximize the signal-to-noise ratio as the SN flux faded.  The \emph{Swift}-UVOT photometry is reported in Table~\ref{tab:PhotomUVOT} in the photometric system described in \citet{Breeveld2011}.

\subsubsection{Challis Observatory Optical Photometry}

We observed SN\,2013ge with the Challis Astronomical Observatory (CAO) on 12 nights spanning 2013 Nov.\ 12 to 2013 Dec.\ 14 ($-$10 to $+$22 days).  CAO is located in central Idaho near the River of No Return Wilderness.  The facility telescope is a 0.4$-$m f/10 Meade LX$-$200, equipped with an Apogee Alta U47-MB camera and UBVRI Bessell filters.

\begin{figure}[!ht]
\begin{center}
\includegraphics[width=\columnwidth]{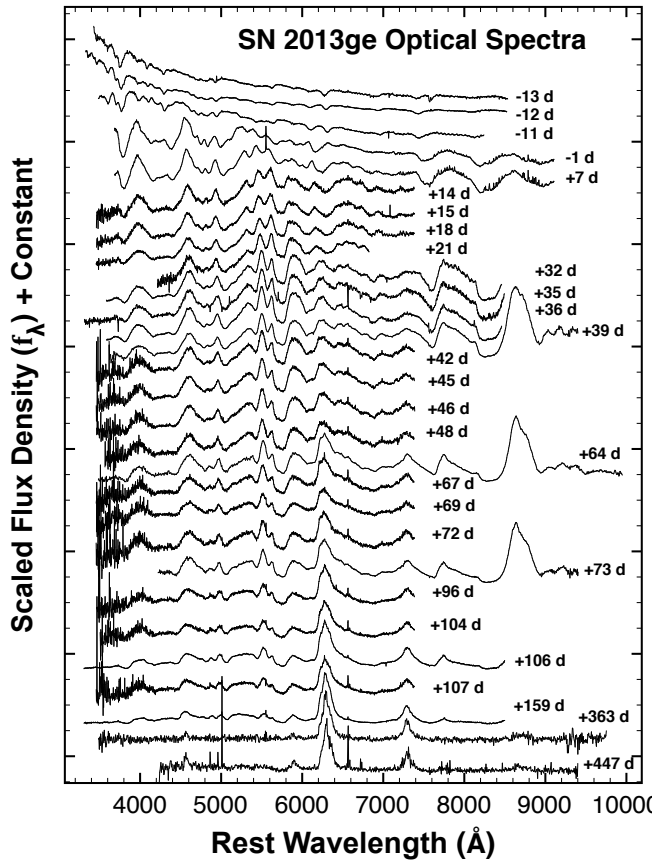}
\caption{Normalized optical spectra spanning $-$13 to $+$447 days with respect to V-band maximum.  \label{fig:opticalspec}}
\end{center}
\end{figure}

For each epoch, 5 $-$ 10 raw images were stacked and PSF photometry was performed in IRAF. Absolute calibration was performed using SDSS observations of field stars, which were converted to the Bessell BVRI system using the equations of \citet{Smith2002}.  A single nightly zeropoint offset was applied, as data was not available to fully calibrate color terms between the CAO and standard Bessell filters.  However, no strong trends between zeropoint and color were evident in observations of $\sim$15 field stars on multiple epochs, and the observed color of SN\,2013ge is well matched to the range of colors of the field stars used for calibration.  We estimate that the error in our calibration (assuming a range of possible color terms) is $\lesssim$0.05 mag over the epochs we observed with CAO.  This data is listed in Table~\ref{Tab:IdahoPhotom} in the Bessell photometric system.

\subsubsection{FLWO 1.2-m Optical Photometry}

We obtained 55 epochs of BVri photometry of SN\,2013ge spanning 2013 Dec.\ 7 to 2014 May 3 ($+$15 to $+$162 days) with the Fred Lawrence Whipple Observatory (FLWO) 1.2 meter telescope plus KeplerCam CCD. The KeplerCam data were reduced using IRAF, IDL and PERL procedures as described in \citet{Hicken2012} for the CfA4 Type Ia sample, with the exception that no host galaxy subtraction was performed. BVri instrumental magnitudes were measured using PSF fitting. For calibration, we applied a set of linear transforms, which account for zeropoint, atmospheric and color terms.  These linear transforms were calibrated using \citet{Landolt1992} standards for BV and \citet{Smith2002} standards for r and i bands on 5 photometric nights, while nightly zeropoints were determined by measuring the magnitudes of local comparison stars in the SN\,2013ge field (see \citealt{Hicken2012} for further discussion of the calibration applied).  The resulting BV data in the Bessell photometric system and ri data in the SDSS photometric system are listed in Table~\ref{Tab:FLWOPhotom}. 

\begin{figure}[!ht]
\begin{center}
\includegraphics[width=\columnwidth]{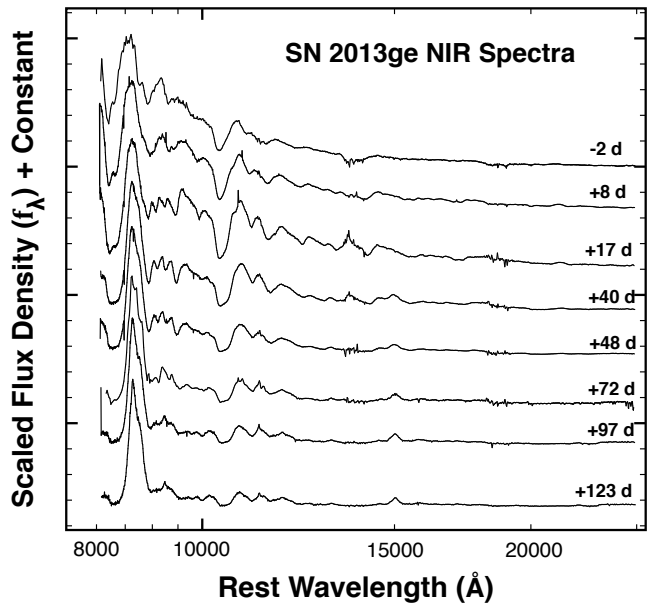}
\caption{Normalized NIR spectra spanning $-$2 to $+$123 days with respect to V-band maximum.  \label{fig:NIRspec}}
\end{center}
\end{figure}

\subsubsection{MMTCam, LBT and IMACS Optical Photometry}

In addition, we obtained eight epochs of ri$-$band photometry and two epochs of z$-$band photometry with the MMTCam instrument mounted on the 6.5m MMT telescope, one epoch of ri$-$band imaging with the Large Binocular Camera (LBC; \citealt{sdg+08}) mounted on the Large Binocular Telescope (LBT), and two epochs of r$-$band imaging with IMACS on Magellan-Baade.  One epoch was obtained on 2013 Nov.\ 19 ($-$3 days), and the other epochs span 2013 Apr.\ 1 to 2015 Apr.\ 16 ($+$129 to $+$510 days). Bias and flat field corrections were made to all images and nightly stacks were produced using standard routines in IRAF.  Dark frame corrections were also applied to images taken with MMTCam. PSF photometry was performed and absolute calibration was carried out using SDSS observations of field stars. A single nightly zeropoint offset was performed. These data listed in the SDSS photometric system in Table~\ref{Tab:MMTPhotom}.

\subsubsection{Combined UV-Optical Light Curve}\label{sec:LCcomb}

As a result of the extensive observations described above, we possess \emph{Swift}-UVOT data in the \citet{Breeveld2011} photometric system, CAO BVRI and FLWO BV data in the Bessell photometric system, and FLWO ri, MMTCam riz, LBT ri and IMACS r$-$band data in the SDSS photometric system.  We plot the resulting UV/optical light curve in Figure~\ref{Fig:OpticalPhotom}.  In this Figure color signifies the observed band and the shape of the symbol signifies the source of the photometry (see legend).  In order to place data from different photometric systems on the same plot, we have shifted the photometric zeropoint of all data to the AB magnitude scale (m = $-2.5 \log_{10} (F_{\rm{Jy}}/3630)$) and have corrected for the extinction in each observed band based on the total E(B$-$V)$=$0.067 mag derived in Section~\ref{sec:ext}, below. We find good agreement between our data from different sources. Slight variations, on the order of 0.05$-$0.1 mag, are observed between the \emph{Swift} bv data and CAO BV data, as expected for the different filter curves. We note that the overlapping epochs of CAO/FLWO V$-$band data and FWLO/MMTCam ri$-$band data agree within the quoted uncertainties, consistent with the conclusion that the color corrections between the CAO/MMTCam filters and the standard Bessell/SDSS filters are small.  

The unfiltered discovery photometry and pre-explosion limit are shown as a hexagon with an asterisk and an open downward facing triangle in Figure~\ref{Fig:OpticalPhotom}, respectively.  These points have been processed in the same manner as the R$-$band data for placement on this plot, but we caution that some variations likely exist between this data and the standard Bessell R-band at early times when the SN is very blue.  We do not attempt to transform all of our data to a single photometric/filter system.  In the analysis below, each point is treated appropriately for the filter in which it was originally observed.

\subsection{Chandra X-ray Observations}

We obtained deep X-ray limits for SN\,2013ge with the \emph{Chandra} X-ray Observatory on 2013 Dec.\ 7 under an approved Director Discretionary Time Proposal (PI: Margutti). The total exposure time was 18.8 ks. \emph{Chandra} ACIS-S data were reduced with the {\tt CIAO} software package (v4.5) and relevant calibration files, applying standard filtering criteria. Using {\tt wavedetect} we find no evidence for X-ray emission at the location of SN\,2013ge. The $3\,\sigma$ count-rate upper limit is $3.1\times10^{-4}$ s$^{-1}$ in the 0.5-8 keV energy band. 

The neutral hydrogen column density in the direction of the SN is $\rm{NH_{\rm{gal}}}=1.9\times 10^{20}$ cm$^{-2}$ \citep{Kalberla05}.  From our optical spectra we estimate E(B-V)$_{\rm{host}}=0.047$ mag (Section~\ref{sec:ext}). For a Galactic dust to gas ratio, this value corresponds to an intrinsic neutral hydrogen column density of $\rm{NH_{\rm{host}}}\sim3\times 10^{20}\,\rm{cm^{-2}}$. For an assumed simple power-law spectral model with spectral photon index $\Gamma=2$ we find an unabsorbed flux limit $3.4\times10^{-15}$ erg s$^{-1}$ cm$^{-2}$ (0.3-10 keV). At the distance of 23.7 Mpc (Section~\ref{sec:ext}), this flux translates into a luminosity limit of $2.3\times 10^{38}$ erg s$^{-1}$. 

\subsection{Very Large Array Radio Observations}

We obtained three epochs of deep radio limits for the emission from SN\,2013ge at 4.8 and 7.1 GHz with the Very Large Array (VLA)\footnote{The National Radio Astronomy Observatory is a facility of the National Science Foundation operated under cooperative agreement by Associated Universities, Inc.}. Observations were obtained on 2013 Nov.\ 16, 25, and 2014 Jan.\ 14 ($-$6, $+$3, and $+$53 days), when the VLA was in B configuration, under program 13A-270.

All observations were taken in standard continuum observing mode with a bandwidth of $16 \times 64 \times 2$ MHz. During the reduction we split the data in two basebands of approximately 1 GHz each\footnote{8 IF $\times$ 64 (channels/IF) $\times$ 2 (MHz/channel)}.  We used 3C286 for flux calibration, and calibrator J1018+3542 for phase referencing. The data were reduced using standard packages within the Astronomical Image Processing System (AIPS). No radio emission was detected from SN\,2013ge in any of these observations. We measured the RMS noise at the location of SN\,2013ge in each image using the task `JMFIT' in AIPS.  The resulting 3$\sigma$ upper limits for each frequency and epoch are listed in Table~\ref{Tab:VLA}.

\subsection{Optical Spectroscopy}

We obtained 29 epochs of low resolution optical spectra for SN\,2013ge, spanning $-$13 to $+$447 days. In addition, one moderate resolution (R$\sim$3400) spectrum was obtained at $-$12 days in the region of Na ID.  All optical spectra are listed in Table~\ref{tab:spectra} and shown in Figure~\ref{fig:opticalspec}. All long slit observations were carried out with the slit oriented at the parallactic angle, with the exception of those obtained on Magellan/IMACS, which possesses an atmospheric dispersion corrector.

Initial reduction (overscan correction, flat fielding, extraction, wavelength calibration) for all long slit spectra was carried out using the standard packages in IRAF.  The MODS/LBT spectrum was taken in the dual channel mode with gratings G400L and G670L, and we used the modsCCDRed suite of python programs to perform bias subtraction, flat-field correction, and adjust for bad columns, before extracting the spectrum using standard packages in IRAF.  Flux calibration and telluric correction for all long slit spectra were performed using a set of custom idl scripts (see, e.g., \citealt{Matheson2008,Blondin2012}) and standard star observations obtained the same night as the science exposures.  Spectra obtained with the Hectospec multi-fiber spectrograph \citep{Fabricant2005} were reduced using the IRAF package ``hectospec'' and the CfA pipeline designed for this instrument. In all cases, when possible, spectroscopic flux calibration has been checked against observed photometry.

\subsection{Infrared Spectroscopy}

We obtained 8 epochs of NIR spectra for SN\,2013ge with the Folded-port InfraRed Echellette spectrograph (FIRE; \citealt{Simcoe2013}) on Magellan-Baade spanning $-$2 to $+$123 days.  All observations were obtained in longslit mode with the slit orientated at the paratactic angle. These data were reduced following standard procedures \citep[see, e.g.][]{Hsiao2015} using an IDL pipeline designed specifically for FIRE (\texttt{firehose})\footnote{Available at http://web.mit.edu/~rsimcoe/www/FIRE/}.  A0V standards observed with each science exposure were used to correct for telluric absorption using the IDL tool \texttt{xtellcor} \citep{Vacca2003}. A model spectrum of Vega was used to remove stellar absorption features from the telluric standards and the resulting spectra were also used for flux calibration. All NIR spectra are listed in Table~\ref{tab:spectra} and shown in Figure~\ref{fig:NIRspec}. 

\subsection{Distance and Reddening}\label{sec:ext}

SN\,2013ge exploded on the outskirts of NGC\,3287.  In this paper we adopt of distance of 23.7 $\pm$ 1.7 Mpc, corresponding to the NED distance after correction for Virgo, Great Attractor, and Shapley Supercluster infall, assuming H$_0$ $=$ 73 km s$^{-1}$ Mpc$^{-1}$ \citep{mhf+00}.

\begin{figure}[!ht]
\begin{center}
\includegraphics[width=\columnwidth]{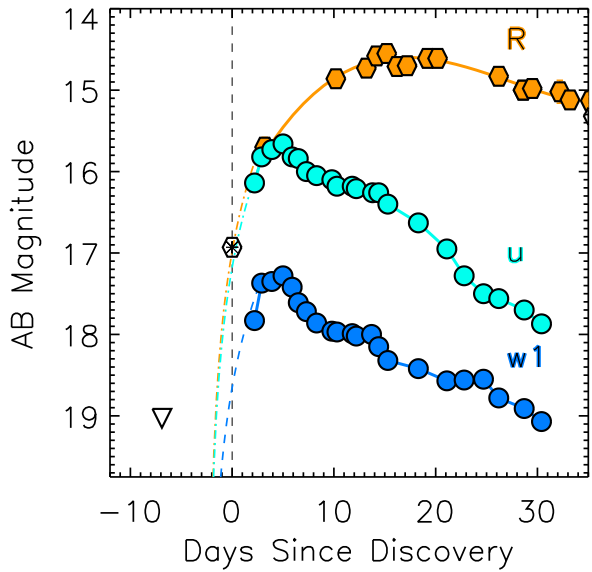}
\caption{Characterization of the epoch of first light.  Shown are the R-band (orange), u-band (cyan), and w1-band (blue) light curves, shifted to the AB system in the same manner as for Figure~\ref{Fig:OpticalPhotom} (see Section~\ref{sec:LCcomb}).  The rapid rise observed in the u$-$ and w1$-$band light curves puts a tight constraint on the epoch of first light.  Dashed colored lines are power law fits of the form t$^{1.5}$ to the rising phase of each band, all of which imply an epoch of first light 2 days prior to first detection.  The unfiltered discovery and pre-explosion limit are shown as a hexagon with an asterisk and an open triangle, respectively. \label{fig:ExEpoch}}
\end{center}
\end{figure}

Milky Way reddening in the direction of SN\,2013ge is E(B$-$V)$_{\rm{MW}}$ $=$ 0.020 mag \citep{Schlafly2011}.  To estimate the intrinsic absorption due to dust within NGC\,3287, we examine our moderate resolution spectra obtained on 2014 Nov.\ 10 for narrow Na~ID absorption lines.  We identify two Na~ID $\lambda \lambda$5889.9, 5895.9 doublets, one corresponding to Milky Way absorption, and one at the redshift of NGC\,3287.  The total equivalent width of the Na~ID doublet at the redshift of NGC\,3287 is $\sim$0.45 \AA.  Using the empirical relation of \citet{Poznanski2012} this implies a host galaxy contribution to the total reddening of E(B$-$V)$_{\rm{host}}$ $=$ 0.047 mag.  As a check, we apply the same procedure to the observed Milky Way Na~ID absorption which yields an E(B$-$V)$_{\rm{MW}}$ value of 0.037 mag.  This value is slightly higher than that found by \citet{Schlafly2011}, but within the quoted errors of the relation from \citet{Poznanski2012}.  Throughout this paper we adopt an R $=$ A$_V$/E(B$-$V) $=$ 3.1 Milky Way extinction curve with a total reddening of E(B$-$V)$_{\rm{tot}}$ $=$ 0.067 mag.  This value is consistent with that derived from the V$-$R color method described in \citet{Drout2011}.  

\section{Photometric Properties}\label{Sec:Photom}

\subsection{Light Curve Evolution}

A deep pre-explosion limit for SN\,2013ge was obtained 7 days prior to its initial discovery.  However, the rapid rise observed in the \emph{Swift} u$-$ and w1$-$bands (photometric coverage in these bands began $\sim$2 days after first detection; see Figure~\ref{fig:ExEpoch}) indicates that our constraints on the epoch of first light may be more stringent than allowed by this non-detection alone.  Extrapolating polynomial fits of the rising phase of these light curves backward in time, we infer an epoch of first light only $\sim$2 days prior to first detection.  Power-law fits of the form L$_\lambda$ $\propto$ t$^{1.5}$ and L$_\lambda$ $\propto$ t$^{2}$ yield similar results. The former power law is expected for the rising phase of cooling envelope emission \citep[e.g.][]{Piro2013} and the latter is expected for early radioactive heating in the fireball model \citep{Nugent2011}. Throughout this paper we adopt an epoch of first light of 2014 Nov.\ 6.5 $=$ MJD 56602.5 $\pm$ 2 days.  An epoch of first light prior to this date would be inconsistent with the early u$-$ and w1$-$band light curves unless the rate of rise \emph{increased} some time post-explosion.  Such behavior is not typically observed in SN light curves, and would have implications for the early emission source, which will be discussed in Section~\ref{Sec:Discuss}.  The possibility of a ``dark period'' separating the epoch of first light from the explosion epoch \citep{Piro2013} will be also discussed in Section~\ref{Sec:Discuss}.

In Table~\ref{tab:PhotomProps} we list basic properties for the UV and optical light curves of SN\,2013ge, which were measured based on low order polynomial fits\footnote{Polynomial fits for each band were performed based on photometry for a single source, as listed in Table~\ref{tab:PhotomProps}.}.  Throughout this paper, the phase of the SN will be given with respect to V-band maximum light: MJD 56618.6.  This time of maximum implies a rise time in the V-band of $\sim$16 days.  As is observed in other Type I SN, the time of maximum light cascades through the UV and optical bands (as shown by the dashed dotted line in Figure~\ref{Fig:OpticalPhotom}). SN\,2013ge peaks at an absolute magnitude of approximately $-$17.3 mag (AB) in the optical, and declines between 1.1 mag and 0.38 mag in the first 15 days post maximum light (b$-$through$-$I bands, respectively). This places SN\,2013ge at both the low luminosity and slowly-evolving end of Type I SN.

\begin{figure}[!ht]
\begin{center}
\includegraphics[width=\columnwidth]{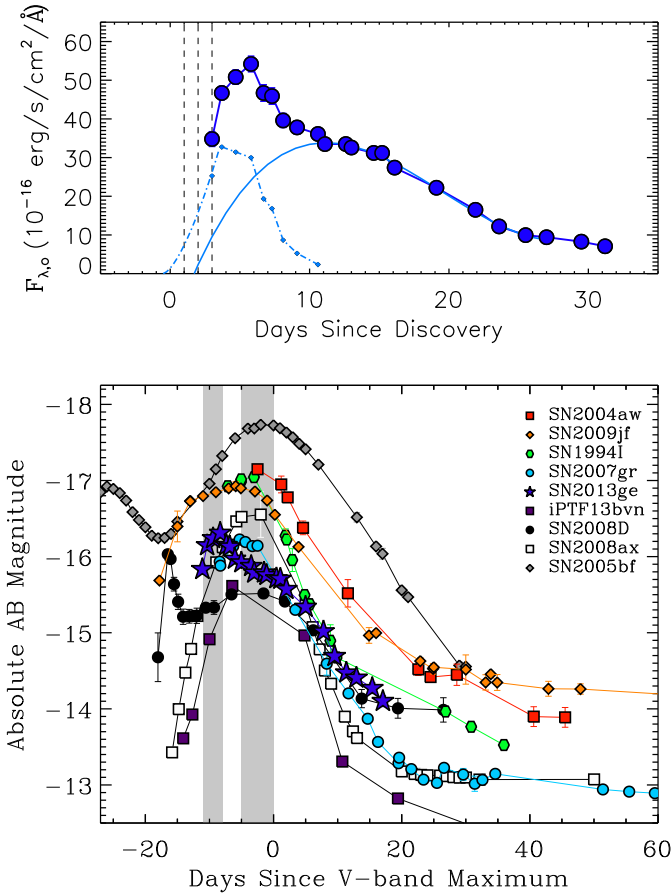}
\caption{\emph{Top:} Example decomposition of the u-band light curve into two components.  This should be taken as representative only as we do not constrain the rising behavior of the second component. Dashed vertical lines represent our first three epochs of spectroscopy, which occur on the early rising portion of the first light curve component.  \emph{Bottom:} Comparison of the u-band light curve of SN\,2013ge to other stripped core-collapse SN.  Shaded regions highlight the time of maximum for both components of SN\,2013ge.  Most events peak on a timescale similar to that observed for the second component in SN\,2013ge.  The light curve of SN\,2009jf shows a plateau that may be indicative of an early component similar to that observed in SN\,2013ge. \emph{References:} SN\,2005bf \citep{Folatelli2006}, SN\,2004aw \citep{Taubenberger2006}, SN\,2009jf \citep{Valenti2011}, SN\,2007gr \citep{Hunter2009}, SN\,2008ax \citep{Pastorello2008}, SN\,2008D \citep{Soderberg2008}, iPTF\,13bvn \citep{Fremling2014,Cao2013}.  Photometry for all objects was converted to the AB system zeropoint before plotting. \label{fig:Uband}}
\end{center}
\end{figure}

\begin{figure}[!ht]
\begin{center}
\includegraphics[width=\columnwidth]{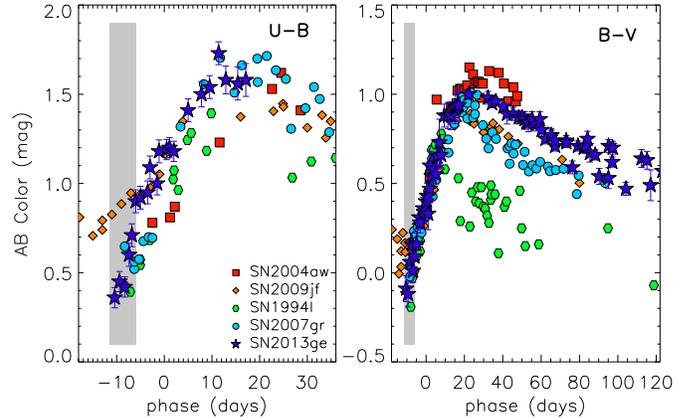}
\caption{u$-$b and B$-$V color evolution of SN\,2013ge in comparison to other Type Ib/c SN. Shaded regions indicate the time of the first light curve component.  During the declining portion of the first light curve component ($-$8 to $-$5 days) the u$-$b color reddens drastically.  The B$-$V evolution of SN\,2013ge is very similar to other Type Ib/c SN. Photometry for all objects was converted to the AB system zeropoint before plotting.  SN\,2013ge data contains u$-$b and b$-$v observations from \emph{Swift}-UVOT as well as Bessell B$-$V data from CAO and FLWO. All literature objects were observed with Bessell filters (see Figure~\ref{fig:Uband} for references). \label{fig:ColorEv}}
\end{center}
\end{figure}

The BVRI light curves are characterized by a smooth rise to maximum light, followed by an initial decline and then a shallowing of the slope between 20 and 30 days post maximum light.  Linear fits to the BVRI light curves between 60 and 120 days post-maximum reveal linear decline rates between 0.01 and 0.02 mag day$^{-1}$.  This light curve morphology is typical for Type I SN. 

\subsection{Early UV Light Curves}

In contrast, the u$-$ and w1$-$band light curves observed for SN\,2013ge show a distinctly non-standard evolution.  They display an early ``bump'' which is characterized by a rapid rise and decline over the first week of observations, before plateauing and then falling off rapidly again.  In the top panel of Figure~\ref{fig:Uband} we demonstrate that the u$-$band light curve can be decomposed into two components: a main component which peaks $\sim$11 days after discovery superimposed with an early peak of emission which rises rapidly over $\sim$4 days and declines rapidly after $\sim$1 week.  This particular decomposition is for illustrative purposes only, as we do not constrain the rising behavior of the second component.  Vertical dashed lines mark our first three epochs of spectroscopy, which probe the early rising portion of the first component.

Although well sampled u$-$band light curves between $-$14 and $-$7 days are still quite rare for Type I SN, no previous object has shown two distinct components with these timescales.  This is demonstrated in the lower panel of Figure~\ref{fig:Uband} where we compile u$-$band light curves from the literature.  A majority of the events peak $\sim$4 days prior to V$-$band maximum and then decline rapidly, consistent with our inferred time of maximum and subsequent evolution for the \emph{second} component in SN\,2013ge.  

Prior to this maximum, the evolution of the literature events is varied.  Both SN\,2008D and SN\,2005bf also show double-peaked u$-$band light curves, but with timescales significantly different than those observed in SN\,2013ge. The first peak in the light curve of SN\,2008D rises on a timescale of $\sim$1 day from the observed X-ray flash, while the first light curve component of the peculiar Type Ib SN\,2005bf has a rise time $\gtrsim$13 days. The early emission from SN\,2008D has been interpreted as both adiabatic cooling emission and as emission associated with a double-peaked $^{56}$Ni distribution. In contrast, in the model of \citet{Maeda2007}, the first peak of SN\,2005bf is powered by the radioactive decay of $^{56}$Ni while the second peak requires an additional power source such as a central engine. Intriguingly, the U$-$band light curve of the Type Ib SN\,2009jf shows a plateau which could be consistent with two components similar to those in SN\,2013ge if the luminosities were comparable.

Finally, we note that the w1$-$band light curve of SN\,2013ge shows tentative evidence for multiple changes in slope over $\sim$30 days. However, additional observations would be necessary to discern if the behavior between $+$20 and $+$30 days is due to a change in input power source or variations in spectral features/line blanketing.

\subsection{Color Evolution and SEDs}

In Figure~\ref{fig:ColorEv} we plot the U$-$B and B$-$V colors of SN\,2013ge and several other Type Ib/c SN. The grey shaded region highlights the time of the u$-$band excess described above.  During these epochs, the u$-$b color of SN\,2013ge is relatively flat for 3 days before rapidly reddening from $\sim$0.4 mag to $\sim$0.9 mag over 4 days. The $B-V$ color evolution of SN\,2013ge is similar to other Type Ib/c SN.  

\begin{figure}[!ht]
\begin{center}
\includegraphics[width=\columnwidth]{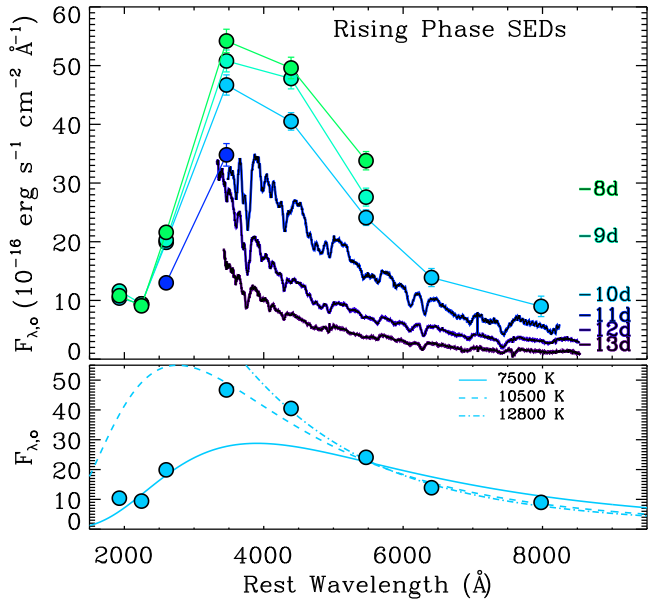}
\vspace{0.05in}
\caption{\emph{Top:} Spectra and broadband photometry obtained on the rising phase of the first light curve component.  Our earliest spectra, obtained before broadband photometry, show a steep blue continuum with narrow superimposed spectral features.  Our first epoch of \emph{Swift}-UVOT photometry ($-$11 days) constrains the peak of the SED to be around 3500 \AA.  Early spectra have been multiplied by a constant for clarity. \emph{Bottom:} The $-$10 day SED with superimposed blackbody fits.  It is not possible to fit both the slope of the SED in the optical and the depressed UV flux with a single temperature blackbody, indicating that line blanketing may be important even at these early epochs.  \label{fig:SED}}
\end{center}
\end{figure}

In Figure~\ref{fig:SED} we plot the spectral energy distribution (SED) of SN\,2013ge for epochs on the rising phase of the initial u$-$band peak (including our first three epochs of spectroscopy).  The first spectroscopic observations are characterized by a very blue continuum, while from day $-$11 onward the \emph{Swift}-UVOT photometry shows that the SED falls off significantly in the UV$-$bands.  As shown in the lower panel of Figure~\ref{fig:SED}, it is not possible to fit both the slope of the optical SED and the low flux level in the UV$-$bands with a single blackbody.  Such behavior may indicate that line blanketing is significantly depressing the UV$-$flux, even at these very early epochs.  

\begin{figure}[!ht]
\begin{center}
\includegraphics[width=\columnwidth]{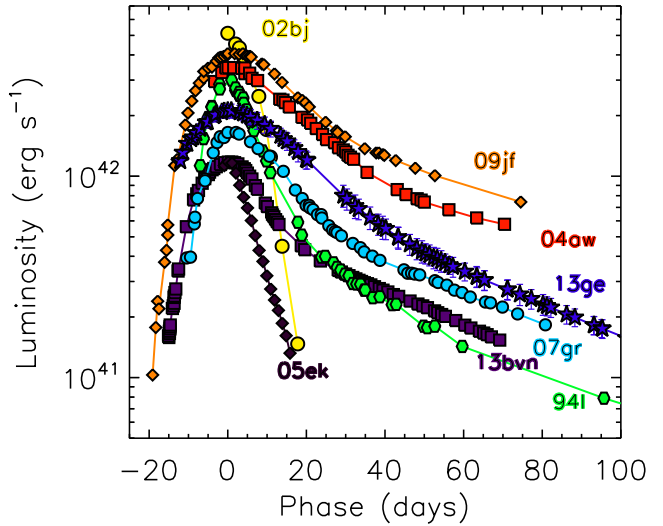}
\caption{Pseudo-bolometric Light Curve of SN\,2013ge in comparison to other Type Ib/c SN.  SN\,2013ge is relatively faint and slowly evolving.  Two distinct peaks are not visible in the pseudo-bolometric light curve, but the morphology of the early UV light curves results in a ``shoulder'' of excess emission at early times. References for comparison objects: SN\,2005ek \citep{Drout2013}, SN\,2002bj \citep{Poznanski2010}, SN\,2004aw \citep{Taubenberger2006}, SN\,2009jf \citep{Valenti2011}, SN\,2007gr \citep{Hunter2009}, SN\,1994I \citep{Richmond1996}, iPTF13bvn \citep{Fremling2014}. U-band photometry was available for all comparison objects with the exception of SN\,2002bj.  When NIR photometry was available, UBVRIJHK data was summed to produce the comparison light curves. When NIR photometry was not available, a black body tail was added to an integration of the UBVRI data. The comparison light curves shown here do not account for UV flux emitted blueward of the U-band. 
\label{fig:Bolo}}
\end{center}
\end{figure}

Fitting blackbodies to the BVRI$-$bands only we find color temperatures which fall from around 13,000 K at $-$12 days to around 7000 K at V$-$band maximum. The precise temperature evolution during the initial u-band peak is difficult to assess due to (1) the depressed UV flux and (2) our lack of dense R$-$ and I-band photometry at the earliest epochs. 

\subsection{Pseudo-Bolometric Light Curve}

To create a pseudo-bolometric light curve, we first sum our UV-optical photometry by means of a trapezoidal integration.  To account for missing IR flux, we attach a blackbody tail from the best-fit blackbody to the BVRI data.  This method of accounting for IR flux is equivalent to adding an IR contribution which monotonically increases from 15\% at early times to $\sim$50\% at late times.  This is in line with what is found for Type I SN with well-observed IR light curves (see, e.g., \citealt{Valenti2008,Lyman2014}). To account for UV contributions at later epochs (beyond our \emph{Swift}-UVOT coverage) we extrapolate the observed w2, m2, w1, and u$-$band light curves until they account for less than 0.5\% of the pseudo-bolometric luminosity, at which point they are dropped from the integration. We note that the flux contained in the w2/m2$-$bands and w1/u$-$bands accounts for $\lesssim$ 1\% and only a few \% of the bolometric luminosity at the epoch of their final observed data points, respectively. Thus, ambiguity in the morphology of the late-time UV light curve should not significantly impact the derived light curve.

The resulting pseudo-bolometric light curve is shown in Figure~\ref{fig:Bolo}, along with the pseudo-bolometric light curves of other SN.  Although two distinct peaks are not evident, an excess ``shoulder'' of emission can be seen at early times corresponding to the first u$-$band component described above. The bolometric light curve peaks at 2.1 $\pm$ 0.1 $\times$ 10$^{42}$ erg s$^{-1}$ and the total radiated energy between $-$12 and $+$120 days is 8.1 $\pm$ 0.3 $\times$ 10$^{48}$~erg.  

\begin{figure}[!ht]
\begin{center}
\includegraphics[width=\columnwidth]{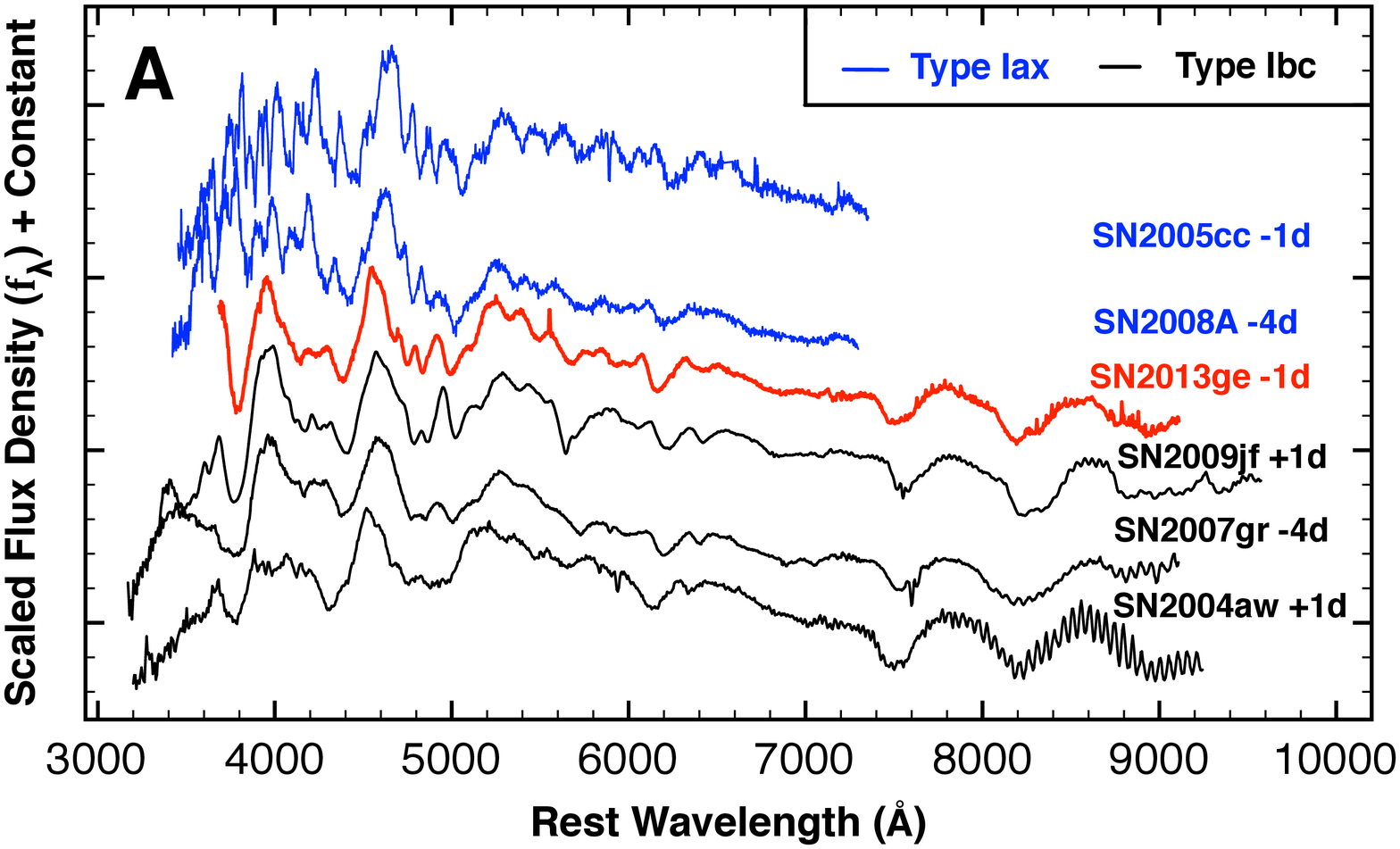}
\includegraphics[width=\columnwidth]{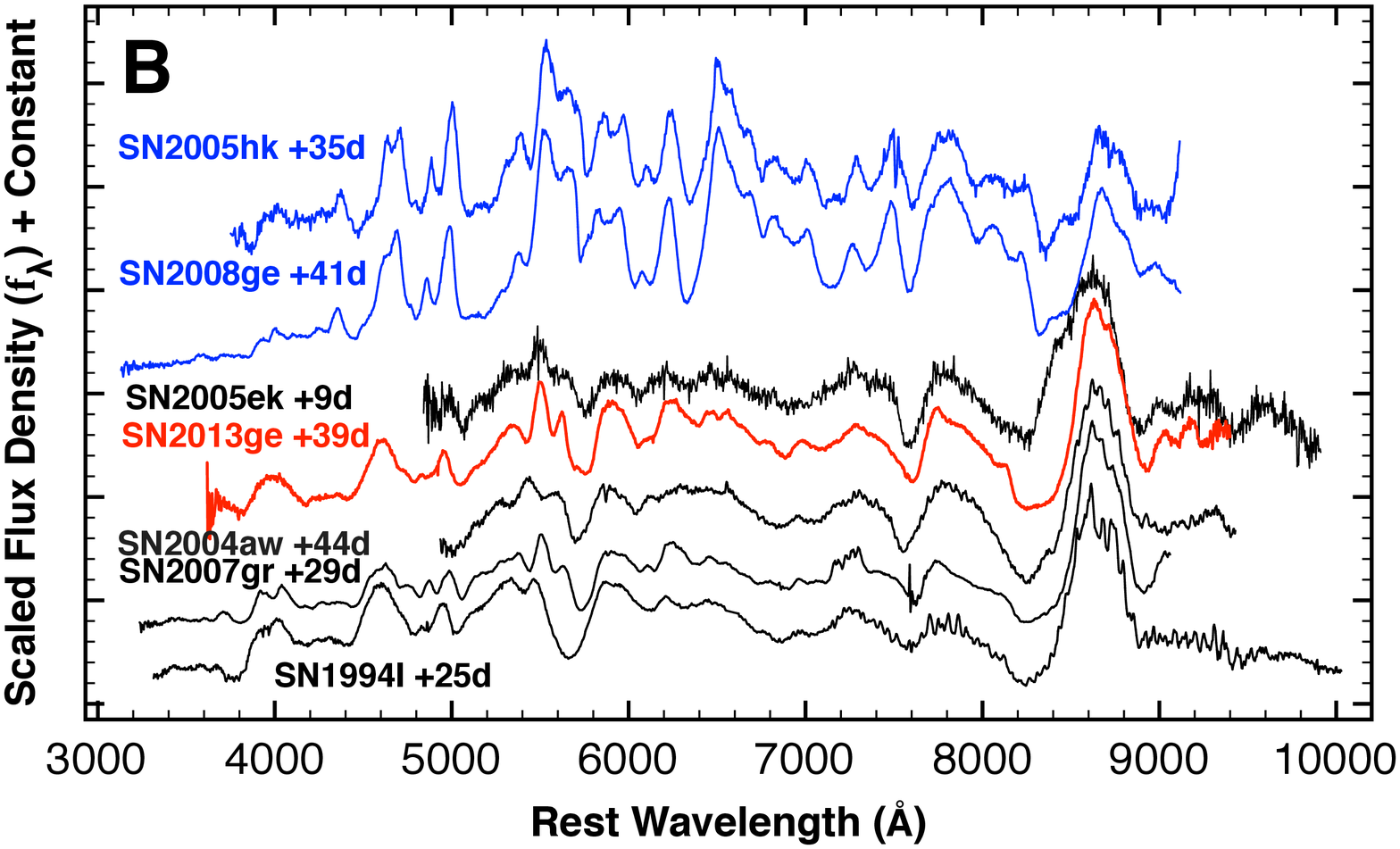}
\includegraphics[width=\columnwidth]{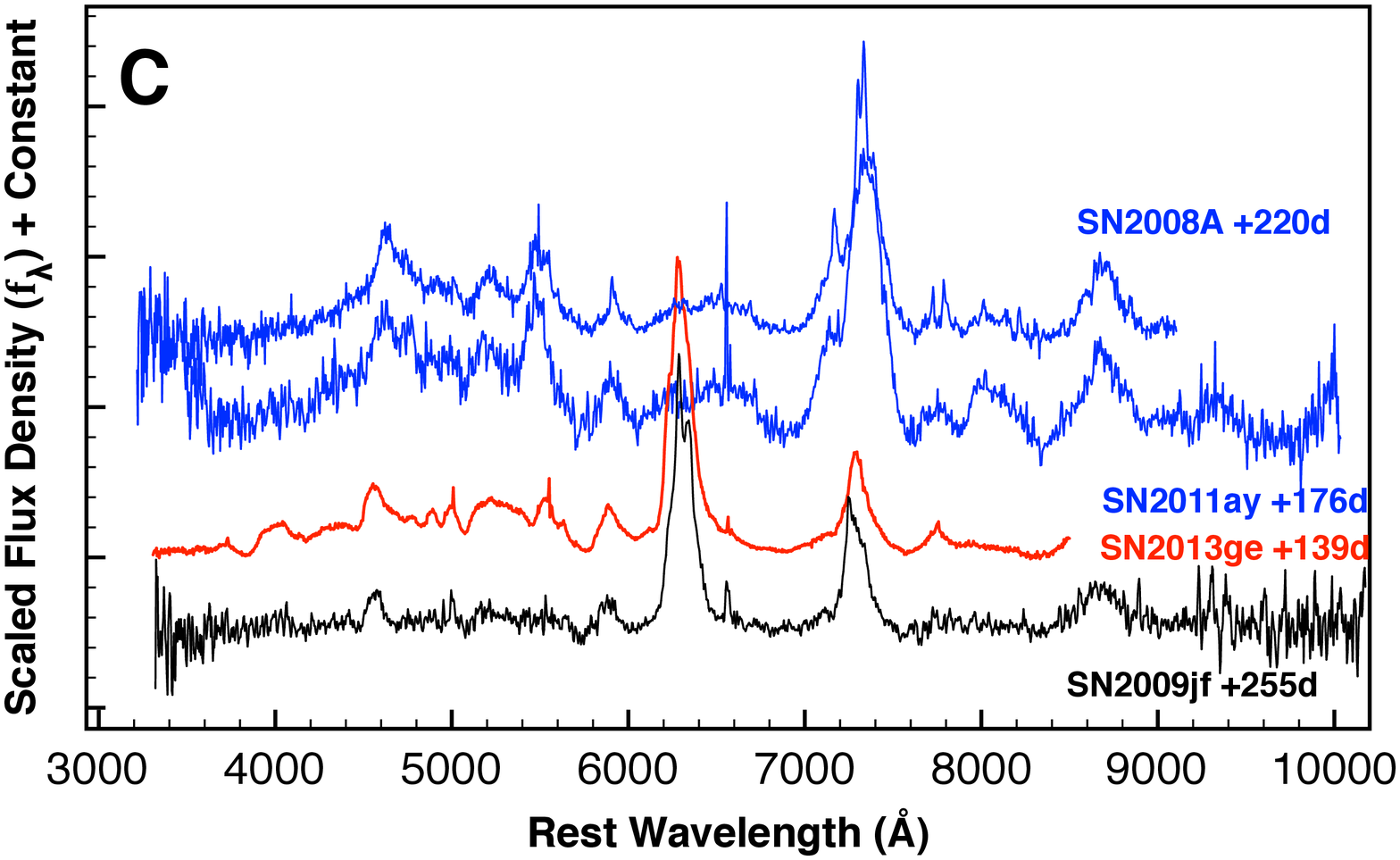}
\caption{Comparison of the spectrum SN\,2013ge at maximum light (panel A), intermediate phases (panel B) and nebular epochs (panel C) to spectra of Type Ib/c SN (black) and Type Iax SN (blue).  Although there is significant overlap between both classes of objects at maximum light, at intermediate and late phases the evolution of SN\,2013ge is similar to Type Ib/c SN. \emph{References:} SN\,2005cc and SN\,2008A \citep{Blondin2012}, SN\,2005hk \citep{Phillips2007}, SN\,2008ge \citep{Silverman2012}, SN\,2011ay \citep{Foley2013a}, SN\,2009jf \citep{Valenti2011}, SN\,2007gr \citep{Valenti2008}, SN\,2004aw \citep{Taubenberger2006}, SN\,2005ek \citep{Drout2013}, SN\,1994I \citep{Filippenko1995}. \label{fig:SpecEvol}}
\end{center}
\end{figure}

Assuming SN\,2013ge is powered mainly by the radioactive decay of $^{56}$Ni, we use the analytic models of \citet{Arnett1982} and \citet{Valenti2008}, with the corrections of \citet{Wheeler2015}, to extract estimates of the explosion parameters from this pseudo-bolometric light curve.  We assume that from $-$8 to $+$20 days SN\,2013ge is in the optically thick photospheric phase (we neglect the earliest data, during the ``UV bump'', when performing our fit), and utilize a constant opacity of 0.07 cm$^{2}$ g$^{-1}$ (corresponding to the case of electron scattering).  Under these assumptions we find best-fit explosion parameters of M$_{\rm{Ni}}$ $\approx$ 0.12 M$_\odot$, M$_{\rm{ej}}$ $=$ 2 $-$ 3 M$_\odot$, and E$_{\rm{K}}$ $=$ 1 $-$ 2 $\times$ 10$^{51}$ erg.  We have assumed a photospheric velocity at maximum of 10,000 $-$ 11,000 km s$^{-1}$ in order to break the degeneracy between M$_{\rm{ej}}$ and E$_{\rm{K}}$ (see Section~\ref{sec:spec}).    

\begin{figure}[!ht]
\begin{center}
\includegraphics[width=\columnwidth]{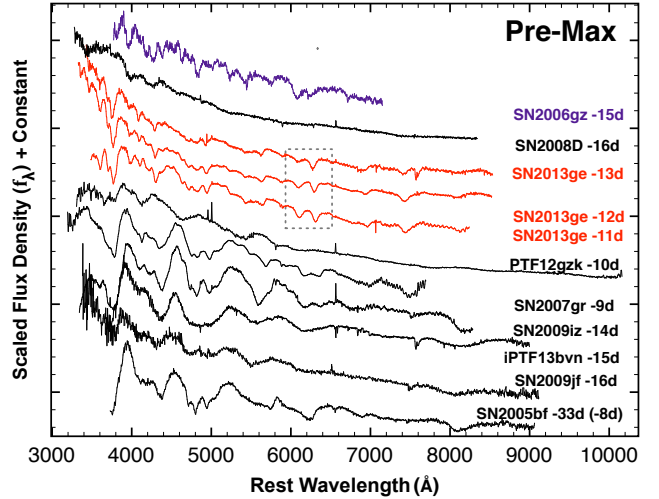}
\caption{Very early spectra of SN\,2013ge (red) compared to other Type Ib/c SN (black) and one ``super-Chandra'' Type Ia SN (purple).  The spectra of SN\,2013ge are characterized by blue continua and \emph{narrow} spectroscopic features.  In contrast, most other Type Ib/c SN with spectra at similar epochs show very broad features.  One exception is the Type Ib SN\,2005bf which during its first light curve peak showed both broad underlying features and narrow high-velocity features of \ion{Fe}{2} and \ion{Ca}{2} (phases are listed for SN\,2005bf with respect to \emph{both} light curve peaks).  Interestingly, the ``super-Chandra'' Type Ia SN\,2006gz also showed narrow spectroscopic features with blue-shifted absorption minima---similar to SN\,2013ge---at early times. The dashed grey box highlights two features in the spectra of SN\,2013ge whose ratio shows significant evolution over $\sim$2 days. \emph{References:} SN\,2006gz \citep{Hicken2007}, SN\,2008D \citep{Modjaz2009}, PTF\,12gzk \citep{Ben-Ami2012}, SN\,2007gr \citep{Valenti2008}, SN\,2009iz \citep{Modjaz2014}, iPTF\,13bvn \citep{Milisavljevic2013c}, SN\,2009jf \citep{Sahu2011}, SN\,2005bf \citep{Folatelli2006}. \label{fig:SpecEarly}}
\end{center}
\end{figure}

Inserting these best-fit parameters from the photospheric phase in the nebular model from \citet{Valenti2008} we find a predicted luminosity which is relatively consistent with our late-time ($>$ 60 days post-maximum) pseudo-bolometric light curve, although the predicted model declines more rapidly than the observed light curve.  Such a trend has been noted in the literature when attempting to model Type Ib/c SN with single zone models (see, e.g., \citealt{Maeda2003} and  \citealt{Valenti2008}), and may indicate that SN\,2013ge could be better described by a two-zone model with a high-density inner region and a low-density outer region.  Alternatively, this could be evidence for an asymmetry in the explosion.

\section{Spectroscopic Properties}\label{sec:spec}

\subsection{Type Ib/c versus Type Iax}

 SN\,2013ge can be immediately identified as a Type I SN since it lacks \emph{conspicuous} hydrogen emission, and can further be sub-classified as a Type Ib/c from the lack of a strong \ion{Si}{2}~$\lambda$6355 feature \citep{Wheeler1995}.  Near maximum light, the spectral features are relatively narrow and resemble maximum light spectra of both normal (not broad-lined) Type Ib/c SN and Type Iax SN \citep{Foley2013a}.  In order to resolve this degeneracy, in Figure~\ref{fig:SpecEvol} we compare spectra of SN\,2013ge (red) to spectra of several normal Type Ib/c SN (black) and Type Iax SN (blue) at three phases: maximum (panel A), intermediate/transitional (panel B) and nebular (panel C).

Taken in conjunction, it is clear that from maximum light onwards SN\,2013ge follows a spectral evolution typical for normal Type Ib/c SN.  Near maximum light, iron peak elements are visible at bluer wavelengths as well as \ion{O}{1} and the \ion{Ca}{2} NIR triplet in the red.  By $+$39 days, SN\,2013ge has entered a transitional phase, marked by the onset of increased emission in the \ion{Ca}{2} NIR triplet.  This growth of the \ion{Ca}{2} NIR feature has been observed in numerous Type Ib/c SN.  Finally, at nebular phases SN\,2013ge is dominated by emission from forbidden transitions of intermediate mass elements such as [O~I], [Ca~II], and Mg~I].

This is in stark contrast to Type Iax SN, which do not show growth of the \ion{Ca}{2} NIR triplet at intermediate phases and whose late-time spectra are dominated by forbidden \ion{Ca}{2} and \ion{Fe}{2} lines with no obvious contribution from [O~I].  However, it is worth emphasizing the strong similarity between some Type Iax spectra and normal Type Ic spectra at maximum light---especially in the wavelength range 4200 $-$ 7500 \AA.  The main differentiating spectral features seem to be at wavelengths shorter than 4200 \AA, where SN Iax show a plethora of lines which are not evident in normal Type Ib/c SN.  Caution should be taken in classifying a Type Iax SN from a single maximum light spectrum.

\subsection{Early Spectra}

Despite evolving similarly to normal Type Ic SN from maximum onward, the early spectra of SN\,2013ge are unusual for Type Ib/c SN.  Our earliest three spectra, obtained between $-$13 and $-$11 days, are characterized by a blue continuum, superimposed with relatively shallow and \emph{narrow} features (FWHM $\lesssim$ 3500 km s$^{-1}$). These spectra were obtained during the early rising phase of the first u$-$band component described in Section~\ref{Sec:Photom}.  Spectroscopic observations of Type Ib/c SN at such early epochs are rare, but a majority of events show \emph{broad} high-velocity features at early times, which narrow as the photospheric velocity deceases.  Indeed, these spectra vary significantly from the earliest spectra obtained for SN\,2008D, SN\,2009jf, PTF\,12gzk, and iPTF\,13bvn, all of which showed broad spectral features at similar epochs (see Figure~\ref{fig:SpecEarly}).  

During its first light curve peak, the spectrum of the peculiar Type Ib SN\,2005bf did show several narrow spectroscopic features between 4500 \AA\ and 5300 \AA\ (see Figure~\ref{fig:SpecEarly}; phases are given with respect to both the first and second peak). However, features in the rest of the spectrum were broader and more comparable to the other Type Ib/c SN shown in Figure~\ref{fig:SpecEarly}.  \citet{Folatelli2006} interpret this as the combination of an underlying photosphere (broad) and high-velocity iron and calcium features (narrow).  In contrast, in the early spectra of SN\,2013ge, \emph{all} of the features observed have similarly narrow widths.  Intriguingly, the ``super-Chandra'' Type Ia SN\,2006gz \citep{Hicken2007} also showed narrow spectroscopic features with blue-shifted absorption minima similar to SN\,2013ge at early times. However, by maximum light SN\,2006gz developed a very prominent \ion{Si}{2}~$\lambda$6355 feature, indicative of Type Ia SN, which was not observed in SN\,2013ge.

In addition, we observe rapid evolution in both the slope of the continuum and photospheric velocity over the two days in which these early spectra of SN\,2013ge were obtained.  Two absorption features between 6000 and 6500 \AA\ also undergo a large change in their ratio over this time period.  These features are highlighted by a grey box in Figure~\ref{fig:SpecEarly}.

\begin{figure}[!ht]
\begin{center}
\includegraphics[width=\columnwidth]{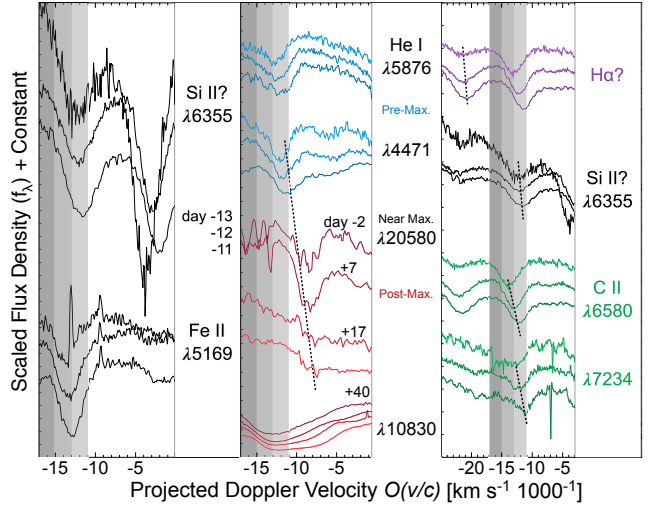}
\caption{Projected Doppler velocities around key features in the spectra of SN\,2013ge. These are used to assess the presence of unburned material (hydrogen, helium, and carbon). Grey regions mark particular velocities and are placed to guide the eye. Dashed lines highlight the velocity evolution of certain features with time. \emph{Left:} The regions around \ion{Si}{2} $\lambda$6355 and \ion{Fe}{2} $\lambda$5169 in the $-$13, $-$12, and $-$11 day spectra.  \emph{Middle:} Assessing the presence of helium in the early optical spectra (top; blue) and maximum light NIR spectra (bottom; red).  See text for details.  \emph{Right:} Assessing the degeneracy between H$\alpha$, \ion{Si}{2} $\lambda$6355, and the presence of \ion{C}{2} $\lambda$6580.  See text for details.\label{fig:13geDop}}
\end{center}
\end{figure}

\subsection{Assessing the Presence of Unburned Material}

For many models, fully stripping the H/He layers from putative Type Ib/c progenitors has proved challenging. This makes possible identifications of trace amounts of these elements in Type~Ib/c SN both important and long debated (see \citealt{Parrent14} for a review). Unfortunately, the identification of such contaminants is particularly complicated in the optical photospheric spectra of Type Ib/c SN where a number of degeneracies exist, most notably between \ion{H}{1}~$\lambda$6563, \ion{C}{2}~$\lambda$6580, and \ion{Si}{2}~$\lambda$6355, and \ion{He}{1}~$\lambda$5876 and \ion{Na}{1}~$\lambda$5889.  These degeneracies can be partially alleviated if NIR spectra are available.

Early phase spectra probe the outermost regions of the ejecta where unburned material such as hydrogen and helium are most likely to be present, if at all. We therefore begin by examining our earliest three optical spectra (days $-$13, $-$12, $-$11) as well as our IR spectra at later epochs. In Figure~\ref{fig:13geDop} we shift the spectra from rest frame wavelength to the projected Doppler velocities of commonly found lines in order to examine preliminary matches between spectral features and dominant candidate lines of hydrogen, helium, and carbon.  Each panel will be discussed in turn, below.

\subsubsection{Hydrogen}

When we associate lines with \ion{Si}{2}~$\lambda$6355 and \ion{Fe}{2} in the left panel of Figure~\ref{fig:13geDop}, we see that the velocity of the absorption minima is $\sim$14,000 km~s$^{-1}$ (grey vertical bands are placed to guide the eye, and correspond the the same velocities in each panel). At the top of the right panel of Figure~\ref{fig:13geDop} we show the spectral region around H$\alpha$ (purple).  While the feature observed at $\sim$6300 \AA\ appears at similar velocities to \ion{Fe}{2} when interpreted as H$\alpha$, we find that the velocity overlap and implied velocity evolution is more consistent with the interpretation of \ion{C}{2}~$\lambda$6580 (see below for further discussion of carbon). However, this does not preclude very weak contributions from H$\alpha$ to the observed feature.

We next investigate the possibility of high-velocity H$\alpha$ in these early spectra.  In particular, we note that the feature most often associated with \ion{Si}{2}~$\lambda$6355 appears at slightly lower velocities than \ion{Fe}{2}, which would seem to suggest that photons from deeper layers are escaping opaque resonance line regions of iron.  This model-independent discrepancy between the line velocities of \ion{Si}{2} and \ion{Fe}{2} absorption minima is frequently encountered for Type Ib/c SN (e.g. \citealt{Branch06}; \citealt{Elmhamdi07}; \citealt{Milisavljevic2015}; see \citealt{Parrent2015} for a comprehensive examination), and may indicate that another ion is contributing to the observed line(s).  In core-collapse SN, a candidate line in the wavelength region around \ion{Si}{2}~$\lambda$6355 is high-velocity H$\alpha$ \citep{Wheeler1995,Benetti11,Parrent2015}. In the top of the right panel of Figure~\ref{fig:13geDop} we show the velocities of \ion{Si}{2} (black) versus \ion{H}{1} (purple) which would be necessary to overlap with the feature observed at $\sim$ 6100 \AA\ (dashed lines indicate the relevant feature in each case). If H$\alpha$ contributes, it requires velocities in excess of 20,000 km s$^{-1}$.  Unfortunately, we lack certain detections of either \ion{Si}{2}~$\lambda$5972 or H$\beta$ in the early spectra of SN\,2013ge, and therefore we cannot definitely confirm the presence or relative contributions of \ion{H}{1} versus \ion{Si}{2} to the observed feature. 

\subsubsection{Helium}

In the middle panel of Figure~\ref{fig:13geDop}, we examine the case for \ion{He}{1} in the spectra of SN\,2013ge. In the top portion of the panel, we show the earliest optical spectra in the region of \ion{He}{1}~$\lambda\lambda$5876,~4471 and find evidence for both lines at  velocities which are consistent with the composite \ion{Fe}{2} feature. A slight notch is also present at $\sim$6400~\AA, consistent with \ion{He}{1}~$\lambda$6678 at a similar velocity (see Figure~\ref{fig:13geLines}).

In the lower portion of the panel, we examine four NIR spectra ($-$2 to $+$40 days; red) and focus on the features near 1$\mu$m and 2$\mu$m. The large 1$\mu$m feature is significantly broader than the 2$\mu$m feature, and is often attributed to a blend of multiple ions including \ion{C}{1} and \ion{Si}{1} \citep{Millard1999,Taubenberger2006}. We detect a feature near 2$\mu$m, which is consistent with faint \ion{He}{1} at $\sim$8000$-$9000 km s$^{-1}$. This is comparable to the photospheric velocities at these epochs (the dotted line is placed to guide the eye between the relevant features). Thus, we find that there is room for the existence of \ion{He}{1} $\lambda\lambda$4471, 5876, 6678, 10830, and 20580 in the spectra of SN\,2013ge. In addition, the identification of weak optical signatures of \ion{He}{1} is consistent with either weak or absent signatures of \ion{He}{1}~$\lambda$6678 and 7065 \citep{Branch02,Hachinger2012}. 

We therefore conclude that our data is consistent with the identification of weak \ion{He}{1} in the early spectra of SN\,2013ge. We emphasize that these lines are weak and fade with time; they are distinct from the strong He I features observed in many SN classified as Type Ib, which are strongest a few weeks past maximum \citep[e.g.][]{Filippenko1997,Liu2015}.  \emph{SN\,2013ge may therefore represent the detection of weak helium features in an event which would generally be classified as Type Ic.}  

\begin{figure}[!ht]
\begin{center}
\includegraphics[width=\columnwidth]{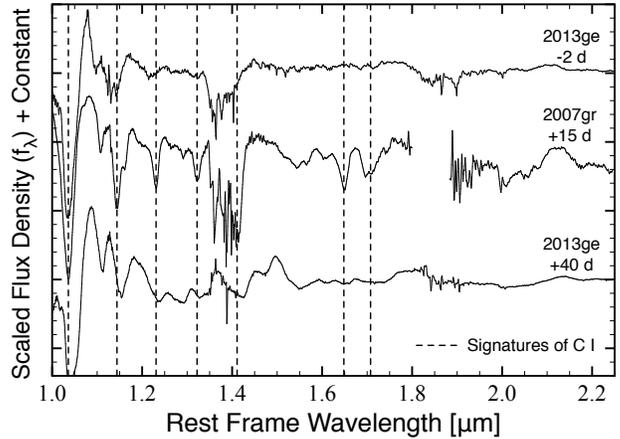}
\caption{A comparison between the NIR spectra of SN\,2007gr and SN\,2013ge. Vertical dashed lines denote \ion{C}{1} lines marked by Hunter et al. 2009 in their Fig.~12. Lines of a comparable strength are not observed in SN\,2013ge.  \label{fig:NIR_Hunter_CI}}
\end{center}
\end{figure}

\subsubsection{Carbon}

Finally, in the lower portion of the right panel of Figure~\ref{fig:13geDop} we identify signatures of \ion{C}{2}~$\lambda\lambda$6580, 7234 in the early spectra of SN\,2013ge (green). We also examine our NIR spectra for evidence of \ion{C}{1} features at later epochs (Figure~\ref{fig:NIR_Hunter_CI}). While we cannot rule out weak contamination from \ion{C}{1}, the influence must be substantially weaker than that observed in the NIR spectra of the ``carbon-rich'' SN\,2007gr \citep{Hunter2009}. 

\subsection{Species Identification and Velocity Estimation}

In order to further constrain the full set of ions present in the spectra of SN\,2013ge, as well as the evolution of the velocity of the line-forming region, we utilize the fast and publicly available spectral synthesis code \texttt{SYN++} \citep{Thomas2011}\footnote{This is an updated version of \texttt{SYNOW}; https://c3.lbl.gov/es/.}. Line formation in \texttt{SYN++} is assumed to be dominated by pure resonant scattering with Boltzmann statistics determining relative lines strengths for a given ion. We use an exponential optical depth profile that begins at the (sharp) photospheric velocity, which is set by hand for each epoch. See \citet{Parrent15} for general fitting methods and procedures.

\begin{figure}[!ht]
\begin{center}
\includegraphics[width=\columnwidth]{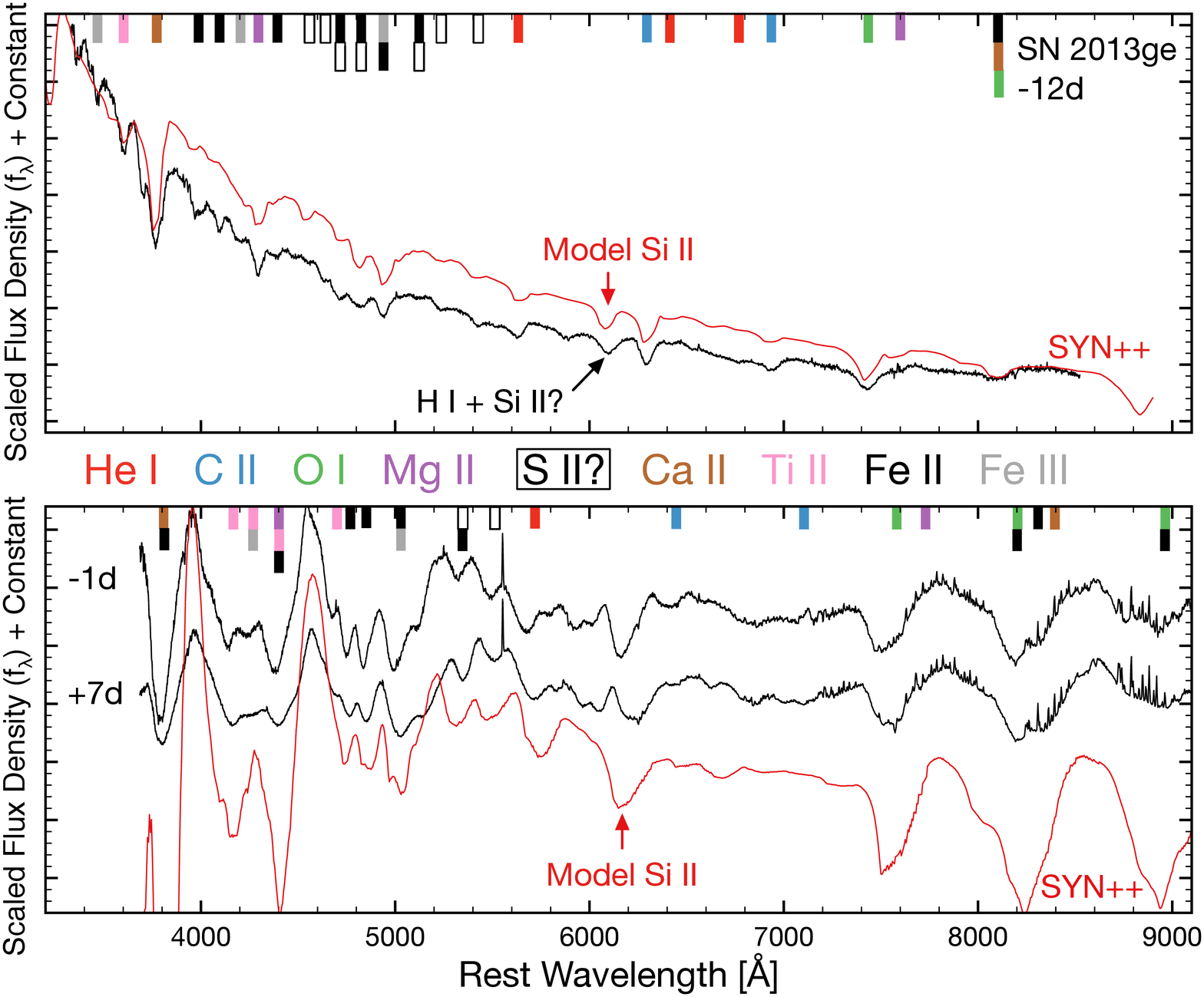}
\caption{Ion identifications in the early (top panel) and maximum light (bottom panel) spectra of SN\,2013ge. Observed spectra are shown in black while SYN++ models are shown in red.  Colored rectangles designate the ions that contribute to each feature in the model spectra, as labeled between the panels.  Species identified include \ion{He}{1}, \ion{C}{2}, and \ion{O}{1} as well as contributions from \ion{Mg}{2}, \ion{Si}{2}, \ion{Ca}{2}, \ion{Ti}{2}, \ion{Fe}{3}, and possibly \ion{S}{2} and \ion{Fe}{3}.  Between $-$12 days and maximum light the blue continuum is significantly depressed.  See text for more details.  \label{fig:13geLines}}
\end{center}
\end{figure}

In Figure~\ref{fig:13geLines} we present detailed line identifications inferred with \texttt{SYN++} and show example \texttt{SYN++} fits (red lines) for both the early phase and maximum light spectra. For the photospheric  spectra of SN\,2013ge, we tested a full list of ionization species (\ion{H}{1} through \ion{Fe}{3}). Colored ticks at the top of each panel mark the wavelengths where a given ion is believed to be influencing the spectrum. Main contributors to the model fits include \ion{He}{1}, \ion{C}{2}, and \ion{O}{1} as well as \ion{Mg}{2}, \ion{Si}{2}, \ion{Ca}{2}, \ion{Ti}{2}, \ion{Fe}{2}, and possible \ion{S}{2} and \ion{Fe}{3}.  In the \texttt{SYN++} models shown, only \ion{Si}{2} contributes to the feature observed at $\sim$6100 \AA\ (we have not included high-velocity hydrogen).

\begin{figure*}[!ht]
\begin{center}
\includegraphics[width=0.85\columnwidth]{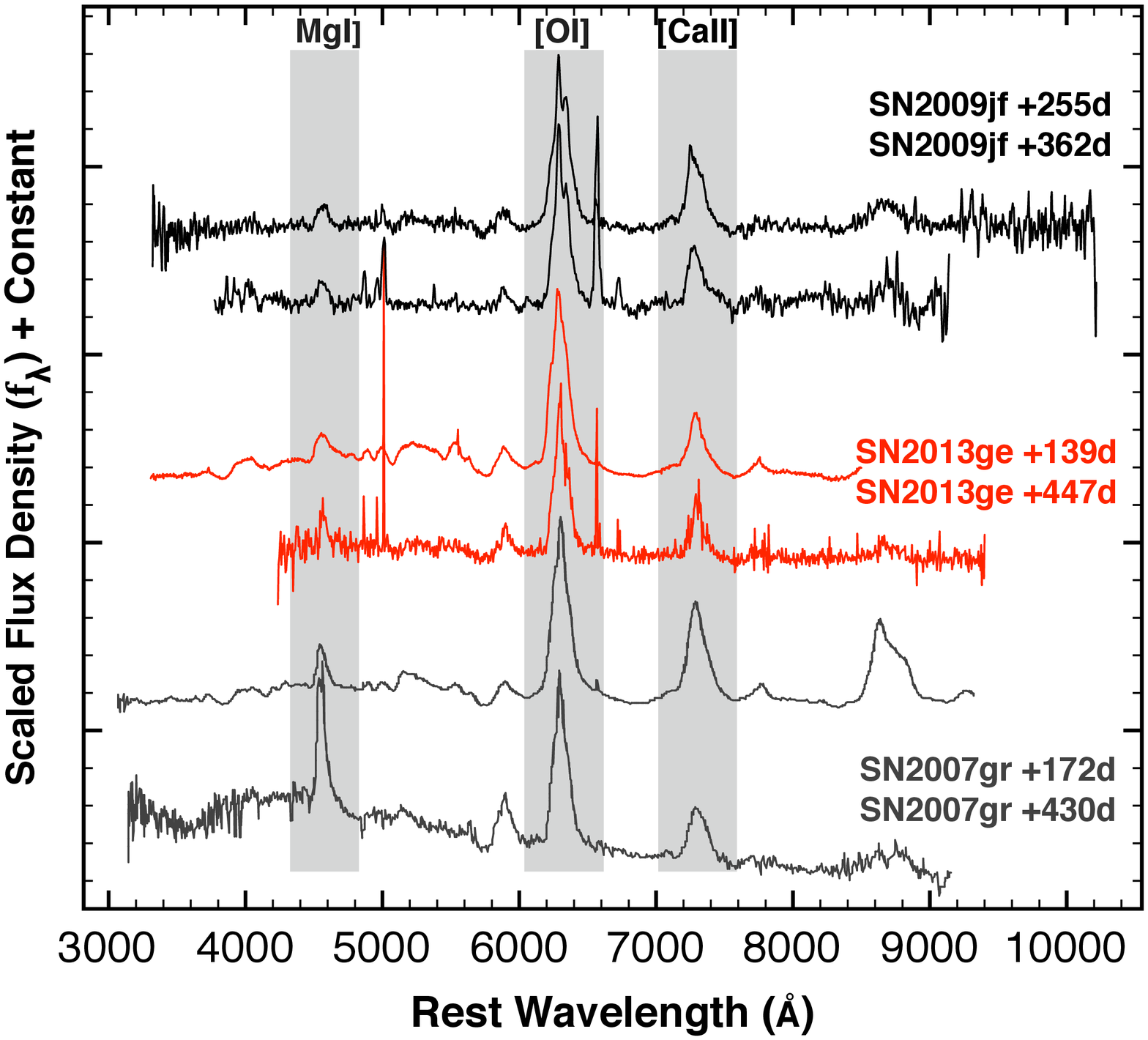}
\includegraphics[width=0.57\columnwidth]{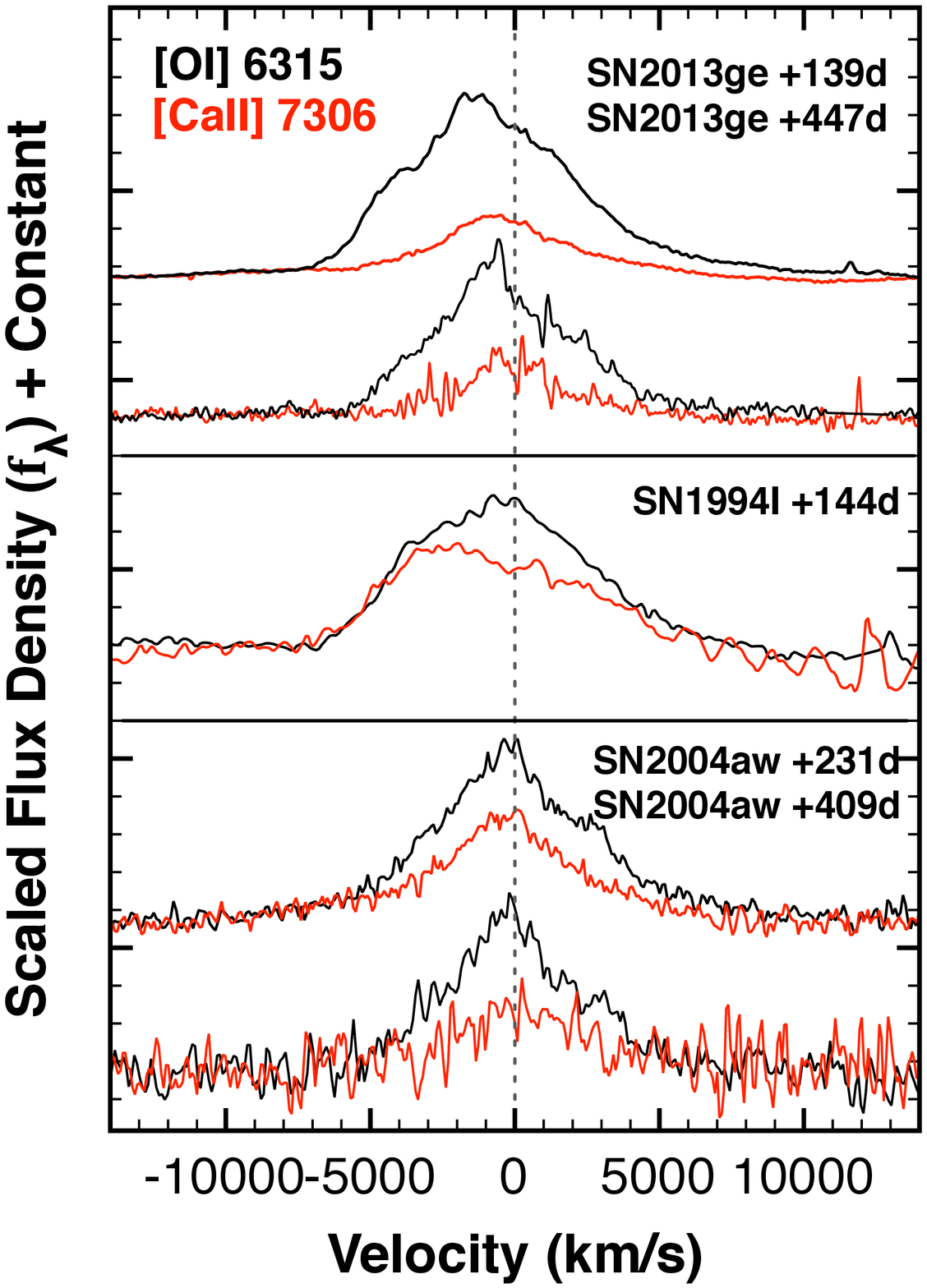}
\includegraphics[width=0.57\columnwidth]{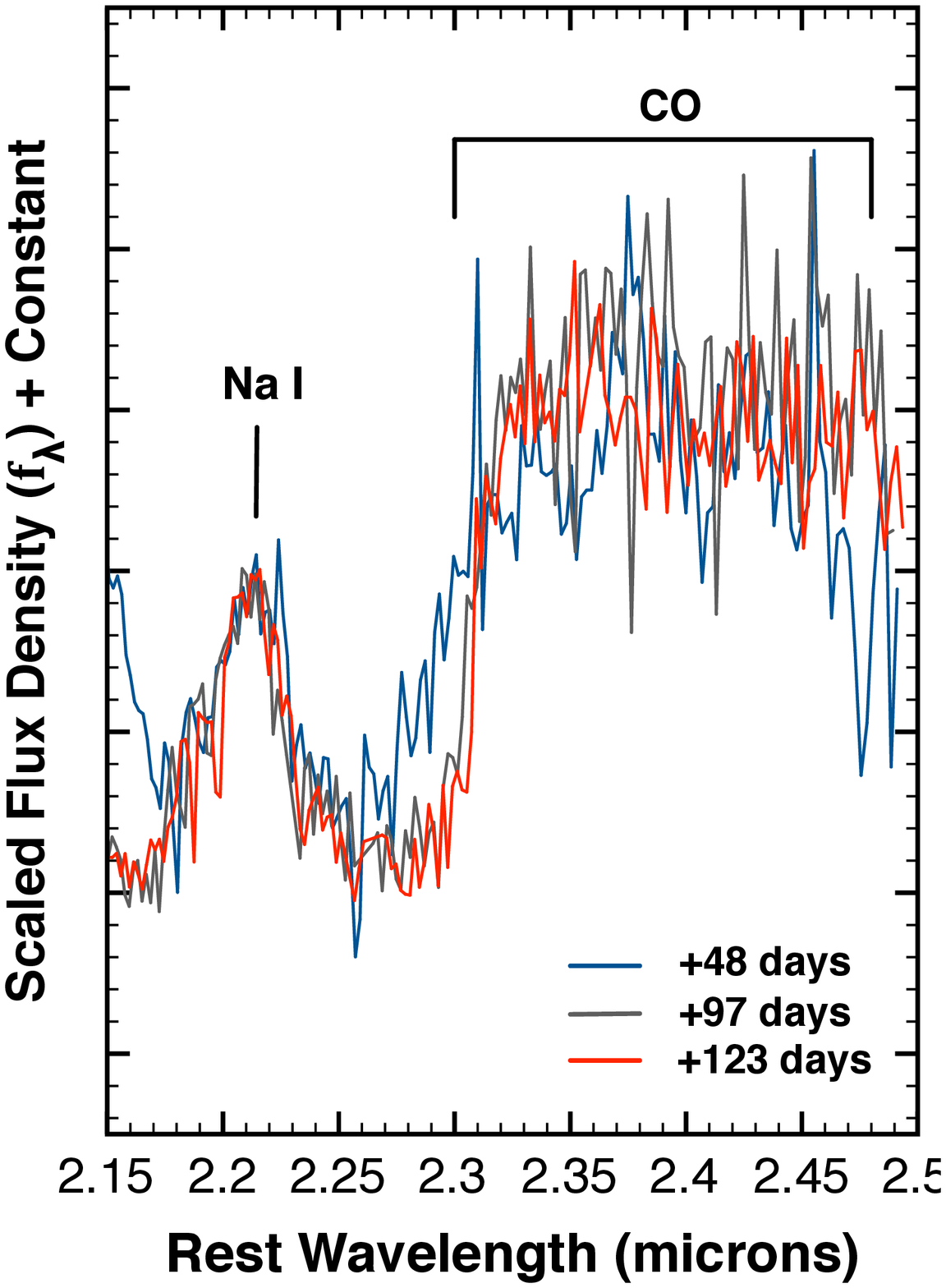}
\caption{\emph{Left: } A comparison between the nebular spectra of SN\,2013ge, SN\,2007gr, and SN\,2009jf at multiple epochs.  At $\sim$150$-$250 days, all have a Mg~I] $\lambda$4571 to [O~I] $\lambda \lambda$6300, 6364 ratios of $\sim$0.15.  However, by $\sim$450 days, the ratio in SN\,2007gr has grown substantially to $\sim$1, while it has remained relatively constant in both SN\,2013ge and SN\,2009jf. \emph{Center:} Nebular line profiles of SN\,2013ge in comparison to SN\,1994I and SN\,2004aw at a variety of epochs.  By late times the [O~I] feature in SN\,2013ge more closely resembles the peaked structure observed in SN\,2004aw.  \emph{Right:} The CO band-head observed in SN\,2013ge at three epochs.  We observe evolution of the profile between $+$48 and $+$97 days.  \emph{References:} SN\,2009jf \citep{Valenti2011}, SN\,2007gr \citep{Chen2014}, SN\,2004aw \citep{Taubenberger2006}, SN\,1994I \citep{Filippenko1995}. \label{fig:NebSpec}}
\end{center}
\end{figure*}

The velocity of the line-forming region is estimated to decrease from $\sim$15,000~km~s$^{-1}$ to $\sim$13,000~km~s$^{-1}$ between $-$13 days and $-$11 days. This is similar to the early velocity evolution observed in SN\,2008D \citep{Modjaz09}. In order to simultaneously reproduce the relatively high blueshift and narrow width of the features at these early epochs with \texttt{SYN++}, we set the minimum velocity parameter, \texttt{v$_{min}$}, for all of the ions to $\sim$15,000~km~s$^{-1}$ while the model photosphere, \texttt{v$_{phot}$}, was set to 10,000~km~s$^{-1}$. Within the semi-empirical parameter space of \texttt{SYN++}, this is equivalent to the process by which one typically adds ``detached'' high-velocity features to a spectrum. We emphasize that in this case this process was applied to \emph{all} of the ions and is not necessarily meant to suggest a region which is physically detached from a luminous source; the physical interpretation of these early spectra will be discussed in Section~\ref{Sec:Discuss}.

A similar set of ions is inferred for the maximum light spectrum, although significant evolution is observed. For instance, the spectrum is now significantly redder, a substantial \ion{Ti}{2} absorption trough has formed between 4000~\AA\ and 4500~\AA, and the strength of the \ion{He}{1} and \ion{C}{2} identified features has decreased significantly with respect to the 6250~\AA\ feature (which may be influenced by both \ion{Si}{2} $\lambda$6355 and trace amounts of hydrogen, as described above). In addition, the velocity of the line-forming region has decreased to $\sim$11,000~km~s$^{-1}$---typical for Type Ib/c SN at maximum light---and the width of the features is relatively well matched to the velocity of their absorption minima. Unfortunately, poor weather prohibited us from obtaining any spectra between $-$11 and $-$1 days. As a result, we were unable to observe the evolution from narrow high-velocity features in the early spectra toward moderate width and velocity features in the maximum light spectra.

\subsection{Nebular Spectra Analysis}\label{sec:neb}

In the left panel of Figure~\ref{fig:NebSpec} we plot the nebular spectra of SN\,2013ge in comparison to SN\,2007gr and SN\,2009jf.  The late-time spectra of SN\,2013ge show conspicuous features due to [O~I], [Ca~II], and Mg~I].  The flux contained in the [O~I] $\lambda \lambda$6300, 6364 feature is significantly larger than that in either [Ca~II] $\lambda \lambda$7291, 7324 or Mg~I] $\lambda$4571, with ratios of $\sim$0.3 and $\sim$0.12, respectively.  Notably, these ratios show very little evolution between $+$150 and $+$450 days. 

The [Ca~II]/[O~I] ratio may be an indicator of progenitor core mass, with lower values implying a larger core mass, although mixing can also play a role \citep{Fransson1989}.  The ratio measured in SN\,2013ge is on the low end of values observed in stripped core-collapse SN, comparable to that observed in SN\,2009jf.  

\citet{Hunter2009} examined the Mg~I]/[O~I] ratio for a large number of stripped core-collapse SN and the ratio observed in SN\,2013ge is on the extreme low end when compared to this sample. It is also notable for its lack of evolution; a majority of events show a Mg~I]/[O~I] ratio that grows with time. For example, in SN\,2007gr the Mg~I]/[O~I] ratio grows from $\sim$0.1 to $\sim$1 between 150 and 450 days post-maximum.  This growth is \emph{not} observed in either SN\,2013ge or SN\,2009jf (see Figure~\ref{fig:NebSpec}).  To explain a weak Mg~I] feature at late times in comparison to other SN we require that either the abundance of Mg produced in SN\,2013ge is lower, or that the Mg~I] line itself is suppressed due to other effects.  For instance, \citet{Hunter2009} invoke mixing in the ejecta to explain the decreasing Mg~I]/[O~I] trend observed in SN\,1998bw and SN\,2006aj.  In addition, in spectral models of Type IIb SN, \citet{Jerkstrand2015} find that the Mg~I] $\lambda$4571 line is very sensitive to clumping in the O-Ne-Mg layer, with denser clumps favoring brighter emission.  In this context, the small Mg~I]/[O~I] ratio observed in SN\,2013ge at late times could indicate a relative lack of high density enhancements due to clumping.

In Figure~\ref{fig:NebSpec} we plot the [O~I] $\lambda \lambda$6300, 6364 and [Ca~II] $\lambda \lambda$7291, 7324 profiles of SN\,2013ge in comparison to those of SN\,2004aw and SN\,1994I.  The shape of the profiles in the $+$139d spectrum of SN\,2013ge appears intermediate between the rounded profiles of SN\,1994I and the highly peaked profiles of SN\,2004aw, while at $+$447 days the spectrum of SN\,2013ge appears more heavily peaked, with slightly blue-shifted velocity. \citet{Mazzali2005} found that a sharp peak in the [O~I] feature is consistent with viewing a jet on axis, although we emphasize that a sharp ejecta density profile can also produce a peaked nebular profile in the absence of a large asymmetry. 

\subsection{The Detection of CO Emission}

We identify CO-band emission from the first overtone ($\Delta v = 2$) at $\sim$2.3$\mu$m in the NIR spectra of SN\,2013ge.  Although CO has been detected in the spectra of a number of Type II SN, SN\,2013ge is only the third Type Ib/c SN in the literature, to date, with reported molecular CO emission.  Previous identifications were reported in SN\,2007gr \citep{Hunter2009} and SN\,2000ew \citep{Gerardy2002}.  In the right panel of Figure~\ref{fig:NebSpec} we plot the spectral region between 2.1 $\mu$m and 2.5 $\mu$m for three epochs of SN\,2013ge NIR spectra ($+$48, $+$97, and $+$123 days). The CO band-head and \ion{Na}{1} are labeled. Only marginal emission above the continuum was present in our previous NIR spectrum, indicating significant growth of the CO emission between $+$40 and $+$48 days.

It is also evident that there is evolution of the emission profile between $+$48 days and $+$97 days.  The feature dramatically narrows, an effect that---to our knowledge---has not previously been observed.  The width of the emission feature observed in SN\,2007gr did not evolve over a similar range of epochs.  This behavior may be linked to different clumping/mixing properties in the ejecta of SN\,2013ge and SN\,2007gr, as evidenced by their nebular line ratios, above. Detailed modeling of the CO-band will be presented in Milisavljevic et al.\ (\emph{in prep}).

\section{Local Environment Properties}\label{Sec:Enviro}

\subsection{Host Galaxy Environment}

SN\,2013ge exploded on the outskirts of NGC\,3287, an SB(s)d galaxy, approximately 51 arcseconds north-east of the galaxy core.  This corresponds to an offset---normalized by the g$-$band half-light radius---of 2.33.  The north-east portion of NGC\,3287 is characterized by a large number of blue stellar knots (see Figure~\ref{fig:finder}).  In our 2015 Jan.\ 15 IMACS spectrum the SN flux had fallen enough to reveal an unresolved knot of star formation (in the form of narrow emission lines) at the SN position. 

We measured the fluxes of these emission lines at the SN explosion site using the MCMC method of \citet{Sanders2012a}.  The measured H$\alpha$ luminosity of $\sim$1.0 $\times$ 10$^{38}$ erg s$^{-1}$ leads to an explosion site star formation rate of $\sim$ 8.0 $\times$ 10$^{-4}$ M$_\odot$ yr$^{-1}$ \citep{Kennicutt1998}.  This value is on the low end of the HII regions associated with core-collapse SN studied by \citet{Crowther2012}.

Using the PP04N2 diagnostic \citep{Pettini2004}, we find an explosion site metallicity for SN\,2013ge of log(O/H) $+$ 12 $=$ 8.40 $\pm$ 0.05.  This value is approximately half-solar (assuming log(O/H)$_{\rm{solar}}$ $+$ 12 $=$ 8.69; \citealt{Asplund2005}) and does not deviate strongly from a metallicity measured with an SDSS spectrum taken near the galaxy core.   This places the host environment metallicity of SN\,2013ge in roughly the bottom 25\% of the distribution measured for Type Ib and Type Ic SN and the top 25\% measured for Type Ic-BL in \citet{Sanders2012a}.

\begin{figure*}[!ht]
\begin{center}
\includegraphics[width=\columnwidth]{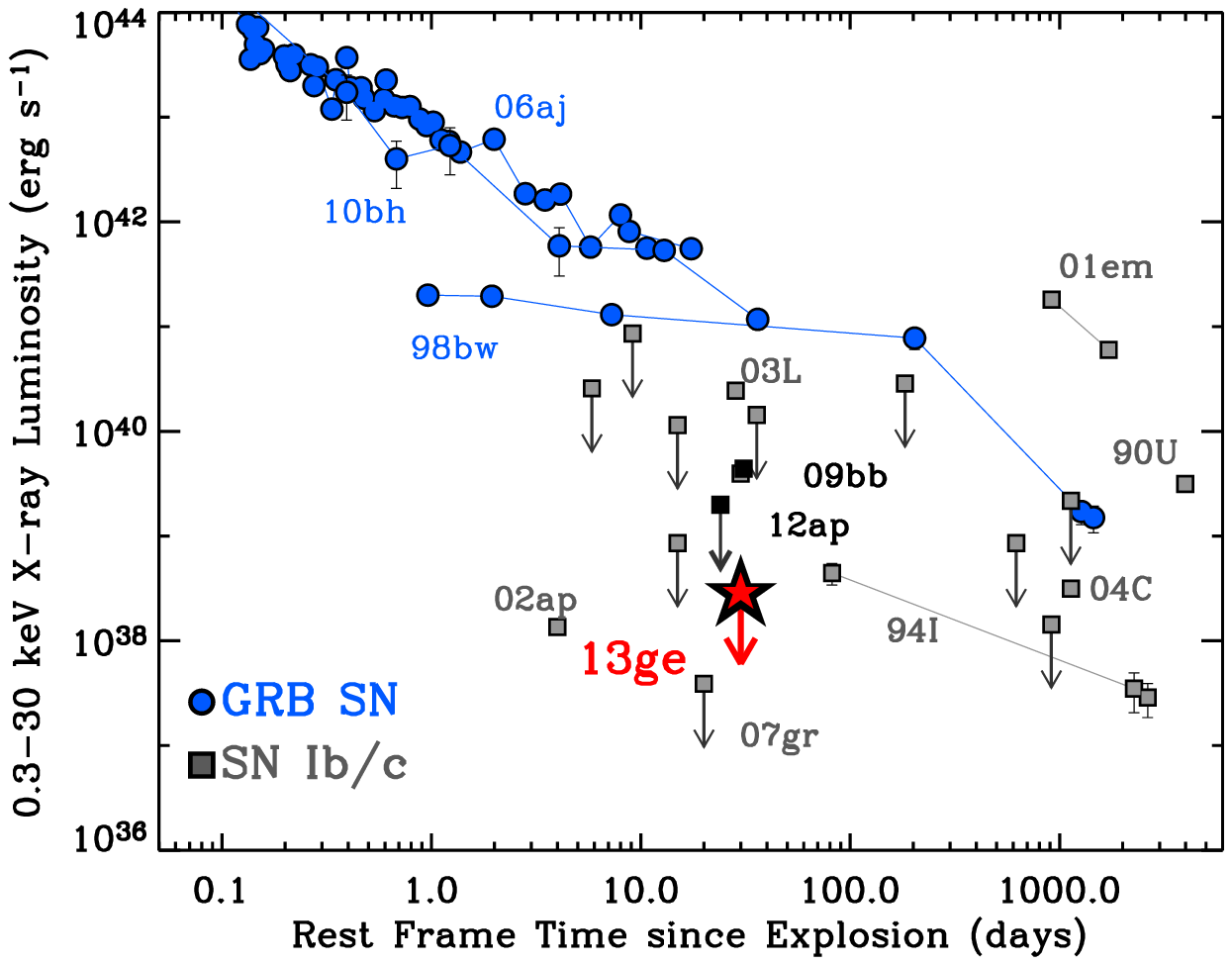}
\includegraphics[width=\columnwidth]{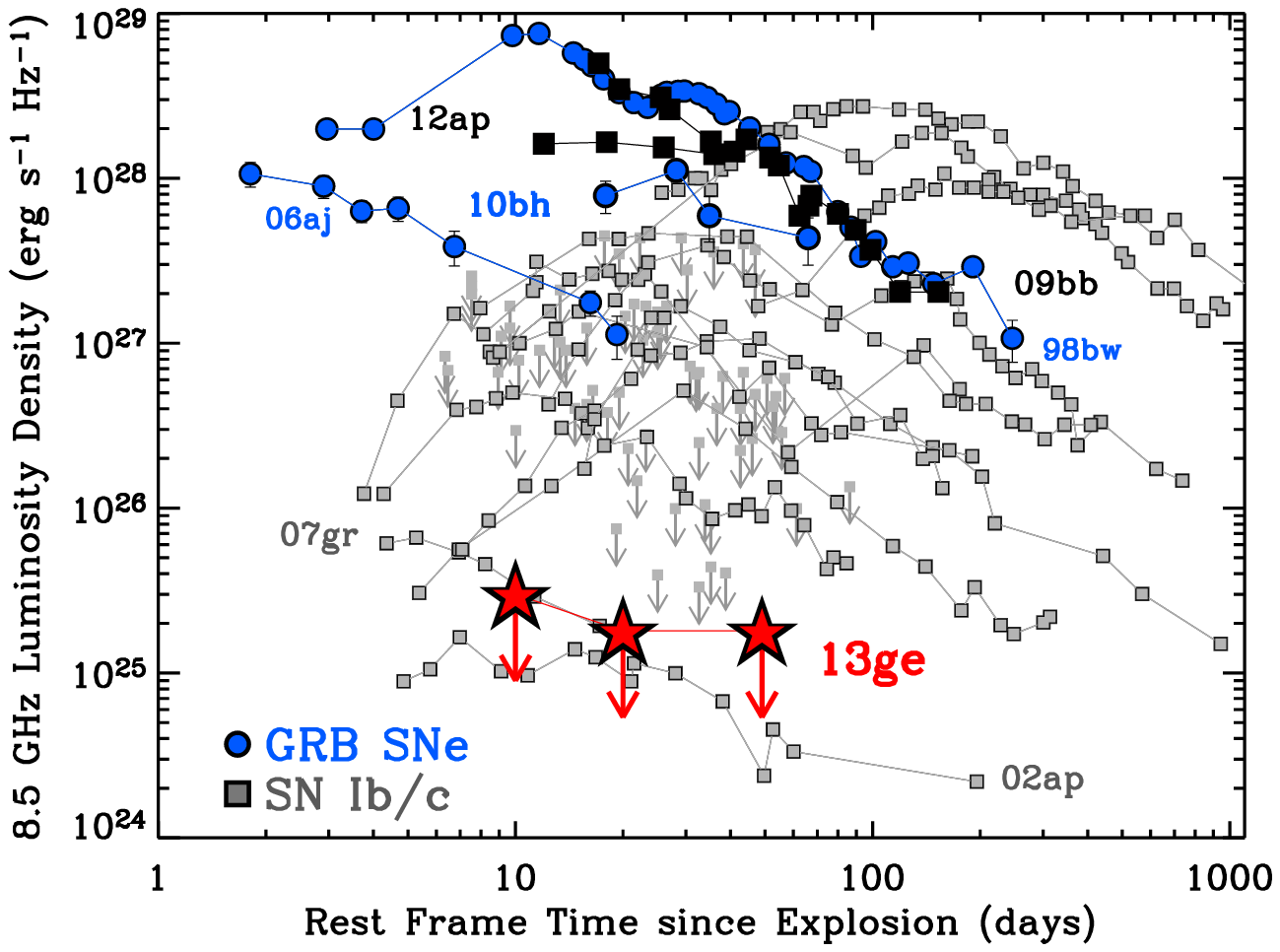}
\caption{X-ray (left) and Radio (right) upper limits obtained for SN\,2013ge (red stars) in comparison to other Type Ib/c SN.  GRB-SN are shown in blue, relativistic SN in black and other Type Ib/c SN in grey (labels for notable objects are shown in the same color as the data points).  The limits obtained for SN\,2013ge are among the deepest to date for a Type Ib/c SN.}
\label{Fig:XrayRadio}
\end{center}
\end{figure*}

\subsection{Non-Thermal Limits on Progenitor Mass-Loss Rate}

We observed SN\,2013ge in both the radio and X-ray bands during the main part of the optical outburst.  Although we obtained only non-detections, these limits are among the deepest ever obtained for a Type Ib/c SN (Figure~\ref{Fig:XrayRadio}).  Only the nearby (d $\sim$ 10 Mpc) Type Ic-BL SN\,2002ap has intrinsically fainter emission which was detected in both regimes.  Particularly notable, our radio observations constrain SN\,2013ge to be fainter than SN\,2007gr at similar epochs.

For SN that explode into a relatively low-density CSM (as is the case for Type Ib/c SN) X-ray emission near maximum light is due to Inverse Compton  (IC) up-scattering of optical photospheric emission by electrons accelerated at the SN shock.  In contrast, radio emission is characterized by a synchrotron self-absorbed spectrum, created when the electrons accelerated by the SN shock interact with shock-amplified magnetic fields \citep{Chevalier2006}.  As such, radio and X-ray emission (or lack thereof) provide \emph{independent} constraints on the density of the CSM surrounding the progenitor star.  

\subsubsection{X-ray IC Limits}

To model the X-ray upper limit in the context of IC up-scattering, we utilize the models of \citet{Margutti2012,Margutti2014b} which are based on the formalism of \citet{Chevalier2006}.  The luminosity of the IC signal is proportional to the bolometric luminosity (L$_{\rm{bol}}$) and additionally depends on the outer density structure of the SN ejecta, the density structure of the CSM, the energy spectrum of electrons which up-scatter the optical photons, the fraction of post-shock energy in relativistic electrons ($\epsilon_{\rm{e}}$), and the explosion properties of the SN (ejecta mass, kinetic energy).   The IC signal does \emph{not} depend on the fraction of energy in magnetic fields ($\epsilon_{\rm{B}}$) and because L$_{\rm{IC}}$ $\propto$ L$_{\rm{bol}}$ the mass-loss rate we derive is independent of any uncertainty in the distance to the SN.

Throughout our analysis we use the bolometric light curve derived in Section~\ref{Sec:Photom} and assume M$_{\rm{ej}} $ $=$ 2.5 M$_\odot$ and E$_{\rm{K}}$ $=$ 1.5 $\times$ 10$^{51}$ erg.  We additionally assume that the accelerated elections possess a power law structure of the form $n(\gamma) \propto \gamma^{-p}$ with p$=$3  ($\gamma$ is the Lorentz factor of the electrons) and that $\epsilon_{\rm{e}}$$=$0.1.  These values are motivated by the study of Type Ib/c SN in the radio \citep[e.g.][]{Chevalier2006,Soderberg2006}.  The outer portion of the SN ejecta is assumed to follow a steep power law of the form $\rho_{\rm{SN}} \propto R^{-n}$ with n$=$10 \citep[e.g.][]{Matzner1999}.  Finally, we consider the case where the density of the CSM can be described as a wind environment with a steady mass-loss rate \.{M} ($\rho$ $=$ \.{M}/4$\pi$r$^{2}$v$_{\rm{w}}$ where v$_{\rm{w}}$ is the wind velocity).  Using these input parameters our X-ray limit leads to an upper limit on the progenitor mass-loss rate of \.{M} $<$ $2.3\times10^{-5}$  ($\frac{\rm{v}_w}{1000 \rm{\,km} \rm{\,s}^{-1}}$) M$_\odot$ yr$^{-1}$.

\subsubsection{Radio Synchrotron Limits}
  
To model the radio upper limits in the context of self-absorbed synchrotron emission, we use the models outlined in \citet{Kamble2014}, which are based on those of \citet{Chevalier1998}.  For the radio spectrum characterized by synchrotron self-absorption (SSA), the peak spectral flux ($F_{\nu_{\rm a}}$) and SSA frequency ($\nu_{\rm a}$) are given by:

\begin{eqnarray*}
F_{\nu_{a}} (\rm mJy) 	&	= 	&	0.16\, A^{1.36}_{*}
								\left( \frac{\epsilon_{B}}{0.1}\right)^{0.64}
								\left( \frac{\beta}{0.15}\right)^{4.14}			
								\\
\nu_{a} (\rm GHz) 		& 	= 	&	6.0\, 	A^{0.64}_{*}
								\left( \frac{t}{10\, \rm day}\right)^{-1.0}
								\left( \frac{\epsilon_{B}}{0.1}\right)^{0.36}
								\left( \frac{\beta}{0.15}\right)^{1.86}		
\label{eqn:radio}								
\end{eqnarray*}

\noindent where $\beta$ is the shock velocity in the units of speed of light and A$_{*}$ is a dimensionless constant used to parameterize the density of the CSM.  As described above, for a stellar wind environment, the density surrounding the progenitor star can be parameterized as $\rho$ $=$ \.{M}/4$\pi$r$^{2}$v$_{\rm{w}}$. Normalizing to a constant mass loss rate and wind velocity of $\dot{M} = 10^{-5}\, \rm M_{\odot} \,yr^{-1}$ and $\rm{v}_{w} = 1000\,\rm km\, s^{-1}$, this can further be expressed as $\rho$ $=$ 5 $\times$ 10$^{11}$ A$_{*}$ r$^{-2}$ g cm$^{-3}$ (e.g. A$_{*}$ $=$ ($\frac{\rm{\dot{M}}}{10^{-5} M_{\odot} \rm{\,yr}^{-1}}$) ($\frac{\rm{v}_w}{1000 \rm{\,km\, s}^{-1}}$)$^{-1}$).   As above, we have also assumed, $n (\gamma) \propto \gamma^{-p}$ and $p=3.0$ for the distribution of relativistic electrons. 

\begin{figure*}[!ht]
\begin{center}
\includegraphics[width=0.7\textwidth]{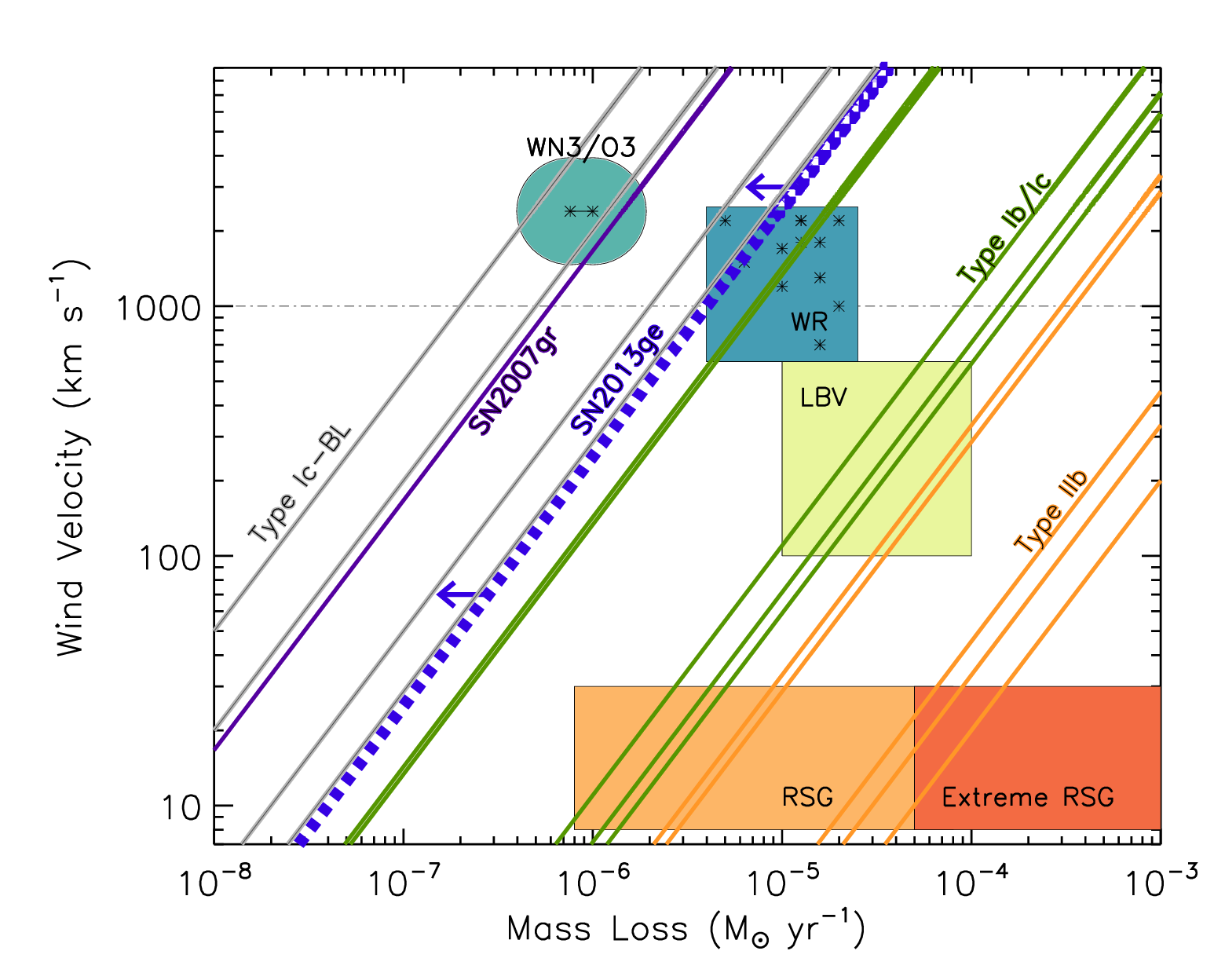}
\caption{Wind Speed versus Mass-loss rate.  Locations of galactic WR stars come from \citet{Crowther2007}, WN3/O3 stars from \citet{Massey2015}, normal and extreme RSGs from \citet{Marshall2004}, \citet{vanLoon2005} and \citet{deJager1988}, and LBV wind (not outburst) properties from \citet{Smith2014}.  Overplotted are measurements of progenitor mass-loss rate as a function of wind speed for SN\,2013ge (this work), SN\,2007gr \citet{Soderberg2010}, SN\,2002ap \citep{Berger2002}, SN\,2006aj \citep{Soderberg2006c}, SN\,2009bb \citep{Soderberg2010b}, SN\,2012ap \citep{Chakraborti2015}, SN\,1994I \citep{Weiler2011}, SN\,1990B \citep{VanDyk1993}, SN\,1983N \citep{Sramek1984}, SN\,2003L \citep{Soderberg2005}, SN\,2008D \citep{Soderberg2008}, SN\,2001ig \citep{Ryder2004}, SN\,2003bg \citep{Soderberg2006b}, SN\,2008ax \citep{Roming2009}, SN\,2011dh \citep{Krauss2012}, and SN\,2013df \citep{Kamble2015}.  We emphasize that the colored boxes represent the wind parameters measured for known, apparently single, evolved massive stars. Binary interaction may play a role in producing the progenitor systems of a significant fraction of stripped SN.\label{fig:MdotWind}}
\end{center}
\end{figure*}

Assuming that the shocked material is in equipartition ($\epsilon_{e} = \epsilon_{B}$) we can calculate the predicted radio flux for various values of \.{M}. Each of our radio upper limits place a constraint on the mass-loss rate of the progenitor system. We find that our 3$\sigma$ upper limit from 9 days post-explosion at 4.8 GHz is the most constraining, leading to an upper limit on the progenitor mass-loss rate of \.{M} $<$ 4.0 $\times 10^{-6}$  ($\frac{\rm{v}_w}{1000 \rm{km} \rm{s}^{-1}}$) ($\frac{\epsilon_{B}}{0.1}$)$^{-0.39}$ M$_\odot$ yr$^{-1}$. We have scaled to a fiducial value of $\epsilon_{B}$ $=$ 0.1.  Combining this with our X-ray limit (which does not depend on $\epsilon_{B}$) we place the following constraints on the mass-loss from the progenitor of SN\,2013ge:

\vspace{0.1in}
\noindent \.{M} $<$ 4.0 $\times 10^{-6}$  v$_{w,1}$ ($\frac{\epsilon_{B}}{0.1}$)$^{-0.39}$ M$_\odot$ yr$^{-1}$ for $\epsilon_{B}$ $>$ 0.001

\vspace{0.1in}
\noindent \.{M} $<$ $2.3\times10^{-5}$  v$_{w,1}$ M$_\odot$ yr$^{-1}$ for $\epsilon_{B}$ $<$ 0.001

\vspace{0.1in}
\noindent where v$_{w,1}$ is v$_w$ normalized by 1000 km s$^{-1}$.

\subsubsection{Comparison to Other Core-Collapse SN}

In Figure~\ref{fig:MdotWind} we plot progenitor mass-loss rate versus wind speed to show how the mass-loss constraints obtained for SN\,2013ge compare to those based on radio observations of 15 other stripped-envelope core-collapse SN (Type IIb, Ib, Ic, and Ic-BL; see caption for references). Radio observations only constrain $\dot{M}/\rm{v}_w$ so each SN appears as a diagonal line in this representation. The SN presented span over four orders of magnitude in $\dot{M}/\rm{v}_w$. A horizontal line designates a wind speed of 1000 km s$^{-1}$, which is often taken as a fiducial value for progenitors of Type Ib/c SN.  Also shown (colored squares) are regions of this phase space occupied for various classes of (apparently single) evolved massive stars.

In interpreting the data presented in Figure~\ref{fig:MdotWind} several caveats must be mentioned.  First, we plot a single value of $\dot{M}/\rm{v}_w$ for each SN, representing the CSM at a physical scales of a few $\times$ 10$^{15}$ cm. Detailed modeling of many Type Ib/c radio light curves reveals a more complex morphology, with some events showing signs of density modulations at larger physical scales or density profiles which vary from the $\rho$ $\propto$ r$^{-2}$ characteristic of a steady wind \citep{Wellons2012}. Similarly, the colored squares represent the locations of known, apparently single, massive stars.  How the values of $\dot{M}$ and $\rm{v}_w$ may change in the final years of a star's life is an open area of study \citep[e.g.][]{Smith2014}, with mass-loss due to binary interaction and eruptive mass-loss events potentially creating complex CSM environments.  This is particularly relevant here, as a significant fraction of the observed population of stripped SN may have progenitors formed via binary interaction \citep[see, e.g.][and references therein]{Smartt2015}, although only a small fraction ($\sim$ 6.5\%; \citealt{Margutti2016}) of Type Ib/c progenitors are expected to undergo mass transfer during the final stages of their evolution (``Case-C'' mass transfer). Finally, the SN shown in Figure~\ref{fig:MdotWind} are events which have been \emph{detected} at radio wavelengths and published in the literature, and therefore may be biased towards brighter events.

Nevertheless, despite these caveats, several trends emerge. The Type IIb SN all appear at the high end of the CSM density range. They have values of $\dot{M}/\rm{v}_w$ that intersect with the red/yellow supergiants, as might be expected if their progenitors have not fully lost their hydrogen envelopes\footnote{The progenitors of several Type IIb SN have been identified as Yellow Supergiants (YGSs) in pre-explosion images \citep{Maund2004,EliasRosa2009,EliasRosa2010,Fraser2010,VanDyk2014}.  The mass-loss rates and wind speeds of YSGs are not well characterized in the literature, but likely fall between RSGs and LBVs.}.  Five of the Type Ib/c SN fall at intermediate densities, while the Type Ic-BL SN appear to preferentially explode in regions of low CSM density.  However, it appears that SN\,2007gr and SN\,2013ge, represent examples of \emph{normal} Type Ib/c SN that exploded into particularly low density environments, similar to those observed for the broad-lined (including engine-driven) SN.

It is possible that this trend is partially due to metallicity. Type Ic-BL events preferentially occur  in low metallicity environments \citep{Sanders2012a} and SN\,2013ge and SN\,2007gr also occurred in low metallicity regions of their hosts\footnote{The explosion site metallicity of SN\,2007gr was log(O/H) + 12 = 8.5 \citep{Modjaz2011}, comparable to the LMC.}.  In addition, there are both empirical observations and theoretical predictions (for line-driven or partially line-driven winds) that WR mass-loss rates are metallicity dependent \citep{Crowther2007}. However, the relativistic\footnote{The term ``relativistic'' SN refers to events whose radio emission reveals a substantial relativistic outflow---likely powered by a central engine---but for which no associated GRB was observed (see \citealt{Soderberg2010b}).} SN\,2009bb and SN\,2012ap were both Type Ic-BL events with low CSM densities that occurred in solar or super-solar environments \citep{Levesque2010,Milisavljevic2015}.

Intriguingly, several Type Ic-BL SN and SN\,2007gr have measured values of $\dot{M}/\rm{v}_w$ which are a factor of 3 to 10 lower than any Galactic or LMC WR star examined in \citet{Crowther2007}.  They are consistent with the mass-loss properties inferred for a new class of WR stars recently discovered in the LMC \citep{Massey2015}.  Dubbed WN3/O3, these stars are both fainter in optical bands than ``normal'' WN/WC stars, and have inferred mass-loss rates an order of magnitude lower.  The formation mechanism of these new WR stars is not fully understood, but they demonstrate that some evolved massive stars in the Local Group have mass-loss properties consistent with the very low densities inferred from radio observations of several Type Ic events.

\section{Discussion}\label{Sec:Discuss}

In this Section we discuss the observations outlined above in the context of their implications for the physical configuration of the progenitor system and explosion mechanism of SN\,2013ge.  From maximum light onward, the optical emission from SN\,2013ge is fairly standard for a Type Ib/c SN. Its explosion parameters (M$_{\rm{ej}}$ $=$ 2$-$3 M$_\odot$, E$_{\rm{K}}$ $=$ 1 $-$ 2 $\times$ 10$^{51}$ erg) are well within the range observed for other SN and its maximum light spectra are characterized by ions of various intermediate mass and iron peak elements at velocities of $\sim$10,000 km s$^{-1}$. 

However, the early UV and spectroscopic observations of SN\,2013ge are unusual.  The u$-$band light curve shows an extra component of emission, which rises on a timescale of $\sim$4$-$5 days.  During the rising phase of this initial peak the optical spectra are characterized by a blue continuum superimposed with a plethora of P Cygni features, which are both rapid ($\sim$15,000 km s$^{-1}$) and \emph{narrow} (FWHM $\lesssim$ 3500 km s$^{-1}$).  Below, we discuss the physical interpretation of these early spectra, possible origins for the excess UV emission, and progenitor scenarios which can explain both these observations and the other properties of SN\,2013ge.

\subsection{Physical Interpretation of the Early Spectra}

When a SN shock reaches the low density outer regions of the progenitor envelope it will accelerate, leading to a high-velocity gradient in the outer regions of the SN ejecta \citep{Matzner1999,Piro2014}. The high velocities, rapid velocity evolution, and rapid evolution of observed line ratios in the early spectra of SN\,2013ge indicate that we are probing these outer regions.  However, for a spherically symmetric explosion in which the differential optical depth of the ejecta decreases monotonically outward from the photosphere, a high-velocity absorption minimum should be accompanied by a broad P Cygni feature.  A departure from this picture (as we see in the early spectra of SN\,2013ge) implies that the line formation is limited in some sense and likely requires either a modification to the geometry of the explosion or to the optical depth profile of the ejecta.  

The high-velocity, narrow, absorption features observed in the early spectra of SN\,2013ge could be understood in terms of an asymmetric explosion in which a fraction of the ejecta was launched at high velocities along the line of sight to the observer.  In this case, the widths of the lines are affected by the opening angle of the ejection.  After some time, this material becomes transparent, revealing the underlying photosphere of the bulk explosion.  \citet{Folatelli2006} suggest a similar model for the double-peaked SN\,2005bf, which displayed both high-velocity \ion{Fe}{2} and \ion{Ca}{2} lines and broader features (associated with the underlying photosphere) at early times. The presence of both features was understood in terms of the asymmetric explosion being close, but slightly off from the observer angle.  In contrast, in SN\,2013ge \emph{all} ions observed in the earliest spectra show narrow, high-velocity features; no underlying photosphere with broader components is visible.  This has implications for both the ions present in such an asymmetric ejection, as well as the angle at which we observe the outflow. 

Alternatively, it may be possible to recreate the spectral features in SN\,2013ge if the line absorption coefficient does not decrease monotonically with radius. In this case, the main line-forming region at early epochs could be ``detached''  above the photosphere.  Physically, this could be due to an actual increase in density or to a change in the ionization state of ejecta at a certain distance above the photosphere, resulting in an increased line opacity.  The latter argument was used by \citet{Tanaka2009} to explain the high-velocity \ion{Ca}{2} and \ion{Fe}{2} lines in SN\,2005bf.  They note that these lines coincide with high-velocity H$\alpha$, indicating that they were formed in a thin hydrogen shell that remained on the progenitor star at the time of explosion.  They argue that the high electron density in this (solar abundance) hydrogen shell enhances the recombination of \ion{Ca}{3} and \ion{Fe}{3} \citep{Mazzali2005,Tanaka2009}, allowing narrow, high-velocity \ion{Ca}{2} and \ion{Fe}{2} lines to be formed in the outer portions of the ejecta.  In the case of SN\,2013ge, a plethora of high-velocity ions are observed at these early epochs which would have implications for the composition of the outer layers of the progenitor star.  

\subsection{The Nature of the early UV emission}

The early rising light curves of Type I SN are powered by a combination of two primary sources: energy deposited by the SN shock and the radioactive decay of $^{56}$Ni. Excess emission can also be produced by external sources, such as the collision of the SN shock with a binary companion. We now examine the likelihood that the excess UV emission observed in the early light curve of SN\,2013ge is powered by each of these sources, and the implications for the explosion in each case.  

\subsubsection{Cooling Envelope Emission}

\begin{figure*}[!ht]
\begin{center}
\includegraphics[width=0.9\columnwidth]{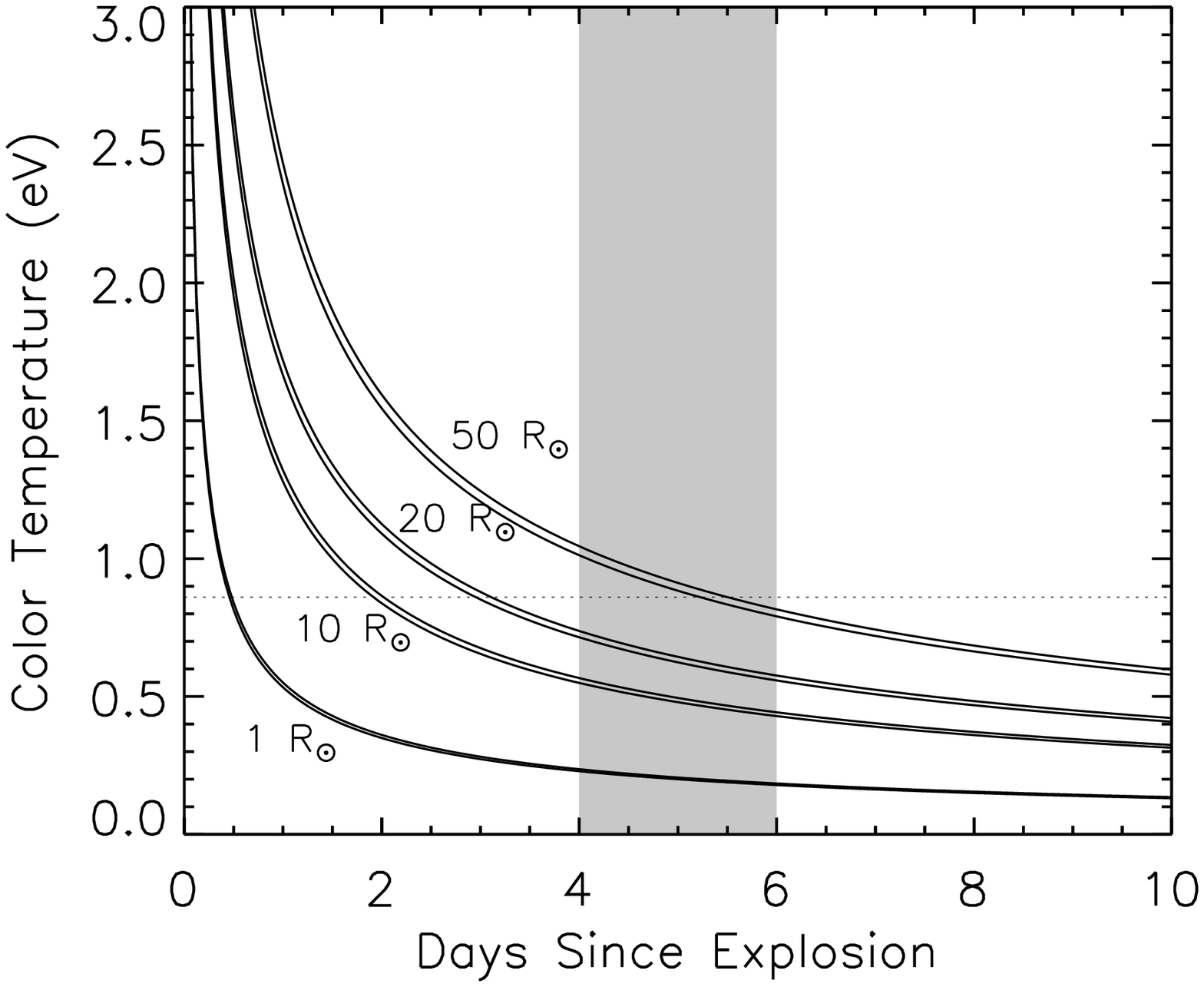}
\includegraphics[width=0.9\columnwidth]{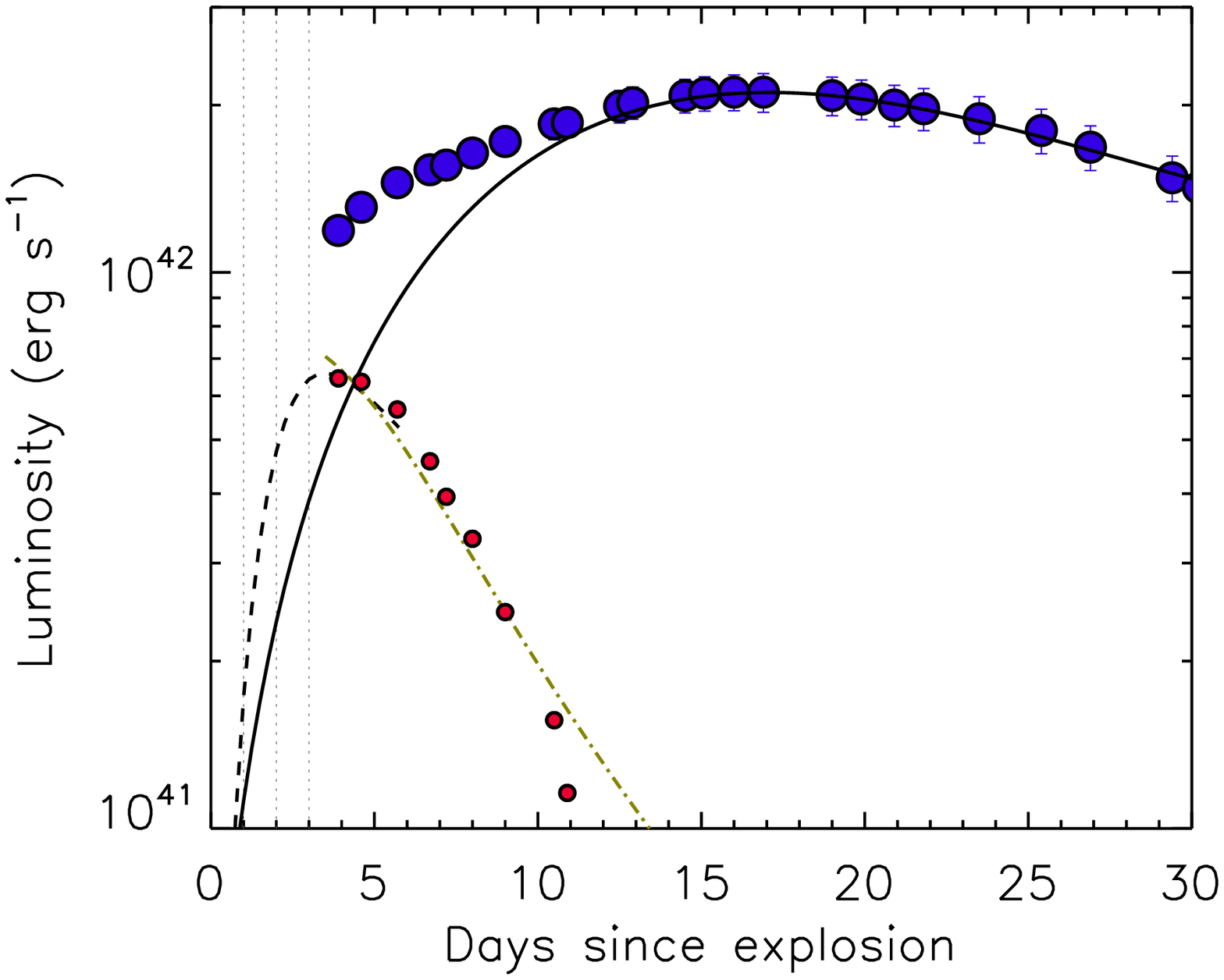}
\vspace{-0.02in}
\caption{\emph{Left:} Color temperature versus time since explosion for cooling envelope emission from hydrogen-poor progenitors with radii between 1 R$_\odot$ and 50 R$_\odot$.  Bands for each radius assume explosion parameters of M$_{\rm{ej}}$ $=$ 2 $-$ 3 M$_\odot$ and E$_{\rm{K}}$ $=$ 1 $-$ 2 $\times$ 10$^{51}$ erg.  If the early emission from SN\,2013ge is due to cooling envelope emission, then the temperature remain must above 10,000 K (0.9 eV) \emph{at a minimum} for 4$-$6 days post-explosion, implying an extended progenitor.  \emph{Right:} Decomposition of the bolometric luminosity of SN\,2013ge (blue) into two components.  The solid black line represents our best fit $^{56}$Ni decay model to the bulk explosion.  Red points show the excess emission above this model at early times. The black dashed line is an Arnett model fit to the rise time and luminosity of this early component and the gold line is a model for the decline phase based on the instantaneous energy deposition from the $^{56}$Ni $\rightarrow$ $^{56}$Co $\rightarrow$ $^{56}$Fe decay chain, allowing for incomplete gamma-ray trapping.  Dotted lines indicate the epochs of our early spectra. \label{fig:coolingenv}}
\end{center}
\end{figure*}

After shock breakout, the shock heated ejecta cool, giving rise to a light curve component independent from that powered by $^{56}$Ni. During this phase, both the bolometric luminosity of the transient and the color temperature should decline with time \citep{Nakar2010}, with the exact values depending on the radius of the progenitor star, the explosion energy, the ejecta mass, and the ejecta opacity.  The luminosity observed in any given optical/UV band will rise as long it is located on the Rayleigh-Jeans tail of the temperature spectrum.  As a result, one expects a rise in the UV/optical bands proportional to t$^{1.5}$ if the emission is powered by shock heated cooling \citep{Piro2013,Nakar2010}.  
 
In the case of SN\,2013ge, \emph{the observed rise time for the first u$-$band component is abnormally large for cooling envelope emission from a stripped progenitor star}. A power-law of the form t$^{1.5}$ can be fit to the initial rise observed in the u$-$ and w1$-$bands (see Figure~\ref{fig:ExEpoch}; although this solution is not unique), yielding a rise time of 4$-$6 days from the epoch of explosion. In contrast, the first light curve components of both SN\,2008D and SN\,2006aj (which some authors associate with cooling envelope emission; \citealt{Soderberg2008,Campana2006}) rise in the UV/optical on timescales $\lesssim$1 day.  By $\sim$5 days post explosion, emission from the radioactive decay of $^{56}$Ni likely comprises a non-negligible fraction of the total light, making it difficult to assess whether the temperature and luminosity evolution of the first component is consistent with cooling envelope emission.  The u$-$b color remains roughly constant during the rising portion of the first u$-$band component (before rapidly reddening during the decline phase), which is consistent with the peak of the blackbody passing through the observed band.  However, we caution that the UV flux was also depressed compared to a single blackbody during this time period, and the rapid reddening could therefore be due to increased UV line blanketing.

Under the assumption that this emission is caused by shock-heated cooling then, when compared with theoretical models, the long rise time implies that SN\,2013ge \emph{cannot have a standard WR progenitor}.  In order to account for the 4$-$6 day rise in the u$-$band, we require that the temperature remain above 10$^{4}$ K ($\sim$ 0.9 eV) \emph{at a minimum} for this time period. In contrast, in the WR model of \citet{Nakar2010} the temperature has already fallen below this level by $\lesssim$1 day post-explosion.  Using the parameterization from \citet{Piro2013} for a progenitor with a hydrogen-free radiative envelope, in Figure~\ref{fig:coolingenv} we plot the color temperature versus time for a range of progenitor radii, assuming M$_{\rm{ej}}$ $=$ 2$-$3 M$_\odot$ and E$_{\rm{K}}$ $=$ 1$-$2 $\times$ 10$^{51}$ erg (as derived from the bulk explosion).  We see that for this envelope structure, we require a progenitor radius of \emph{at least} 50 R$_\odot$ to account for the observed u$-$band rise.  The cooling envelope luminosity at 4 days post-explosion for this model is 4 $-$ 5 $\times$ 10$^{41}$ erg s$^{-1}$, compatible to the shoulder of excess emission observed at early times.

The modest ejecta mass and extended progenitor radius derived above lead us to also consider the models of \citet{Nakar2014} for double-peaked SN.  These models employ a non-standard progenitor envelope structure, in which a massive compact core is surrounded by extended low-mass material.  In this picture, the luminosity of the emission is mainly a function of the radius of the extended material while the time to maximum is related to the total mass in the extended envelope. Both also depend on the velocity and opacity.  For a peak time of 4 days, peak luminosity of 0.6 $-$ 1.2 $\times$ 10$^{42}$ erg s$^{-1}$, a characteristic velocity of $\sim$12,000 km s$^{-1}$, and Eqns.\ 10 and 12 of \citet{Nakar2014} we find an extended envelope mass of $\sim$0.1 M$_\odot$ at a radius of 15 $-$ 25 R$_\odot$.  Although less extreme than the value derived for a standard progenitor envelope structure, this is still more extended than a typical WR progenitor star.

\subsubsection{Outwardly Mixed $^{56}$Ni}

Alternatively, the early emission from SN\,2013ge could be due to $^{56}$Ni mixed outward in the explosion.  In this case, our rising light curve can give constraints on the radial distribution of the outwardly mixed material.

For any given light curve point, there is a degeneracy between the depth of the contributing $^{56}$Ni and the explosion epoch. For deep deposits, there will be a ``dark period'' between the explosion and the epoch of first light \citep{Piro2013}. Our spectroscopic observations from $\sim$3 days after the epoch of first light displayed high velocities and rapid velocity evolution which argue against any significant dark period for SN\,2013ge. This is compatible with our first measurement of the luminosity, temperature and photospheric velocity which, using Eqn.\ 17 of \citet{Piro2013} only require that the explosion was $\sim$2 days prior to our first bolometric light curve point.  In this model, the lack of a significant dark period in SN\,2013ge requires that some amount of  $^{56}$Ni was mixed into the outer portions of the ejecta.

The morphology of the early bolometric light curve also has implications for the distribution of this outwardly mixed $^{56}$Ni.  A radial distribution of $^{56}$Ni which is monotonically decreasing should yield a smoothly rising light curve \citep[e.g.][]{Dessart2012}.  In contrast, the ``shoulder'' of emission as seen in SN\,2013ge may require a distinct deposit of $^{56}$Ni at shallower depths, qualitatively similar to the model used by \citet{Bersten2013} to explain the first light curve component of SN\,2008D.  If a small clump of $^{56}$Ni-rich material was ejected at high velocities, we can obtain a rough order of magnitude estimate for the ejecta mass, nickel mass, and kinetic energy of this material by decomposing the bolometric light curve into two components.  This is done in the right panel of Figure~\ref{fig:coolingenv}, where the solid line is our model for the bulk explosion (Section~\ref{Sec:Photom}) and the red points are the excess above this model at~early times.  

Using the models of \citet{Arnett1982} to fit the rise time and luminosity of this excess---and assuming a velocity of $\sim$12,000 km s$^{-1}$---yields an ejected mass of $\sim$0.06 M$_\odot$, a $^{56}$Ni mass of $\sim$0.012 M$_\odot$, and a kinetic energy of $\sim$6 $\times$ 10$^{49}$ erg.  To investigate if a $^{56}$Ni-powered explosion with these parameters is consistent with the rapid post-maximum decline inferred for the early light curve component we use the model of \citet{Drout2013}, which was developed for the rapidly-declining SN\,2005ek. It fits the entire post-maximum evolution of an explosion based on the instantaneous rate of energy deposition from the $^{56}$Ni $\rightarrow$ $^{56}$Co $\rightarrow$ $^{56}$Fe decay chain, including incomplete trapping of gamma-rays produced during $^{56}$Ni $\rightarrow$ $^{56}$Co decay.  This is appropriate for the declining phase of SN with very low ejecta masses, which become optically thin quickly, making the models of \citet{Arnett1982} \mbox{inapplicable.} 

Using this model with the explosion parameters found above yields the gold curve shown in the right panel of Figure~\ref{fig:coolingenv}, which is well matched to the decline timescale inferred for the early light curve component in SN\,2013ge.  However, we emphasize that these explosion parameters should be taken as order of magnitude estimates only. We do not uniquely decompose the bolometric light curve into multiple components and the analytical models used here do not account for variations in geometry, opacity, or contributing radioactive species.  

\subsubsection{Shock Collision with a Binary Companion}

Finally, we consider an external source for the early UV emission: a collision between the SN ejecta and a binary companion.  During such a collision a bow shock will form, compressing and shock-heating the SN ejecta in the direction of the interaction. This heating can lead to an extra source of UV/optical emission for several days post explosion \citep{Kasen2010}. The observed properties of this emission depend on several parameters (e.g.\ binary separation, ejected mass) and are highly viewing angle dependent. Using a binary population synthesis model for core-collapse SN, \citet{Moriya2015} find that only $\sim$0.53 \% of Type Ib/c SN light curves should have a detectable visible brightening due to this mechanism.  As such, any detection of this collision would be rare.

Using the models of \citet{Kasen2010}, we assess whether both the luminosity and timescale of the early emission observed in SN\,2013ge can be reproduced by this mechanism.   Using their Eqn.\ 22 and the explosion parameters derived in Section~\ref{Sec:Photom} we find that the luminosity of the excess emission in SN\,2013ge would require a binary separation of $\lesssim$10$^{12}$ cm ($\sim$ 15 R$_\odot$). However, reproducing the timescale of the early emission is challenging. The source of the UV/optical emission in this model is shock-heated cooling, analogous to the cooling envelope emission described above.  Thus, in this model, the u-band rise time of 4$-$6 days similarly requires that the u$-$band remain on the Rayleigh-Jeans tail of the temperature spectrum for this time period.  In contrast, none of the models presented in \citet{Kasen2010} have rise times longer than $\sim$2 days.  Using Eqn.\ 15 in \citet{Kasen2010} for the effective temperature of the emission, we find that a separation of \emph{at least} 10$^{13}$ cm is required to have T$_{\rm{eff}}$ $>$ 10$^4$ K at 5 days post-explosion. This is inconsistent with the required separation found above.  Thus, we find that (for the current set of theoretical models) the early observations of SN\,2013ge are inconsistent with the collision of SN ejecta with a companion star.

\subsection{The Progenitor of SN\,2013ge}

SN\,2013ge was the explosion of a stripped massive star with a moderate ejecta mass, \emph{weak} He features in its optical/IR spectra, a low pre-SN mass-loss rate, and a local environment metallicity of $\sim$0.5 solar. However, we are left with two distinct scenarios depending on our interpretation of the early emission. Either SN\,2013ge was the explosion of a star with an extended envelope or it was the result of an explosion in which a clump of $^{56}$Ni was mixed outward in the ejecta, possibly coupled to the ejection of a small amount of mass along the line of sight to the observer.  We now examine the consistency and consequences of each of these progenitor models.  We emphasize that although SN\,2013ge is a relatively unique object, the results presented here potentially have general implication for the progenitors of Type Ib/c SN.  Without either our early spectroscopic observations or early UV coverage \emph{SN\,2013ge may not have been flagged as unusual}.

\subsubsection{Extended Progenitor Surface}

If we interpret the early emission from SN\,2013ge as cooling envelope emission, it implies that shock breakout occurred from an extended surface. Our estimates for the extent of this surface range from 15 $-$ 25 R$_\odot$, for a low-mass envelope on a compact core, to $>$ 50 R$_\odot$ for a standard envelope structure.  In principle, this surface could either be a genuinely extended progenitor envelope, or it could be located within a dense optically thick wind region surrounding the progenitor star.  

While some WR stars have particularly dense wind regions, which extend their photospheric radii by up to a factor of 10 \citep{Li2007}, the interpretation of SN\,2013ge as the explosion of such a star is complicated by the need to reconcile it with the radio observations obtained $\sim$9 days post-explosion.  These observations indicate that by $\sim$10$^{15}$~cm (for a standard SN shock velocity of v$_{\rm{sh}}$ $=$ 0.15$c$) the progenitor of SN\,2013ge is characterized by a \emph{low} density wind region.  Thus, if the shock breakout occurred within a dense CSM,  \emph{the progenitor star must have either experienced a significant change in its mass-loss properties or ejected a portion of its envelope during the final stages of its evolution}. This process must have occurred within the final $\lesssim$100  days before core-collapse in order to be contained within the region probed by our radio observations or the final $\lesssim$0.5 days if the breakout radius we derive is the outer extent of this mass. These timescales are normalized to an ejection velocity of 1000 km s$^{-1}$, and are consistent with models that predict that instabilities and internal gravity waves can be induced during the final advanced nuclear burning stages, possibly leading to enhanced mass-loss/eruptions during the final year(s) before core-collapse (e.g. \citealt{Shiode2014}; \citealt{Smith2014b}). 

Alternatively, SN\,2013ge could be explained by the explosion of a stripped star with a low final mass-loss rate and an inflated stellar envelope. Several of the Type Ib/c binary progenitor models from \citet{Yoon2010} have radii inflated to $\sim$30 R$_\odot$, low pre-SN mass-loss rates ($\dot{M}$ $\lesssim$ 10$^{-6}$ M$_\odot$ yr$^{-1}$), final masses between 3$-$4 M$_\odot$ (consistent with our derived ejecta masses), and sub-solar metallicity.  However, the helium envelopes for these progenitors are relatively large ($\sim$1.5 M$_\odot$; significantly higher than the mass \citealt{Hachinger2012} find is necessary to produce observable features) and are therefore likely inconsistent with the \emph{weak} \ion{He}{1} features observed in SN\,2013ge.  In particular, these low-mass, extended, progenitors are predicted to be more efficient at mixing $^{56}$Ni into the He-rich layers via RT-instabilities \citep{Yoon2010,Hammer2010,Joggerst2009}, which should lead to \emph{stronger} observed He lines. Thus, the observations of SN\,2013ge likely require a progenitor which differs from any presented in \citet{Yoon2010} in having an extended envelope but \emph{low} final He mass.

Finally, we consider if this progenitor scenario can account for the unusual spectra observed during the rise of the first light curve component. In particular, while this scenario requires a high effective temperature, the ions present in these spectra are standard singly ionized species. It is possible that this, as well as the unusual velocity profile of the lines, could be understood if there was a change in ionization state in the outer portion of the ejecta, associated with either the low-mass extended envelope or a density enhancement due to a pre-explosion mass ejection.  In the former case, the presence of even a very small amount of hydrogen could increase the electron density enough to lead to enhanced recombination \citep{Mazzali2005b,Tanaka2008}.

\subsubsection{Asymmetric Ejection}

Alternatively, if we interpret the early emission as heating due to $^{56}$Ni, then an asymmetric ejection of a small amount of $^{56}$Ni-rich material at high velocities could explain both the early light curve peak and the unusual velocity profile in the early spectra.  Intriguingly, our observed velocity ($\sim$15,000 km s$^{-1}$) and estimated mass ($\sim$0.06 M$_\odot$) for this material are comparable to those observed in the high-velocity clumps in the northwest portion of the Cassiopeia A (Cas A) SN remnant \citep{Fesen2001,Laming2006}.  This material in Cas A has an opening angle of $\sim$45 degrees and has been argued by some to originate in the stellar core \citep{Hwang2004,Laming2006,Milisavljevic2013d}.

While the moderately peaked nebular emission line profiles and low-CSM density observed for SN\,2013ge also show similarities to jet-driven explosions in the literature, the observations obtained do not necessarily \emph{require} a highly asymmetric explosion mechanism.  They could also potentially be understood if viewing one of the nickel and silicon-rich plumes of material observed in 3D simulations of mixing instabilities in neutrino driven explosions \citep[e.g.][]{Hammer2010}. In these models, fast clumps of metal-containing material are able to penetrate through the outer layers of the ejecta, possibly leading to asymmetric variations in the ejecta velocity.  Indeed, velocity variations of $\sim$4000 km s$^{-1}$ were detected in light echoes from the explosion of Cas A, depending on viewing angle \citep{Rest2011}. However, resolved imaging of titanium in Cas A indicates that the explosion mechanism was dominated by slightly asymmetric, low-mode, convection, as opposed to a highly asymmetric/bipolar explosion mechanism \citep{Grefenstette2014}.  

In this scenario, the weak helium features observed in the spectra of SN\,2013ge also have implications for the true helium abundance of the progenitor star.  In particular, they suggest a different scenario than that observed in SN\,2005bf, where \citet{Tanaka2009} proposed that a $^{56}$Ni-rich plume penetrated only slightly into a nearly intact helium envelope.  In this case, both the strength and velocity of the observed helium features grew with time as more of the helium envelope fell within a $\gamma$-ray optical depth of the $^{56}$Ni deposit.  In the models of  \citep{Dessart2012}, the asymmetric ejection of a single blob of $^{56}$Ni-rich material to high-velocities is only predicted to produce weak helium features. However, if this blob is part of a larger scale mixing instability, as described above, then it would favor a scenario where the progenitor of SN\,2013ge was genuinely He-poor, containing only a thin layer at the time of explosion. 

\subsection{Comparison of the Early Emission to the Rapidly-Declining SN\,2002bj}

\begin{figure}[!t]
\begin{center}
\includegraphics[width=\columnwidth]{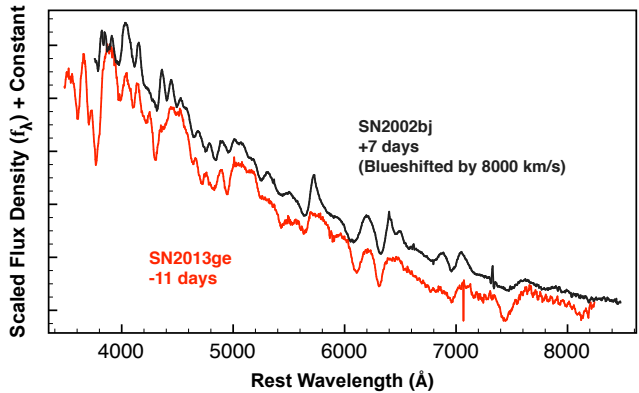}
\caption{A comparison of an early spectrum of SN\,2013ge to a spectrum of the rapidly evolving SN\,2002bj \citep{Poznanski2010}.  The spectrum of SN\,2002bj has been linearly blueshifted by 8000 km s$^{-1}$.  \label{fig:02bj}}
\end{center}
\end{figure}

While investigating the early spectra of SN\,2013ge, we found that the first spectrum obtained for the rapidly declining SN\,2002bj \citep{Poznanski2010} displayed a similar blue continuum and narrow spectroscopic features. In Figure~\ref{fig:02bj} we show this spectrum along with the $-$11 day spectrum of SN\,2013ge.  The spectrum of SN\,2002bj was obtained at $+$7 days, when its photospheric velocity was only $\sim$4000 km s$^{-1}$, and we have linearly blue-shifted it by 8000 km s$^{-1}$ for comparison with SN\,2013ge.  We emphasize that, unlike in the early spectra of SN\,2013ge, there is no mismatch between the widths the features in this spectrum and the blueshifts of their absorption minima.  From modeling its SED as a blackbody, \citet{Poznanski2010} found that the velocity of SN\,2002bj was higher at earlier epochs---consistent with a photosphere rapidly receding into a low-mass envelope. Unfortunately, no spectrum of SN\,2002bj is available to assess whether its spectroscopic features were similarly narrow at earlier epochs.  However, the similarity between the spectrum of SN\,2002bj and spectra obtained during the first emission component of SN\,2013ge is still striking.

The nature of the explosion which produced SN\,2002bj is still a mystery.  The rapid light curve and unusual spectrum lead \citet{Poznanski2010} to hypothesize that it was due to the detonation of a He shell on the surface of a WD.  Comparing the early emission of SN\,2013ge (as shown in Figure~\ref{fig:coolingenv}) to the light curve of SN\,2002bj (see Figure~\ref{fig:Bolo}) we find that the first emission component in SN\,2013ge is nearly an order of magnitude fainter and also declines a factor of $\sim$1.4 faster than the bolometric light curve of SN\,2002bj.  Thus, even neglecting the second (main) light curve component of SN\,2013ge (which is entirely lacking in SN\,2002bj) the energetics of these explosions are very different. However, Figure~\ref{fig:02bj} demonstrates that the ions and ionization state present in the ejecta of SN\,2002bj can also be produced during the core-collapse of a massive star.  It has already been shown that massive stars may be able to produce rapidly-evolving Type I SN, either due to very low ejecta masses \citep{Drout2013,Tauris2013} or through the combination of a large progenitor radius and a lack of ejected radioactive elements \citep{Kleiser2014}.  More detailed modeling attempting to ascertain whether SN\,2002bj may be a more extreme example of the first emission component in SN\,2013ge in an explosion which lacks the second, main, light curve component powered by $^{56}$Ni would be warranted.

\section{Summary and Conclusions}

We have presented extensive observations of the Type Ib/c SN\,2013ge beginning $\sim$2 days post-explosion, when the light curve is particularity sensitive to both the progenitor configuration and mixing within the ejecta.  Here we summarize our main conclusions.

{\bf \emph{Early Emission:}} The rapid velocity evolution and rapid rise observed in the early spectra and UV light curves, respectively, indicate that our first observations probe the outer regions of the ejecta shortly after explosion.  The early u$-$band and UV light curves show two distinct components.  The first component has a rise time of $\sim$4$-$5 days and is visible for the first week post-explosion.  This manifests itself as a ``shoulder'' of excess emission in the bolometric light curve with a luminosity of $\sim$ 6 $\times$ 10$^{41}$ erg s$^{-1}$.  Spectra of the first component display a blue continuum and are unusual in possessing features which are both moderately high-velocity ($\sim$15,000 km s$^{-1}$) and \emph{narrow} ($\sim$3500 km s$^{-1}$).  This indicates that the line formation region is limited in some sense, possibly due to an asymmetric geometry or a change in ionization state in the outer portions of the ejecta.  

{\bf \emph{Bulk Explosion:}}  With M$_{\rm{R,peak}}$ $=$ $-$17.5 and $\Delta$m$_{\rm{15,R}}$ $=$ 0.4, SN\,2013ge is relatively faint and slowly evolving, but the derived ejecta mass (2 $-$ 3 M$_\odot$) and kinetic energy (1 $-$ 2 $\times$ 10$^{51}$ erg) are well within the distribution observed for Type Ib/c SN.  \emph{Weak} \ion{He}{1} lines, which fade with time, are detected in early optical and NIR spectra. These are distinct from the conspicuous \ion{He}{1} lines which are usually used to classify Type Ib SN. Near maximum light the spectra are dominated by a plethora of intermediate mass and iron peak elements.  The late-time spectral evolution of SN\,2013ge is also distinctive, showing a lack of evolution in the Mg~I]/[O~I] ratio and a shifting CO-emission profile.

{\bf \emph{Environment Properties:}}  SN\,2013ge exploded on the outskirts of a star forming galaxy. There is an unresolved HII region at the explosion site, which has a metallicity of $\sim$0.5 solar.  The radio and X-ray limits for SN\,2013ge are among the deepest ever obtained for a stripped-envelope SN and constrain the progenitor mass-loss rate to be \.{M} $<$ 4.0 $\times 10^{-6}$ M$_\odot$ yr$^{-1}$ for $\epsilon_{\rm{B}}$ $=$ 0.1.

{\bf \emph{Power Sources and Progenitors:}}  SN\,2013ge was the explosion of a stripped massive star with a moderate ejecta mass. However, we are left with two distinct progenitor scenarios depending on our interpretation of the early emission.  In both cases, we find it likely that the progenitor of SN\,2013ge had only a thin layer of helium remaining at the time of core-collapse.
\begin{enumerate}
\item If the early emission is due to post-shock-breakout cooling envelope emission, then its relatively long rise time ($\sim$4$-$6 days) requires that the progenitor of SN\,2013ge either had an extended envelope or that it ejected a small portion of its envelope in the final $\lesssim$1 year before core-collapse.  
\item If the early emission is due to outwardly mixed $^{56}$Ni then we require that a distinct clump of $^{56}$Ni was mixed into the very outer portions of the ejecta. Coupled with the early spectra, this may imply an asymmetric ejection of a small amount of nickel-rich material at high-velocities. 
\end{enumerate}

More detailed modeling beyond the scope of this work will be necessary to fully distinguish between or rule out one of these two progenitor scenarios.  In particular, it would be useful to ascertain whether either scenario can actually reproduce the plethora of high-velocity and narrow features observed in the early spectra, with the cooling envelope model facing the additional challenge of explaining the depressed UV flux at early epochs.  We note that if the epoch of first light is earlier than that derived in Section~\ref{Sec:Photom} from power law fits to the early UV light curves, then the cooling envelope/extended progenitor scenario would be put under additional tension, or ruled out entirely.  In contrast, the model of a high-velocity clump could be naturally extended to explain a rise in the UV light curve after the epoch of first light.  Finally, we find that current theoretical models for the collision of a SN shock with a binary companion cannot reproduce both the luminosity and timescale of the early emission observed in SN\,2013ge.

{\bf \emph{Rapidly-Evolving SN\,2002bj:}}  The early spectra of SN\,2013ge are similar to the spectrum of the rapidly-evolving SN\,2002bj, demonstrating that the ions and ionization state present in the ejecta of SN\,2002bj can also be produced by the core-collapse of a massive star.  

\vspace{0.1in}

We thank the anonymous referee for numerous comments that improved this manuscript.  M. R. D. thanks L. Z. Kelley, D. Kasen, and E. Ramirez-Ruiz for useful discussions.  We thank N. Morrell for obtaining some of the observations reported here. M. R. D. is supported in part by the NSF Graduate Research Fellowship.  M.L.G.Õs position in the supernova research group at U.C. Berkeley is supported by Gary and Cynthia Bengier and NSF grant AST-1211916.  E.~Y.~H. acknowledges the generous support provided by the Danish Agency for Science and Technology and Innovation through a Sapere Aude Level 2 grant.

This paper includes data gathered with the 6.5 m Magellan Telescopes located at Las Campanas Observatory, Chile. Some observations reported here were obtained at the MMT observatory, a joint facility of the Smithsonian Institution and the University of Arizona. This paper uses data taken with the MODS spectrographs built with funding from NSF grant AST-9987045 and the NSF Telescope System Instrumentation Program (TSIP), with additional funds from the Ohio Board of Regents and the Ohio State University Office of Research.

\emph{Facilities:} \emph{Swift}-UVOT, Magellan:Baade (IMACS, FIRE), Magellan:Clay (LDSS3), MMT (Blue Channel spectrograph, Hectospec, MMTCam), LBT (MODS), CAO, FLWO (FAST, KeplerCam), Chandra, VLA, Lick:Shane (Kast), MDM (OSMOS)


\begin{thebibliography}{161}
\expandafter\ifx\csname natexlab\endcsname\relax\def\natexlab#1{#1}\fi

\bibitem[{{Allington-Smith} {et~al.}(1994){Allington-Smith}, {Breare}, {Ellis},
  {Gellatly}, {Glazebrook}, {Jorden}, {Maclean}, {Oates}, {Shaw}, {Tanvir},
  {Taylor}, {Taylor}, {Webster}, \& {Worswick}}]{Allington-Smith1994}
{Allington-Smith}, J., {Breare}, M., {Ellis}, R., {et~al.} 1994, \pasp, 106,
  983

\bibitem[{{Arcavi} {et~al.}(2011){Arcavi}, {Gal-Yam}, {Yaron}, {Sternberg},
  {Rabinak}, {Waxman}, {Kasliwal}, {Quimby}, {Ofek}, {Horesh}, {Kulkarni},
  {Filippenko}, {Silverman}, {Cenko}, {Li}, {Bloom}, {Sullivan}, {Nugent},
  {Poznanski}, {Gorbikov}, {Fulton}, {Howell}, {Bersier}, {Riou},
  {Lamotte-Bailey}, {Griga}, {Cohen}, {Hachinger}, {Polishook}, {Xu},
  {Ben-Ami}, {Manulis}, {Walker}, {Maguire}, {Pan}, {Matheson}, {Mazzali},
  {Pian}, {Fox}, {Gehrels}, {Law}, {James}, {Marchant}, {Smith}, {Mottram},
  {Barnsley}, {Kandrashoff}, \& {Clubb}}]{Arcavi2011}
{Arcavi}, I., {Gal-Yam}, A., {Yaron}, O., {et~al.} 2011, \apjl, 742, L18

\bibitem[{Arnett(1982)}]{Arnett1982}
Arnett, W.~D. 1982, ApJ, 253, 785

\bibitem[{{Asplund} {et~al.}(2005){Asplund}, {Grevesse}, \&
  {Sauval}}]{Asplund2005}
{Asplund}, M., {Grevesse}, N., \& {Sauval}, A.~J. 2005, in Astronomical Society
  of the Pacific Conference Series, Vol. 336, Cosmic Abundances as Records of
  Stellar Evolution and Nucleosynthesis, ed. T.~G. {Barnes}, III \& F.~N.
  {Bash}, 25

\bibitem[{{Begelman} \& {Sarazin}(1986)}]{Begelman1986}
{Begelman}, M.~C., \& {Sarazin}, C.~L. 1986, \apjl, 302, L59

\bibitem[{{Ben-Ami} {et~al.}(2012){Ben-Ami}, {Gal-Yam}, {Filippenko},
  {Mazzali}, {Modjaz}, {Yaron}, {Arcavi}, {Cenko}, {Horesh}, {Howell},
  {Graham}, {Horst}, {Im}, {Jeon}, {Kulkarni}, {Leonard}, {Perley}, {Pian},
  {Sand}, {Sullivan}, {Becker}, {Bersier}, {Bloom}, {Bottom}, {Brown}, {Clubb},
  {Dilday}, {Dixon}, {Fortinsky}, {Fox}, {Gonzalez}, {Harutyunyan}, {Kasliwal},
  {Li}, {Malkan}, {Manulis}, {Matheson}, {Moskovitz}, {Muirhead}, {Nugent},
  {Ofek}, {Quimby}, {Richards}, {Ross}, {Searcy}, {Silverman}, {Smith},
  {Vanderburg}, \& {Walker}}]{Ben-Ami2012}
{Ben-Ami}, S., {Gal-Yam}, A., {Filippenko}, A.~V., {et~al.} 2012, \apjl, 760,
  L33

\bibitem[{{Benetti} {et~al.}(2011){Benetti}, {Turatto}, {Valenti},
  {Pastorello}, {Cappellaro}, {Botticella}, {Bufano}, {Ghinassi},
  {Harutyunyan}, {Inserra}, {Magazz{\~a}{\sup1}}, {Patat}, {Pumo}, \&
  {Taubenberger}}]{Benetti11}
{Benetti}, S., {Turatto}, M., {Valenti}, S., {et~al.} 2011, \mnras, 411, 2726

\bibitem[{{Berger} {et~al.}(2002){Berger}, {Kulkarni}, \&
  {Chevalier}}]{Berger2002}
{Berger}, E., {Kulkarni}, S.~R., \& {Chevalier}, R.~A. 2002, \apjl, 577, L5

\bibitem[{{Bersten} {et~al.}(2013){Bersten}, {Tanaka}, {Tominaga}, {Benvenuto},
  \& {Nomoto}}]{Bersten2013}
{Bersten}, M.~C., {Tanaka}, M., {Tominaga}, N., {Benvenuto}, O.~G., \&
  {Nomoto}, K. 2013, \apj, 767, 143

\bibitem[{{Bersten} {et~al.}(2012){Bersten}, {Benvenuto}, {Nomoto}, {Ergon},
  {Folatelli}, {Sollerman}, {Benetti}, {Botticella}, {Fraser}, {Kotak},
  {Maeda}, {Ochner}, \& {Tomasella}}]{Bersten2012}
{Bersten}, M.~C., {Benvenuto}, O.~G., {Nomoto}, K., {et~al.} 2012, \apj, 757,
  31

\bibitem[{Blondin {et~al.}(2012)Blondin, Matheson, Kirshner, Mandel, Berlind,
  Calkins, Challis, Garnavich, Jha, Modjaz, Riess, \& Schmidt}]{Blondin2012}
Blondin, S., Matheson, T., Kirshner, R.~P., {et~al.} 2012, AJ, 143, 126

\bibitem[{{Branch} {et~al.}(2006){Branch}, {Jeffery}, {Young}, \&
  {Baron}}]{Branch06}
{Branch}, D., {Jeffery}, D.~J., {Young}, T.~R., \& {Baron}, E. 2006, \pasp,
  118, 791

\bibitem[{{Branch} {et~al.}(2002){Branch}, {Benetti}, {Kasen}, {Baron},
  {Jeffery}, {Hatano}, {Stathakis}, {Filippenko}, {Matheson}, {Pastorello},
  {Altavilla}, {Cappellaro}, {Rizzi}, {Turatto}, {Li}, {Leonard}, \&
  {Shields}}]{Branch02}
{Branch}, D., {Benetti}, S., {Kasen}, D., {et~al.} 2002, \apj, 566, 1005

\bibitem[{{Breeveld} {et~al.}(2011){Breeveld}, {Landsman}, {Holland}, {Roming},
  {Kuin}, \& {Page}}]{Breeveld2011}
{Breeveld}, A.~A., {Landsman}, W., {Holland}, S.~T., {et~al.} 2011, in American
  Institute of Physics Conference Series, Vol. 1358, American Institute of
  Physics Conference Series, ed. J.~E. {McEnery}, J.~L. {Racusin}, \&
  N.~{Gehrels}, 373--376

\bibitem[{{Brown} {et~al.}(2009){Brown}, {Holland}, {Immler}, {Milne},
  {Roming}, {Gehrels}, {Nousek}, {Panagia}, {Still}, \& {Vanden
  Berk}}]{Brown09}
{Brown}, P.~J., {Holland}, S.~T., {Immler}, S., {et~al.} 2009, \aj, 137, 4517

\bibitem[{{Burrows} {et~al.}(2007){Burrows}, {Dessart}, {Livne}, {Ott}, \&
  {Murphy}}]{Burrows2007}
{Burrows}, A., {Dessart}, L., {Livne}, E., {Ott}, C.~D., \& {Murphy}, J. 2007,
  \apj, 664, 416

\bibitem[{{Campana} {et~al.}(2006){Campana}, {Mangano}, {Blustin}, {Brown},
  {Burrows}, {Chincarini}, {Cummings}, {Cusumano}, {Della Valle}, {Malesani},
  {M{\'e}sz{\'a}ros}, {Nousek}, {Page}, {Sakamoto}, {Waxman}, {Zhang}, {Dai},
  {Gehrels}, {Immler}, {Marshall}, {Mason}, {Moretti}, {O'Brien}, {Osborne},
  {Page}, {Romano}, {Roming}, {Tagliaferri}, {Cominsky}, {Giommi}, {Godet},
  {Kennea}, {Krimm}, {Angelini}, {Barthelmy}, {Boyd}, {Palmer}, {Wells}, \&
  {White}}]{Campana2006}
{Campana}, S., {Mangano}, V., {Blustin}, A.~J., {et~al.} 2006, \nat, 442, 1008

\bibitem[{{Cao} {et~al.}(2013){Cao}, {Kasliwal}, {Arcavi}, {Horesh}, {Hancock},
  {Valenti}, {Cenko}, {Kulkarni}, {Gal-Yam}, {Gorbikov}, {Ofek}, {Sand},
  {Yaron}, {Graham}, {Silverman}, {Wheeler}, {Marion}, {Walker}, {Mazzali},
  {Howell}, {Li}, {Kong}, {Bloom}, {Nugent}, {Surace}, {Masci}, {Carpenter},
  {Degenaar}, \& {Gelino}}]{Cao2013}
{Cao}, Y., {Kasliwal}, M.~M., {Arcavi}, I., {et~al.} 2013, \apjl, 775, L7

\bibitem[{{Chakraborti} {et~al.}(2015){Chakraborti}, {Soderberg}, {Chomiuk},
  {Kamble}, {Yadav}, {Ray}, {Hurley}, {Margutti}, {Milisavljevic},
  {Bietenholz}, {Brunthaler}, {Pignata}, {Pian}, {Mazzali}, {Fransson},
  {Bartel}, {Hamuy}, {Levesque}, {MacFadyen}, {Dittmann}, {Krauss}, {Briggs},
  {Connaughton}, {Yamaoka}, {Takahashi}, {Ohno}, {Fukazawa}, {Tashiro},
  {Terada}, {Murakami}, {Goldsten}, {Barthelmy}, {Gehrels}, {Cummings},
  {Krimm}, {Palmer}, {Golenetskii}, {Aptekar}, {Frederiks}, {Svinkin}, {Cline},
  {Mitrofanov}, {Golovin}, {Litvak}, {Sanin}, {Boynton}, {Fellows}, {Harshman},
  {Enos}, {von Kienlin}, {Rau}, {Zhang}, \& {Savchenko}}]{Chakraborti2015}
{Chakraborti}, S., {Soderberg}, A., {Chomiuk}, L., {et~al.} 2015, \apj, 805,
  187

\bibitem[{{Chen} {et~al.}(2014){Chen}, {Wang}, {Ganeshalingam}, {Silverman},
  {Filippenko}, {Li}, {Chornock}, {Li}, \& {Steele}}]{Chen2014}
{Chen}, J., {Wang}, X., {Ganeshalingam}, M., {et~al.} 2014, \apj, 790, 120

\bibitem[{{Chevalier}(1998)}]{Chevalier1998}
{Chevalier}, R.~A. 1998, \apj, 499, 810

\bibitem[{{Chevalier} \& {Fransson}(2006)}]{Chevalier2006}
{Chevalier}, R.~A., \& {Fransson}, C. 2006, \apj, 651, 381

\bibitem[{Corsi {et~al.}(2012)Corsi, Ofek, Gal-Yam, Frail, Poznanski, Mazzali,
  Kulkarni, Kasliwal, Arcavi, Ben-Ami, Cenko, Filippenko, Fox, Horesh, Howell,
  Kleiser, Nakar, Rabinak, Sari, Silverman, Xu, Bloom, Law, Nugent, \&
  Quimby}]{Corsi2012}
Corsi, a., Ofek, E.~O., Gal-Yam, a., {et~al.} 2012, ApJ, 747, L5

\bibitem[{{Crowther}(2007)}]{Crowther2007}
{Crowther}, P.~A. 2007, \araa, 45, 177

\bibitem[{Crowther(2012)}]{Crowther2012}
Crowther, P.~a. 2012, MNRAS, 1943, 1927

\bibitem[{{de Jager} {et~al.}(1988){de Jager}, {Nieuwenhuijzen}, \& {van der
  Hucht}}]{deJager1988}
{de Jager}, C., {Nieuwenhuijzen}, H., \& {van der Hucht}, K.~A. 1988, \aaps,
  72, 259

\bibitem[{{Dessart} {et~al.}(2012){Dessart}, {Hillier}, {Li}, \&
  {Woosley}}]{Dessart2012}
{Dessart}, L., {Hillier}, D.~J., {Li}, C., \& {Woosley}, S. 2012, \mnras, 424,
  2139

\bibitem[{{Dressler} {et~al.}(2006){Dressler}, {Hare}, {Bigelow}, \&
  {Osip}}]{Dressler2006}
{Dressler}, A., {Hare}, T., {Bigelow}, B.~C., \& {Osip}, D.~J. 2006, in Society
  of Photo-Optical Instrumentation Engineers (SPIE) Conference Series, Vol.
  6269, Society of Photo-Optical Instrumentation Engineers (SPIE) Conference
  Series

\bibitem[{Drout {et~al.}(2011)Drout, Soderberg, Gal-Yam, Cenko, Fox, Leonard,
  Sand, Moon, Arcavi, \& Green}]{Drout2011}
Drout, M.~R., Soderberg, A.~M., Gal-Yam, A., {et~al.} 2011, ApJ, 741, 97

\bibitem[{{Drout} {et~al.}(2013){Drout}, {Soderberg}, {Mazzali}, {Parrent},
  {Margutti}, {Milisavljevic}, {Sanders}, {Chornock}, {Foley}, {Kirshner},
  {Filippenko}, {Li}, {Brown}, {Cenko}, {Chakraborti}, {Challis}, {Friedman},
  {Ganeshalingam}, {Hicken}, {Jensen}, {Modjaz}, {Perets}, {Silverman}, \&
  {Wong}}]{Drout2013}
{Drout}, M.~R., {Soderberg}, A.~M., {Mazzali}, P.~A., {et~al.} 2013, \apj, 774,
  58

\bibitem[{{Elias-Rosa} {et~al.}(2009){Elias-Rosa}, {Van Dyk}, {Li}, {Morrell},
  {Gonzalez}, {Hamuy}, {Filippenko}, {Cuillandre}, {Foley}, \&
  {Smith}}]{EliasRosa2009}
{Elias-Rosa}, N., {Van Dyk}, S.~D., {Li}, W., {et~al.} 2009, \apj, 706, 1174

\bibitem[{{Elias-Rosa} {et~al.}(2010){Elias-Rosa}, {Van Dyk}, {Li}, {Miller},
  {Silverman}, {Ganeshalingam}, {Boden}, {Kasliwal}, {Vink{\'o}}, {Cuillandre},
  {Filippenko}, {Steele}, {Bloom}, {Griffith}, {Kleiser}, \&
  {Foley}}]{EliasRosa2010}
---. 2010, \apjl, 714, L254

\bibitem[{{Elmhamdi} {et~al.}(2007){Elmhamdi}, {Danziger}, {Branch}, \&
  {Leibundgut}}]{Elmhamdi07}
{Elmhamdi}, A., {Danziger}, I.~J., {Branch}, D., \& {Leibundgut}, B. 2007, in
  American Institute of Physics Conference Series, Vol. 924, The Multicolored
  Landscape of Compact Objects and Their Explosive Origins, ed. T.~{di Salvo},
  G.~L. {Israel}, L.~{Piersant}, L.~{Burderi}, G.~{Matt}, A.~{Tornambe}, \&
  M.~T. {Menna}, 277--284

\bibitem[{Fabricant {et~al.}(1998)Fabricant, Cheimets, Caldwell, \&
  Geary}]{Fabricant1998}
Fabricant, D., Cheimets, P., Caldwell, N., \& Geary, J. 1998, PASP, 110, 79

\bibitem[{{Fabricant} {et~al.}(2005){Fabricant}, {Fata}, {Roll}, {Hertz},
  {Caldwell}, {Gauron}, {Geary}, {McLeod}, {Szentgyorgyi}, {Zajac}, {Kurtz},
  {Barberis}, {Bergner}, {Brown}, {Conroy}, {Eng}, {Geller}, {Goddard},
  {Honsa}, {Mueller}, {Mink}, {Ordway}, {Tokarz}, {Woods}, {Wyatt}, {Epps}, \&
  {Dell'Antonio}}]{Fabricant2005}
{Fabricant}, D., {Fata}, R., {Roll}, J., {et~al.} 2005, \pasp, 117, 1411

\bibitem[{{Fesen}(2001)}]{Fesen2001}
{Fesen}, R.~A. 2001, \apjs, 133, 161

\bibitem[{Filippenko(1997)}]{Filippenko1997}
Filippenko, A.~V. 1997, ARA\&A

\bibitem[{{Filippenko} {et~al.}(1995){Filippenko}, {Barth}, {Matheson},
  {Armus}, {Brown}, {Espey}, {Fan}, {Goodrich}, {Ho}, {Junkkarinen}, {Koo},
  {Lehnert}, {Martel}, {Mazzarella}, {Miller}, {Smith}, {Tytler}, \&
  {Wirth}}]{Filippenko1995}
{Filippenko}, A.~V., {Barth}, A.~J., {Matheson}, T., {et~al.} 1995, \apjl, 450,
  L11

\bibitem[{{Folatelli} {et~al.}(2006){Folatelli}, {Contreras}, {Phillips},
  {Woosley}, {Blinnikov}, {Morrell}, {Suntzeff}, {Lee}, {Hamuy},
  {Gonz{\'a}lez}, {Krzeminski}, {Roth}, {Li}, {Filippenko}, {Foley},
  {Freedman}, {Madore}, {Persson}, {Murphy}, {Boissier}, {Galaz},
  {Gonz{\'a}lez}, {McCarthy}, {McWilliam}, \& {Pych}}]{Folatelli2006}
{Folatelli}, G., {Contreras}, C., {Phillips}, M.~M., {et~al.} 2006, \apj, 641,
  1039

\bibitem[{Foley {et~al.}(2013)Foley, Challis, Chornock, Ganeshalingam, Li,
  Marion, Morrell, Pignata, Stritzinger, Silverman, Wang, Anderson, Filippenko,
  Freedman, Hamuy, Jha, Kirshner, McCully, Persson, Phillips, Reichart, \&
  Soderberg}]{Foley2013a}
Foley, R.~J., Challis, P.~J., Chornock, R., {et~al.} 2013, ApJ, 767, 57

\bibitem[{{Fransson} \& {Chevalier}(1989)}]{Fransson1989}
{Fransson}, C., \& {Chevalier}, R.~A. 1989, \apj, 343, 323

\bibitem[{{Fraser} {et~al.}(2010){Fraser}, {Tak{\'a}ts}, {Pastorello},
  {Smartt}, {Mattila}, {Botticella}, {Valenti}, {Ergon}, {Sollerman}, {Arcavi},
  {Benetti}, {Bufano}, {Crockett}, {Danziger}, {Gal-Yam}, {Maund},
  {Taubenberger}, \& {Turatto}}]{Fraser2010}
{Fraser}, M., {Tak{\'a}ts}, K., {Pastorello}, A., {et~al.} 2010, \apjl, 714,
  L280

\bibitem[{{Fremling} {et~al.}(2014){Fremling}, {Sollerman}, {Taddia}, {Ergon},
  {Valenti}, {Arcavi}, {Ben-Ami}, {Cao}, {Cenko}, {Filippenko}, {Gal-Yam}, \&
  {Howell}}]{Fremling2014}
{Fremling}, C., {Sollerman}, J., {Taddia}, F., {et~al.} 2014, \aap, 565, A114

\bibitem[{{Gehrels} {et~al.}(2004){Gehrels}, {Chincarini}, {Giommi}, {Mason},
  {Nousek}, {Wells}, {White}, {Barthelmy}, {Burrows}, {Cominsky}, {Hurley},
  {Marshall}, {M{\'e}sz{\'a}ros}, {Roming}, {Angelini}, {Barbier}, {Belloni},
  {Campana}, {Caraveo}, {Chester}, {Citterio}, {Cline}, {Cropper}, {Cummings},
  {Dean}, {Feigelson}, {Fenimore}, {Frail}, {Fruchter}, {Garmire}, {Gendreau},
  {Ghisellini}, {Greiner}, {Hill}, {Hunsberger}, {Krimm}, {Kulkarni}, {Kumar},
  {Lebrun}, {Lloyd-Ronning}, {Markwardt}, {Mattson}, {Mushotzky}, {Norris},
  {Osborne}, {Paczynski}, {Palmer}, {Park}, {Parsons}, {Paul}, {Rees},
  {Reynolds}, {Rhoads}, {Sasseen}, {Schaefer}, {Short}, {Smale}, {Smith},
  {Stella}, {Tagliaferri}, {Takahashi}, {Tashiro}, {Townsley}, {Tueller},
  {Turner}, {Vietri}, {Voges}, {Ward}, {Willingale}, {Zerbi}, \&
  {Zhang}}]{Gehrels2004}
{Gehrels}, N., {Chincarini}, G., {Giommi}, P., {et~al.} 2004, \apj, 611, 1005

\bibitem[{{Gerardy} {et~al.}(2002){Gerardy}, {Fesen}, {Nomoto}, {Maeda},
  {Hoflich}, \& {Wheeler}}]{Gerardy2002}
{Gerardy}, C.~L., {Fesen}, R.~A., {Nomoto}, K., {et~al.} 2002, \pasj, 54, 905

\bibitem[{{Gorbikov} {et~al.}(2014){Gorbikov}, {Gal-Yam}, {Ofek}, {Vreeswijk},
  {Nugent}, {Chotard}, {Kulkarni}, {Cao}, {De Cia}, {Yaron}, {Tal}, {Arcavi},
  {Kasliwal}, {Cenko}, {Sullivan}, \& {Chen}}]{Gorbikov2014}
{Gorbikov}, E., {Gal-Yam}, A., {Ofek}, E.~O., {et~al.} 2014, \mnras, 443, 671

\bibitem[{{Grefenstette} {et~al.}(2014){Grefenstette}, {Harrison}, {Boggs},
  {Reynolds}, {Fryer}, {Madsen}, {Wik}, {Zoglauer}, {Ellinger}, {Alexander},
  {An}, {Barret}, {Christensen}, {Craig}, {Forster}, {Giommi}, {Hailey},
  {Hornstrup}, {Kaspi}, {Kitaguchi}, {Koglin}, {Mao}, {Miyasaka}, {Mori},
  {Perri}, {Pivovaroff}, {Puccetti}, {Rana}, {Stern}, {Westergaard}, \&
  {Zhang}}]{Grefenstette2014}
{Grefenstette}, B.~W., {Harrison}, F.~A., {Boggs}, S.~E., {et~al.} 2014, \nat,
  506, 339

\bibitem[{Hachinger {et~al.}(2012)Hachinger, Mazzali, Taubenberger,
  Hillebrandt, Nomoto, \& Sauer}]{Hachinger2012}
Hachinger, S., Mazzali, P.~a., Taubenberger, S., {et~al.} 2012, MNRAS, 422, 70

\bibitem[{{Hammer} {et~al.}(2010){Hammer}, {Janka}, \&
  {M{\"u}ller}}]{Hammer2010}
{Hammer}, N.~J., {Janka}, H.-T., \& {M{\"u}ller}, E. 2010, \apj, 714, 1371

\bibitem[{{Hicken} {et~al.}(2007){Hicken}, {Garnavich}, {Prieto}, {Blondin},
  {DePoy}, {Kirshner}, \& {Parrent}}]{Hicken2007}
{Hicken}, M., {Garnavich}, P.~M., {Prieto}, J.~L., {et~al.} 2007, \apjl, 669,
  L17

\bibitem[{{Hicken} {et~al.}(2012){Hicken}, {Challis}, {Kirshner}, {Rest},
  {Cramer}, {Wood-Vasey}, {Bakos}, {Berlind}, {Brown}, {Caldwell}, {Calkins},
  {Currie}, {de Kleer}, {Esquerdo}, {Everett}, {Falco}, {Fernandez},
  {Friedman}, {Groner}, {Hartman}, {Holman}, {Hutchins}, {Keys}, {Kipping},
  {Latham}, {Marion}, {Narayan}, {Pahre}, {Pal}, {Peters}, {Perumpilly},
  {Ripman}, {Sipocz}, {Szentgyorgyi}, {Tang}, {Torres}, {Vaz}, {Wolk}, \&
  {Zezas}}]{Hicken2012}
{Hicken}, M., {Challis}, P., {Kirshner}, R.~P., {et~al.} 2012, \apjs, 200, 12

\bibitem[{{Hsiao} {et~al.}(2015){Hsiao}, {Burns}, {Contreras}, {H{\"o}flich},
  {Sand}, {Marion}, {Phillips}, {Stritzinger}, {Gonz{\'a}lez-Gait{\'a}n},
  {Mason}, {Folatelli}, {Parent}, {Gall}, {Amanullah}, {Anupama}, {Arcavi},
  {Banerjee}, {Beletsky}, {Blanc}, {Bloom}, {Brown}, {Campillay}, {Cao}, {De
  Cia}, {Diamond}, {Freedman}, {Gonzalez}, {Goobar}, {Holmbo}, {Howell},
  {Johansson}, {Kasliwal}, {Kirshner}, {Krisciunas}, {Kulkarni}, {Maguire},
  {Milne}, {Morrell}, {Nugent}, {Ofek}, {Osip}, {Palunas}, {Perley}, {Persson},
  {Piro}, {Rabus}, {Roth}, {Schiefelbein}, {Srivastav}, {Sullivan}, {Suntzeff},
  {Surace}, {Wo{\'z}niak}, \& {Yaron}}]{Hsiao2015}
{Hsiao}, E.~Y., {Burns}, C.~R., {Contreras}, C., {et~al.} 2015, \aap, 578, A9

\bibitem[{Hunter {et~al.}(2009)Hunter, Valenti, Kotak, Meikle, Taubenberger,
  Pastorello, Benetti, Stanishev, Smartt, Trundle, Arkharov, Bufano,
  Cappellaro, {Di Carlo}, Dolci, Elias-Rosa, Frandsen, Fynbo, Hopp, Larionov,
  Laursen, Mazzali, Navasardyan, Ries, Riffeser, Rizzi, Tsvetkov, Turatto, \&
  Wilke}]{Hunter2009}
Hunter, D.~J., Valenti, S., Kotak, R., {et~al.} 2009, A\&A, 508, 371

\bibitem[{{Hwang} {et~al.}(2004){Hwang}, {Laming}, {Badenes}, {Berendse},
  {Blondin}, {Cioffi}, {DeLaney}, {Dewey}, {Fesen}, {Flanagan}, {Fryer},
  {Ghavamian}, {Hughes}, {Morse}, {Plucinsky}, {Petre}, {Pohl}, {Rudnick},
  {Sankrit}, {Slane}, {Smith}, {Vink}, \& {Warren}}]{Hwang2004}
{Hwang}, U., {Laming}, J.~M., {Badenes}, C., {et~al.} 2004, \apjl, 615, L117

\bibitem[{{Jerkstrand} {et~al.}(2015){Jerkstrand}, {Ergon}, {Smartt},
  {Fransson}, {Sollerman}, {Taubenberger}, {Bersten}, \&
  {Spyromilio}}]{Jerkstrand2015}
{Jerkstrand}, A., {Ergon}, M., {Smartt}, S.~J., {et~al.} 2015, \aap, 573, A12

\bibitem[{{Joggerst} {et~al.}(2009){Joggerst}, {Woosley}, \&
  {Heger}}]{Joggerst2009}
{Joggerst}, C.~C., {Woosley}, S.~E., \& {Heger}, A. 2009, \apj, 693, 1780

\bibitem[{{Kalberla} {et~al.}(2005){Kalberla}, {Burton}, {Hartmann}, {Arnal},
  {Bajaja}, {Morras}, \& {P{\"o}ppel}}]{Kalberla05}
{Kalberla}, P.~M.~W., {Burton}, W.~B., {Hartmann}, D., {et~al.} 2005, \aap,
  440, 775

\bibitem[{{Kamble} {et~al.}(2014){Kamble}, {Soderberg}, {Chomiuk}, {Margutti},
  {Medvedev}, {Milisavljevic}, {Chakraborti}, {Chevalier}, {Chugai},
  {Dittmann}, {Drout}, {Fransson}, {Nakar}, \& {Sanders}}]{Kamble2014}
{Kamble}, A., {Soderberg}, A.~M., {Chomiuk}, L., {et~al.} 2014, \apj, 797, 2

\bibitem[{{Kamble} {et~al.}(2015){Kamble}, {Margutti}, {Soderberg},
  {Chakraborti}, {Fransson}, {Chevalier}, {Powell}, {Milisavljevic}, {Parrent},
  \& {Bietenholz}}]{Kamble2015}
{Kamble}, A., {Margutti}, R., {Soderberg}, A.~M., {et~al.} 2015, ArXiv e-prints

\bibitem[{{Kasen}(2010)}]{Kasen2010}
{Kasen}, D. 2010, \apj, 708, 1025

\bibitem[{Kennicutt(1998)}]{Kennicutt1998}
Kennicutt, R.~C. 1998, ARA\&A, 36, 189

\bibitem[{{Kifonidis} {et~al.}(2006){Kifonidis}, {Plewa}, {Scheck}, {Janka}, \&
  {M{\"u}ller}}]{Kifonidis2006}
{Kifonidis}, K., {Plewa}, T., {Scheck}, L., {Janka}, H.-T., \& {M{\"u}ller}, E.
  2006, \aap, 453, 661

\bibitem[{{Kleiser} \& {Kasen}(2014)}]{Kleiser2014}
{Kleiser}, I.~K.~W., \& {Kasen}, D. 2014, \mnras, 438, 318

\bibitem[{{Krauss} {et~al.}(2012){Krauss}, {Soderberg}, {Chomiuk}, {Zauderer},
  {Brunthaler}, {Bietenholz}, {Chevalier}, {Fransson}, \& {Rupen}}]{Krauss2012}
{Krauss}, M.~I., {Soderberg}, A.~M., {Chomiuk}, L., {et~al.} 2012, \apjl, 750,
  L40

\bibitem[{{Laming} {et~al.}(2006){Laming}, {Hwang}, {Radics}, {Lekli}, \&
  {Tak{\'a}cs}}]{Laming2006}
{Laming}, J.~M., {Hwang}, U., {Radics}, B., {Lekli}, G., \& {Tak{\'a}cs}, E.
  2006, \apj, 644, 260

\bibitem[{{Landolt}(1992)}]{Landolt1992}
{Landolt}, A.~U. 1992, \aj, 104, 340

\bibitem[{Levesque {et~al.}(2010)Levesque, Soderberg, Foley, Berger, Kewley,
  Chakraborti, Ray, Torres, Challis, Kirshner, Barthelmy, Bietenholz, Chandra,
  Chaplin, Chevalier, Chugai, Connaughton, Copete, Fox, Fransson, Grindlay,
  Hamuy, Milne, Pignata, Stritzinger, \& Wieringa}]{Levesque2010}
Levesque, E.~M., Soderberg, a.~M., Foley, R.~J., {et~al.} 2010, ApJ, 709, L26

\bibitem[{{Li}(2007)}]{Li2007}
{Li}, L.-X. 2007, \mnras, 375, 240

\bibitem[{{Liu} {et~al.}(2015){Liu}, {Modjaz}, {Bianco}, \& {Graur}}]{Liu2015}
{Liu}, Y.-Q., {Modjaz}, M., {Bianco}, F.~B., \& {Graur}, O. 2015, ArXiv
  e-prints

\bibitem[{{Lucy}(1991)}]{Lucy1991}
{Lucy}, L.~B. 1991, \apj, 383, 308

\bibitem[{{Lyman} {et~al.}(2014){Lyman}, {Bersier}, {James}, {Mazzali},
  {Eldridge}, {Fraser}, \& {Pian}}]{Lyman2014}
{Lyman}, J., {Bersier}, D., {James}, P., {et~al.} 2014, ArXiv e-prints

\bibitem[{{Maeda} {et~al.}(2003){Maeda}, {Mazzali}, {Deng}, {Nomoto}, {Yoshii},
  {Tomita}, \& {Kobayashi}}]{Maeda2003}
{Maeda}, K., {Mazzali}, P.~A., {Deng}, J., {et~al.} 2003, \apj, 593, 931

\bibitem[{{Maeda} {et~al.}(2002){Maeda}, {Nakamura}, {Nomoto}, {Mazzali},
  {Patat}, \& {Hachisu}}]{Maeda2002}
{Maeda}, K., {Nakamura}, T., {Nomoto}, K., {et~al.} 2002, \apj, 565, 405

\bibitem[{{Maeda} {et~al.}(2007){Maeda}, {Tanaka}, {Nomoto}, {Tominaga},
  {Kawabata}, {Mazzali}, {Umeda}, {Suzuki}, \& {Hattori}}]{Maeda2007}
{Maeda}, K., {Tanaka}, M., {Nomoto}, K., {et~al.} 2007, \apj, 666, 1069

\bibitem[{{Marek} \& {Janka}(2009)}]{Marek2009}
{Marek}, A., \& {Janka}, H.-T. 2009, \apj, 694, 664

\bibitem[{{Margutti} {et~al.}(2014){Margutti}, {Parrent}, {Kamble},
  {Soderberg}, {Foley}, {Milisavljevic}, {Drout}, \&
  {Kirshner}}]{Margutti2014b}
{Margutti}, R., {Parrent}, J., {Kamble}, A., {et~al.} 2014, \apj, 790, 52

\bibitem[{{Margutti} {et~al.}(2012){Margutti}, {Soderberg}, {Chomiuk},
  {Chevalier}, {Hurley}, {Milisavljevic}, {Foley}, {Hughes}, {Slane},
  {Fransson}, {Moe}, {Barthelmy}, {Boynton}, {Briggs}, {Connaughton}, {Costa},
  {Cummings}, {Del Monte}, {Enos}, {Fellows}, {Feroci}, {Fukazawa}, {Gehrels},
  {Goldsten}, {Golovin}, {Hanabata}, {Harshman}, {Krimm}, {Litvak},
  {Makishima}, {Marisaldi}, {Mitrofanov}, {Murakami}, {Ohno}, {Palmer},
  {Sanin}, {Starr}, {Svinkin}, {Takahashi}, {Tashiro}, {Terada}, \&
  {Yamaoka}}]{Margutti2012}
{Margutti}, R., {Soderberg}, A.~M., {Chomiuk}, L., {et~al.} 2012, \apj, 751,
  134

\bibitem[{{Margutti} {et~al.}(2015){Margutti}, {Guidorzi}, {Lazzati},
  {Milisavljevic}, {Kamble}, {Laskar}, {Parrent}, {Gehrels}, \&
  {Soderberg}}]{Margutti2015}
{Margutti}, R., {Guidorzi}, C., {Lazzati}, D., {et~al.} 2015, \apj, 805, 159

\bibitem[{{Margutti} {et~al.}(2016){Margutti}, {Kamble}, {Milisavljevic}, {De
  Mink}, {Zapartas}, {Drout}, {Chornock}, {Risaliti}, {Zauderer}, {Bietenholz},
  {Cantiello}, {Chakraborti}, {Chomiuk}, {Fong}, {Grefenstette}, {Guidorzi},
  {Kirshner}, {Parrent}, {Patnaude}, {Soderberg}, {Gehrels}, \&
  {Harrison}}]{Margutti2016}
{Margutti}, R., {Kamble}, A., {Milisavljevic}, D., {et~al.} 2016, ArXiv
  e-prints

\bibitem[{{Marshall} {et~al.}(2004){Marshall}, {van Loon}, {Matsuura}, {Wood},
  {Zijlstra}, \& {Whitelock}}]{Marshall2004}
{Marshall}, J.~R., {van Loon}, J.~T., {Matsuura}, M., {et~al.} 2004, \mnras,
  355, 1348

\bibitem[{{Martini} {et~al.}(2011){Martini}, {Stoll}, {Derwent}, {Zhelem},
  {Atwood}, {Gonzalez}, {Mason}, {O'Brien}, {Pappalardo}, {Pogge}, {Ward}, \&
  {Wong}}]{Martini2011}
{Martini}, P., {Stoll}, R., {Derwent}, M.~A., {et~al.} 2011, \pasp, 123, 187

\bibitem[{{Massey} {et~al.}(2015){Massey}, {Neugent}, {Morrell}, \& {John
  Hillier}}]{Massey2015}
{Massey}, P., {Neugent}, K.~F., {Morrell}, N., \& {John Hillier}, D. 2015, in
  IAU Symposium, Vol. 307, IAU Symposium, 64--69

\bibitem[{Matheson {et~al.}(2008)Matheson, Kirshner, Challis, Jha, Garnavich,
  Berlind, Calkins, Blondin, Balog, Bragg, Caldwell, Concannon, Falco, Graves,
  Huchra, Kuraszkiewicz, Mader, Mahdavi, Phelps, Rines, Song, \&
  Wilkes}]{Matheson2008}
Matheson, T., Kirshner, R.~P., Challis, P., {et~al.} 2008, AJ, 135, 1598

\bibitem[{{Matzner} \& {McKee}(1999)}]{Matzner1999}
{Matzner}, C.~D., \& {McKee}, C.~F. 1999, \apj, 510, 379

\bibitem[{{Maund} {et~al.}(2004){Maund}, {Smartt}, {Kudritzki},
  {Podsiadlowski}, \& {Gilmore}}]{Maund2004}
{Maund}, J.~R., {Smartt}, S.~J., {Kudritzki}, R.~P., {Podsiadlowski}, P., \&
  {Gilmore}, G.~F. 2004, \nat, 427, 129

\bibitem[{{Mazzali} {et~al.}(2005{\natexlab{a}}){Mazzali}, {Benetti}, {Stehle},
  {Branch}, {Deng}, {Maeda}, {Nomoto}, \& {Hamuy}}]{Mazzali2005b}
{Mazzali}, P.~A., {Benetti}, S., {Stehle}, M., {et~al.} 2005{\natexlab{a}},
  \mnras, 357, 200

\bibitem[{{Mazzali} {et~al.}(2005{\natexlab{b}}){Mazzali}, {Kawabata}, {Maeda},
  {Nomoto}, {Filippenko}, {Ramirez-Ruiz}, {Benetti}, {Pian}, {Deng},
  {Tominaga}, {Ohyama}, {Iye}, {Foley}, {Matheson}, {Wang}, \&
  {Gal-Yam}}]{Mazzali2005}
{Mazzali}, P.~A., {Kawabata}, K.~S., {Maeda}, K., {et~al.} 2005{\natexlab{b}},
  Science, 308, 1284

\bibitem[{{Mazzali} {et~al.}(2008){Mazzali}, {Valenti}, {Della Valle},
  {Chincarini}, {Sauer}, {Benetti}, {Pian}, {Piran}, {D'Elia}, {Elias-Rosa},
  {Margutti}, {Pasotti}, {Antonelli}, {Bufano}, {Campana}, {Cappellaro},
  {Covino}, {D'Avanzo}, {Fiore}, {Fugazza}, {Gilmozzi}, {Hunter}, {Maguire},
  {Maiorano}, {Marziani}, {Masetti}, {Mirabel}, {Navasardyan}, {Nomoto},
  {Palazzi}, {Pastorello}, {Panagia}, {Pellizza}, {Sari}, {Smartt},
  {Tagliaferri}, {Tanaka}, {Taubenberger}, {Tominaga}, {Trundle}, \&
  {Turatto}}]{Mazzali2008b}
{Mazzali}, P.~A., {Valenti}, S., {Della Valle}, M., {et~al.} 2008, Science,
  321, 1185

\bibitem[{{Milisavljevic} \& {Fesen}(2013)}]{Milisavljevic2013d}
{Milisavljevic}, D., \& {Fesen}, R.~A. 2013, \apj, 772, 134

\bibitem[{{Milisavljevic} {et~al.}(2013){Milisavljevic}, {Fesen}, {Pickering},
  {Miszalski}, {Buckley}, {Parrent}, {Marion}, {Silverman}, {Vinko}, {Wheeler},
  {Quimby}, {Jha}, {Mohamed}, {Kasliwal}, \& {Soderberg}}]{Milisavljevic2013c}
{Milisavljevic}, D., {Fesen}, R., {Pickering}, T., {et~al.} 2013, The
  Astronomer's Telegram, 5142, 1

\bibitem[{{Milisavljevic} {et~al.}(2015){Milisavljevic}, {Margutti}, {Parrent},
  {Soderberg}, {Fesen}, {Mazzali}, {Maeda}, {Sanders}, {Cenko}, {Silverman},
  {Filippenko}, {Kamble}, {Chakraborti}, {Drout}, {Kirshner}, {Pickering},
  {Kawabata}, {Hattori}, {Hsiao}, {Stritzinger}, {Marion}, {Vinko}, \&
  {Wheeler}}]{Milisavljevic2015}
{Milisavljevic}, D., {Margutti}, R., {Parrent}, J.~T., {et~al.} 2015, \apj,
  799, 51

\bibitem[{{Millard} {et~al.}(1999){Millard}, {Branch}, {Baron}, {Hatano},
  {Fisher}, {Filippenko}, {Kirshner}, {Challis}, {Fransson}, {Panagia},
  {Phillips}, {Sonneborn}, {Suntzeff}, {Wagoner}, \& {Wheeler}}]{Millard1999}
{Millard}, J., {Branch}, D., {Baron}, E., {et~al.} 1999, \apj, 527, 746

\bibitem[{{Miller} \& {Stone}(1993)}]{Miller93}
{Miller}, J.~S., \& {Stone}, R.~P.~S. 1993, Lick Observatory Technical Reports,
  Vol.~66 (Santa Cruz, CA: Lick Obs.)

\bibitem[{{Modjaz} {et~al.}(2011){Modjaz}, {Kewley}, {Bloom}, {Filippenko},
  {Perley}, \& {Silverman}}]{Modjaz2011}
{Modjaz}, M., {Kewley}, L., {Bloom}, J.~S., {et~al.} 2011, \apjl, 731, L4

\bibitem[{{Modjaz} {et~al.}(2009{\natexlab{a}}){Modjaz}, {Li}, {Butler},
  {Chornock}, {Perley}, {Blondin}, {Bloom}, {Filippenko}, {Kirshner},
  {Kocevski}, {Poznanski}, {Hicken}, {Foley}, {Stringfellow}, {Berlind},
  {Barrado y Navascues}, {Blake}, {Bouy}, {Brown}, {Challis}, {Chen}, {de
  Vries}, {Dufour}, {Falco}, {Friedman}, {Ganeshalingam}, {Garnavich},
  {Holden}, {Illingworth}, {Lee}, {Liebert}, {Marion}, {Olivier}, {Prochaska},
  {Silverman}, {Smith}, {Starr}, {Steele}, {Stockton}, {Williams}, \&
  {Wood-Vasey}}]{Modjaz2009}
{Modjaz}, M., {Li}, W., {Butler}, N., {et~al.} 2009{\natexlab{a}}, \apj, 702,
  226

\bibitem[{{Modjaz} {et~al.}(2009{\natexlab{b}}){Modjaz}, {Li}, {Butler},
  {Chornock}, {Perley}, {Blondin}, {Bloom}, {Filippenko}, {Kirshner},
  {Kocevski}, {Poznanski}, {Hicken}, {Foley}, {Stringfellow}, {Berlind},
  {Barrado y Navascues}, {Blake}, {Bouy}, {Brown}, {Challis}, {Chen}, {de
  Vries}, {Dufour}, {Falco}, {Friedman}, {Ganeshalingam}, {Garnavich},
  {Holden}, {Illingworth}, {Lee}, {Liebert}, {Marion}, {Olivier}, {Prochaska},
  {Silverman}, {Smith}, {Starr}, {Steele}, {Stockton}, {Williams}, \&
  {Wood-Vasey}}]{Modjaz09}
---. 2009{\natexlab{b}}, \apj, 702, 226

\bibitem[{{Modjaz} {et~al.}(2014){Modjaz}, {Blondin}, {Kirshner}, {Matheson},
  {Berlind}, {Bianco}, {Calkins}, {Challis}, {Garnavich}, {Hicken}, {Jha},
  {Liu}, \& {Marion}}]{Modjaz2014}
{Modjaz}, M., {Blondin}, S., {Kirshner}, R.~P., {et~al.} 2014, \aj, 147, 99

\bibitem[{{Moriya} {et~al.}(2015){Moriya}, {Liu}, \& {Izzard}}]{Moriya2015}
{Moriya}, T.~J., {Liu}, Z.-W., \& {Izzard}, R.~G. 2015, \mnras, 450, 3264

\bibitem[{{Mould} {et~al.}(2000){Mould}, {Huchra}, {Freedman}, {Kennicutt},
  {Ferrarese}, {Ford}, {Gibson}, {Graham}, {Hughes}, {Illingworth}, {Kelson},
  {Macri}, {Madore}, {Sakai}, {Sebo}, {Silbermann}, \& {Stetson}}]{mhf+00}
{Mould}, J.~R., {Huchra}, J.~P., {Freedman}, W.~L., {et~al.} 2000, \apj, 529,
  786

\bibitem[{{Nakar}(2015)}]{Nakar2015}
{Nakar}, E. 2015, ArXiv e-prints

\bibitem[{{Nakar} \& {Piro}(2014)}]{Nakar2014}
{Nakar}, E., \& {Piro}, A.~L. 2014, \apj, 788, 193

\bibitem[{Nakar \& Sari(2010)}]{Nakar2010}
Nakar, E., \& Sari, R. 2010, ApJ, 725, 904

\bibitem[{{Nugent} {et~al.}(2011){Nugent}, {Sullivan}, {Cenko}, {Thomas},
  {Kasen}, {Howell}, {Bersier}, {Bloom}, {Kulkarni}, {Kandrashoff},
  {Filippenko}, {Silverman}, {Marcy}, {Howard}, {Isaacson}, {Maguire},
  {Suzuki}, {Tarlton}, {Pan}, {Bildsten}, {Fulton}, {Parrent}, {Sand},
  {Podsiadlowski}, {Bianco}, {Dilday}, {Graham}, {Lyman}, {James}, {Kasliwal},
  {Law}, {Quimby}, {Hook}, {Walker}, {Mazzali}, {Pian}, {Ofek}, {Gal-Yam}, \&
  {Poznanski}}]{Nugent2011}
{Nugent}, P.~E., {Sullivan}, M., {Cenko}, S.~B., {et~al.} 2011, \nat, 480, 344

\bibitem[{{Parrent} {et~al.}(2014){Parrent}, {Friesen}, \&
  {Parthasarathy}}]{Parrent14}
{Parrent}, J., {Friesen}, B., \& {Parthasarathy}, M. 2014, \apss, 351, 1

\bibitem[{{Parrent}(2014)}]{Parrent15}
{Parrent}, J.~T. 2014, ArXiv e-prints

\bibitem[{{Parrent} {et~al.}(2015){Parrent}, {Milisavljevic}, {Soderberg}, \&
  {Parthasarathy}}]{Parrent2015}
{Parrent}, J.~T., {Milisavljevic}, D., {Soderberg}, A.~M., \& {Parthasarathy},
  M. 2015, ArXiv e-prints

\bibitem[{Pastorello {et~al.}(2008)Pastorello, Kasliwal, Crockett, Valenti,
  Arbour, Itagaki, Kaspi, Gal-Yam, Smartt, Griffith, Maguire, Ofek, Seymour,
  Stern, \& Wiethoff}]{Pastorello2008}
Pastorello, a., Kasliwal, M.~M., Crockett, R.~M., {et~al.} 2008, MNRAS, 389,
  955

\bibitem[{Pettini \& Pagel(2004)}]{Pettini2004}
Pettini, M., \& Pagel, B. E.~J. 2004, MNRAS, 348, L59

\bibitem[{{Phillips} {et~al.}(2007){Phillips}, {Li}, {Frieman}, {Blinnikov},
  {DePoy}, {Prieto}, {Milne}, {Contreras}, {Folatelli}, {Morrell}, {Hamuy},
  {Suntzeff}, {Roth}, {Gonz{\'a}lez}, {Krzeminski}, {Filippenko}, {Freedman},
  {Chornock}, {Jha}, {Madore}, {Persson}, {Burns}, {Wyatt}, {Murphy}, {Foley},
  {Ganeshalingam}, {Serduke}, {Krisciunas}, {Bassett}, {Becker}, {Dilday},
  {Eastman}, {Garnavich}, {Holtzman}, {Kessler}, {Lampeitl}, {Marriner},
  {Frank}, {Marshall}, {Miknaitis}, {Sako}, {Schneider}, {van der Heyden}, \&
  {Yasuda}}]{Phillips2007}
{Phillips}, M.~M., {Li}, W., {Frieman}, J.~A., {et~al.} 2007, \pasp, 119, 360

\bibitem[{{Piro} \& {Morozova}(2014)}]{Piro2014}
{Piro}, A.~L., \& {Morozova}, V.~S. 2014, \apjl, 792, L11

\bibitem[{{Piro} \& {Nakar}(2013)}]{Piro2013}
{Piro}, A.~L., \& {Nakar}, E. 2013, \apj, 769, 67

\bibitem[{{Podsiadlowski} {et~al.}(1992){Podsiadlowski}, {Joss}, \&
  {Hsu}}]{Podsiadlowski1992}
{Podsiadlowski}, P., {Joss}, P.~C., \& {Hsu}, J.~J.~L. 1992, \apj, 391, 246

\bibitem[{{Pogge} {et~al.}(2010){Pogge}, {Atwood}, {Brewer}, {Byard},
  {Derwent}, {Gonzalez}, {Martini}, {Mason}, {O'Brien}, {Osmer}, {Pappalardo},
  {Steinbrecher}, {Teiga}, \& {Zhelem}}]{pab+10}
{Pogge}, R.~W., {Atwood}, B., {Brewer}, D.~F., {et~al.} 2010, in Society of
  Photo-Optical Instrumentation Engineers (SPIE) Conference Series, Vol. 7735,
  Society of Photo-Optical Instrumentation Engineers (SPIE) Conference Series,
  0

\bibitem[{{Poznanski} {et~al.}(2012){Poznanski}, {Prochaska}, \&
  {Bloom}}]{Poznanski2012}
{Poznanski}, D., {Prochaska}, J.~X., \& {Bloom}, J.~S. 2012, \mnras, 426, 1465

\bibitem[{Poznanski {et~al.}(2010)Poznanski, Chornock, Nugent, Bloom,
  Filippenko, Ganeshalingam, Leonard, Li, \& Thomas}]{Poznanski2010}
Poznanski, D., Chornock, R., Nugent, P.~E., {et~al.} 2010, Science, 327, 58

\bibitem[{{Rabinak} \& {Waxman}(2011)}]{Rabinak2011}
{Rabinak}, I., \& {Waxman}, E. 2011, \apj, 728, 63

\bibitem[{{Rest} {et~al.}(2011){Rest}, {Foley}, {Sinnott}, {Welch}, {Badenes},
  {Filippenko}, {Bergmann}, {Bhatti}, {Blondin}, {Challis}, {Damke}, {Finley},
  {Huber}, {Kasen}, {Kirshner}, {Matheson}, {Mazzali}, {Minniti}, {Nakajima},
  {Narayan}, {Olsen}, {Sauer}, {Smith}, \& {Suntzeff}}]{Rest2011}
{Rest}, A., {Foley}, R.~J., {Sinnott}, B., {et~al.} 2011, \apj, 732, 3

\bibitem[{Richmond {et~al.}(1996)Richmond, van Dyk, Ho, Peng, Paik, Treffers,
  Filippenko, Bustamante-Donas, Moeller, Pawellek, Tartara, \&
  Spence}]{Richmond1996}
Richmond, M.~W., van Dyk, S.~D., Ho, W., {et~al.} 1996, AJ, 111, 327

\bibitem[{{Roming} {et~al.}(2005){Roming}, {Kennedy}, {Mason}, {Nousek}, {Ahr},
  {Bingham}, {Broos}, {Carter}, {Hancock}, {Huckle}, {Hunsberger}, {Kawakami},
  {Killough}, {Koch}, {McLelland}, {Smith}, {Smith}, {Soto}, {Boyd},
  {Breeveld}, {Holland}, {Ivanushkina}, {Pryzby}, {Still}, \&
  {Stock}}]{Roming05}
{Roming}, P.~W.~A., {Kennedy}, T.~E., {Mason}, K.~O., {et~al.} 2005, \ssr, 120,
  95

\bibitem[{{Roming} {et~al.}(2009){Roming}, {Pritchard}, {Brown}, {Holland},
  {Immler}, {Stockdale}, {Weiler}, {Panagia}, {Van Dyk}, {Hoversten}, {Milne},
  {Oates}, {Russell}, \& {Vandrevala}}]{Roming2009}
{Roming}, P.~W.~A., {Pritchard}, T.~A., {Brown}, P.~J., {et~al.} 2009, \apjl,
  704, L118

\bibitem[{{Ryder} {et~al.}(2004){Ryder}, {Sadler}, {Subrahmanyan}, {Weiler},
  {Panagia}, \& {Stockdale}}]{Ryder2004}
{Ryder}, S.~D., {Sadler}, E.~M., {Subrahmanyan}, R., {et~al.} 2004, \mnras,
  349, 1093

\bibitem[{{Sahu} {et~al.}(2011){Sahu}, {Gurugubelli}, {Anupama}, \&
  {Nomoto}}]{Sahu2011}
{Sahu}, D.~K., {Gurugubelli}, U.~K., {Anupama}, G.~C., \& {Nomoto}, K. 2011,
  \mnras, 413, 2583

\bibitem[{Sanders {et~al.}(2012)Sanders, Soderberg, Levesque, Foley, Chornock,
  Milisavljevic, Margutti, Berger, Drout, Czekala, \& Dittmann}]{Sanders2012a}
Sanders, N.~E., Soderberg, a.~M., Levesque, E.~M., {et~al.} 2012, ApJ, 758, 132

\bibitem[{{Scheck} {et~al.}(2006){Scheck}, {Kifonidis}, {Janka}, \&
  {M{\"u}ller}}]{Scheck2006}
{Scheck}, L., {Kifonidis}, K., {Janka}, H.-T., \& {M{\"u}ller}, E. 2006, \aap,
  457, 963

\bibitem[{{Schlafly} \& {Finkbeiner}(2011)}]{Schlafly2011}
{Schlafly}, E.~F., \& {Finkbeiner}, D.~P. 2011, \apj, 737, 103

\bibitem[{{Schmidt} {et~al.}(1989){Schmidt}, {Weymann}, \&
  {Foltz}}]{Schmidt1989}
{Schmidt}, G.~D., {Weymann}, R.~J., \& {Foltz}, C.~B. 1989, \pasp, 101, 713

\bibitem[{{Shiode} \& {Quataert}(2014)}]{Shiode2014}
{Shiode}, J.~H., \& {Quataert}, E. 2014, \apj, 780, 96

\bibitem[{Silverman {et~al.}(2012)Silverman, Foley, Filippenko, Ganeshalingam,
  Barth, Chornock, Griffith, Kong, Lee, Leonard, Matheson, Miller, Steele,
  Barris, Bloom, Cobb, Coil, Desroches, Gates, Ho, Jha, Kandrashoff, Li,
  Mandel, Modjaz, Moore, Mostardi, Papenkova, Park, Perley, Poznanski, Reuter,
  Scala, Serduke, Shields, Swift, Tonry, {Van Dyk}, Wang, \&
  Wong}]{Silverman2012}
Silverman, J.~M., Foley, R.~J., Filippenko, A.~V., {et~al.} 2012, MNRAS, 425,
  1789

\bibitem[{{Simcoe} {et~al.}(2013){Simcoe}, {Burgasser}, {Schechter}, {Fishner},
  {Bernstein}, {Bigelow}, {Pipher}, {Forrest}, {McMurtry}, {Smith}, \&
  {Bochanski}}]{Simcoe2013}
{Simcoe}, R.~A., {Burgasser}, A.~J., {Schechter}, P.~L., {et~al.} 2013, \pasp,
  125, 270

\bibitem[{{Smartt}(2015)}]{Smartt2015}
{Smartt}, S.~J. 2015, PASA, 32, e016

\bibitem[{Smith {et~al.}(2002)Smith, Tucker, Kent, Richmond, Fukugita,
  Ichikawa, Ichikawa, Jorgensen, Uomoto, Gunn, Hamabe, Watanabe, Tolea, Henden,
  Annis, Pier, McKay, Brinkmann, Chen, Holtzman, Shimasaku, \&
  York}]{Smith2002}
Smith, J.~A., Tucker, D.~L., Kent, S., {et~al.} 2002, AJ, 123, 2121

\bibitem[{{Smith}(2014)}]{Smith2014}
{Smith}, N. 2014, \araa, 52, 487

\bibitem[{{Smith} \& {Arnett}(2014)}]{Smith2014b}
{Smith}, N., \& {Arnett}, W.~D. 2014, \apj, 785, 82

\bibitem[{Soderberg {et~al.}(2010)Soderberg, Brunthaler, Nakar, Chevalier, \&
  Bietenholz}]{Soderberg2010}
Soderberg, a.~M., Brunthaler, a., Nakar, E., Chevalier, R.~a., \& Bietenholz,
  M.~F. 2010, ApJ, 725, 922

\bibitem[{{Soderberg} {et~al.}(2006{\natexlab{a}}){Soderberg}, {Chevalier},
  {Kulkarni}, \& {Frail}}]{Soderberg2006b}
{Soderberg}, A.~M., {Chevalier}, R.~A., {Kulkarni}, S.~R., \& {Frail}, D.~A.
  2006{\natexlab{a}}, \apj, 651, 1005

\bibitem[{{Soderberg} {et~al.}(2005){Soderberg}, {Kulkarni}, {Berger},
  {Chevalier}, {Frail}, {Fox}, \& {Walker}}]{Soderberg2005}
{Soderberg}, A.~M., {Kulkarni}, S.~R., {Berger}, E., {et~al.} 2005, \apj, 621,
  908

\bibitem[{{Soderberg} {et~al.}(2006{\natexlab{b}}){Soderberg}, {Nakar},
  {Berger}, \& {Kulkarni}}]{Soderberg2006}
{Soderberg}, A.~M., {Nakar}, E., {Berger}, E., \& {Kulkarni}, S.~R.
  2006{\natexlab{b}}, \apj, 638, 930

\bibitem[{{Soderberg} {et~al.}(2006{\natexlab{c}}){Soderberg}, {Kulkarni},
  {Nakar}, {Berger}, {Cameron}, {Fox}, {Frail}, {Gal-Yam}, {Sari}, {Cenko},
  {Kasliwal}, {Chevalier}, {Piran}, {Price}, {Schmidt}, {Pooley}, {Moon},
  {Penprase}, {Ofek}, {Rau}, {Gehrels}, {Nousek}, {Burrows}, {Persson}, \&
  {McCarthy}}]{Soderberg2006c}
{Soderberg}, A.~M., {Kulkarni}, S.~R., {Nakar}, E., {et~al.}
  2006{\natexlab{c}}, \nat, 442, 1014

\bibitem[{Soderberg {et~al.}(2008)Soderberg, Berger, Page, Schady, Parrent,
  Pooley, Wang, Ofek, Cucchiara, Rau, Waxman, Simon, Bock, Milne, Page,
  Barentine, Barthelmy, Beardmore, Bietenholz, Brown, Burrows, Burrows,
  Bryngelson, Byrngelson, Cenko, Chandra, Cummings, Fox, Gal-Yam, Gehrels,
  Immler, Kasliwal, Kong, Krimm, Kulkarni, Maccarone, M\'{e}sz\'{a}ros, Nakar,
  O'Brien, Overzier, de~Pasquale, Racusin, Rea, \& York}]{Soderberg2008}
Soderberg, a.~M., Berger, E., Page, K.~L., {et~al.} 2008, Nature, 453, 469

\bibitem[{{Soderberg} {et~al.}(2010){Soderberg}, {Chakraborti}, {Pignata},
  {Chevalier}, {Chandra}, {Ray}, {Wieringa}, {Copete}, {Chaplin},
  {Connaughton}, {Barthelmy}, {Bietenholz}, {Chugai}, {Stritzinger}, {Hamuy},
  {Fransson}, {Fox}, {Levesque}, {Grindlay}, {Challis}, {Foley}, {Kirshner},
  {Milne}, \& {Torres}}]{Soderberg2010b}
{Soderberg}, A.~M., {Chakraborti}, S., {Pignata}, G., {et~al.} 2010, \nat, 463,
  513

\bibitem[{{Speziali} {et~al.}(2008){Speziali}, {Di Paola}, {Giallongo},
  {Pedichini}, {Ragazzoni}, {Testa}, {Baruffolo}, {De Santis}, {Diolaiti},
  {Farinato}, {Fontana}, {Gallozzi}, {Gasparo}, {Gentile}, {Grazian},
  {Manzato}, {Pasian}, {Smareglia}, \& {Vernet}}]{sdg+08}
{Speziali}, R., {Di Paola}, A., {Giallongo}, E., {et~al.} 2008, in Society of
  Photo-Optical Instrumentation Engineers (SPIE) Conference Series, Vol. 7014,
  Society of Photo-Optical Instrumentation Engineers (SPIE) Conference Series,
  4

\bibitem[{{Sramek} {et~al.}(1984){Sramek}, {Panagia}, \& {Weiler}}]{Sramek1984}
{Sramek}, R.~A., {Panagia}, N., \& {Weiler}, K.~W. 1984, \apjl, 285, L59

\bibitem[{{Tanaka} {et~al.}(2008){Tanaka}, {Mazzali}, {Benetti}, {Nomoto},
  {Elias-Rosa}, {Kotak}, {Pignata}, {Stanishev}, \& {Hachinger}}]{Tanaka2008}
{Tanaka}, M., {Mazzali}, P.~A., {Benetti}, S., {et~al.} 2008, \apj, 677, 448

\bibitem[{{Tanaka} {et~al.}(2009){Tanaka}, {Kawabata}, {Maeda}, {Iye},
  {Hattori}, {Pian}, {Nomoto}, {Mazzali}, \& {Tominaga}}]{Tanaka2009}
{Tanaka}, M., {Kawabata}, K.~S., {Maeda}, K., {et~al.} 2009, \apj, 699, 1119

\bibitem[{Taubenberger {et~al.}(2006)Taubenberger, Pastorello, Mazzali,
  Valenti, Pignata, Sauer, Arbey, B??rnbantner, Benetti, {Della Valle},
  Deng, Elias-Rosa, Filippenko, Foley, Goobar, Kotak, Li, Meikle, Mendez,
  Patat, Pian, Ries, Ruiz-Lapuente, Salvo, Stanishev, Turatto, \&
  Hillebrandt}]{Taubenberger2006}
Taubenberger, S., Pastorello, A., Mazzali, P.~a., {et~al.} 2006, MNRAS, 371,
  1459

\bibitem[{{Tauris} {et~al.}(2013){Tauris}, {Langer}, {Moriya}, {Podsiadlowski},
  {Yoon}, \& {Blinnikov}}]{Tauris2013}
{Tauris}, T.~M., {Langer}, N., {Moriya}, T.~J., {et~al.} 2013, \apjl, 778, L23

\bibitem[{Thomas {et~al.}(2011)Thomas, Nugent, \& Meza}]{Thomas2011}
Thomas, R.~C., Nugent, P.~E., \& Meza, J.~C. 2011, PASP, 123, 237

\bibitem[{{Vacca} {et~al.}(2003){Vacca}, {Cushing}, \& {Rayner}}]{Vacca2003}
{Vacca}, W.~D., {Cushing}, M.~C., \& {Rayner}, J.~T. 2003, \pasp, 115, 389

\bibitem[{Valenti {et~al.}(2008)Valenti, Elias-Rosa, Taubenberger, Stanishev,
  Agnoletto, Sauer, Cappellaro, Pastorello, Benetti, Riffeser, Hopp,
  Navasardyan, Tsvetkov, Lorenzi, Patat, Turatto, Barbon, Ciroi, {Di Mille},
  Frandsen, Fynbo, Laursen, \& Mazzali}]{Valenti2008}
Valenti, S., Elias-Rosa, N., Taubenberger, S., {et~al.} 2008, ApJ, 673, L155

\bibitem[{Valenti {et~al.}(2011)Valenti, Fraser, Benetti, Pignata, Sollerman,
  Inserra, Cappellaro, Pastorello, Smartt, Ergon, Botticella, Brimacombe,
  Bufano, Crockett, Eder, Fugazza, Haislip, Hamuy, Harutyunyan, Ivarsen,
  Kankare, Kotak, LaCluyze, Magill, Mattila, Maza, Mazzali, Reichart,
  Taubenberger, Turatto, \& Zampieri}]{Valenti2011}
Valenti, S., Fraser, M., Benetti, S., {et~al.} 2011, MNRAS, 416, 3138

\bibitem[{{van Dyk} {et~al.}(1993){van Dyk}, {Sramek}, {Weiler}, \&
  {Panagia}}]{VanDyk1993}
{van Dyk}, S.~D., {Sramek}, R.~A., {Weiler}, K.~W., \& {Panagia}, N. 1993,
  \apj, 409, 162

\bibitem[{{Van Dyk} {et~al.}(2014){Van Dyk}, {Zheng}, {Fox}, {Cenko}, {Clubb},
  {Filippenko}, {Foley}, {Miller}, {Smith}, {Kelly}, {Lee}, {Ben-Ami}, \&
  {Gal-Yam}}]{VanDyk2014}
{Van Dyk}, S.~D., {Zheng}, W., {Fox}, O.~D., {et~al.} 2014, \aj, 147, 37

\bibitem[{{van Loon} {et~al.}(2005){van Loon}, {Cioni}, {Zijlstra}, \&
  {Loup}}]{vanLoon2005}
{van Loon}, J.~T., {Cioni}, M.-R.~L., {Zijlstra}, A.~A., \& {Loup}, C. 2005,
  \aap, 438, 273

\bibitem[{Weiler {et~al.}(2011)Weiler, Panagia, Stockdale, Rupen, Sramek, \&
  Williams}]{Weiler2011}
Weiler, K.~W., Panagia, N., Stockdale, C., {et~al.} 2011, ApJ, 740, 79

\bibitem[{{Wellons} {et~al.}(2012){Wellons}, {Soderberg}, \&
  {Chevalier}}]{Wellons2012}
{Wellons}, S., {Soderberg}, A.~M., \& {Chevalier}, R.~A. 2012, \apj, 752, 17

\bibitem[{{Wheeler} {et~al.}(1995){Wheeler}, {Harkness}, {Khokhlov}, \&
  {Hoeflich}}]{Wheeler1995}
{Wheeler}, J.~C., {Harkness}, R.~P., {Khokhlov}, A.~M., \& {Hoeflich}, P. 1995,
  \physrep, 256, 211

\bibitem[{{Wheeler} {et~al.}(2015){Wheeler}, {Johnson}, \&
  {Clocchiatti}}]{Wheeler2015}
{Wheeler}, J.~C., {Johnson}, V., \& {Clocchiatti}, A. 2015, \mnras, 450, 1295

\bibitem[{{Wheeler} \& {Levreault}(1985)}]{Wheeler1985}
{Wheeler}, J.~C., \& {Levreault}, R. 1985, \apjl, 294, L17

\bibitem[{Wheeler {et~al.}(1993)Wheeler, Barker, Benjamin, Boisseau,
  Clocchiatti, de~Vaucouleurs, Gaffney, Harkness, Khokhlov, Lester, Smith,
  Smith, \& Tomkin}]{Wheeler1993}
Wheeler, J.~C., Barker, E., Benjamin, R., {et~al.} 1993, ApJ, 417, L71

\bibitem[{Woosley \& Weaver(1995)}]{Woosley1995}
Woosley, S., \& Weaver, T. 1995, ApJSS, 101, 181

\bibitem[{Yoon {et~al.}(2010)Yoon, Woosley, \& Langer}]{Yoon2010}
Yoon, S.-C., Woosley, S.~E., \& Langer, N. 2010, ApJ, 725, 940

\end{thebibliography}


\begin{deluxetable}{lc cccccc}
\tabletypesize{\tiny}
\setlength{\tabcolsep}{0.02in}
\tablecaption{\emph{Swift} UVOT Photometry \label{tab:PhotomUVOT}}
\tablewidth{0pt}
\tablehead{
\colhead{UT Date} &
\colhead{MJD} &
\colhead{$uvw2$ (err)} &
\colhead{$uvm2$ (err)} &
\colhead{$uvw1$ (err)} &
\colhead{$u$ (err)} &
\colhead{$b$ (err)} &
\colhead{$v$ (err)} \\
& & \colhead{mag} &\colhead{mag} &\colhead{mag} &\colhead{mag}
&\colhead{mag} &\colhead{mag} 
}
\startdata
2013 Nov 11 & 56607.0 & \nodata & \nodata  & 16.66 (0.08)  & 15.34 (0.06)  & \nodata  & \nodata  \\
2013 Nov 11 & 56607.8 & 17.32 (0.07) & 17.37 (0.06)  & 16.20 (0.06)  & 15.03 (0.04)  & 15.79 (0.04)  & 15.69 (0.06)  \\
2013 Nov 12 & 56608.7 & 17.21 (0.07) & 17.36 (0.07)  & 16.18 (0.06)  & 14.94 (0.04)  & 15.61 (0.04)  & 15.54 (0.06)  \\
2013 Nov 13 & 56609.8 & 17.29 (0.07) & 17.40 (0.07)  & 16.11 (0.05)  & 14.87 (0.04)  & 15.56 (0.04)  & 15.32 (0.05)  \\
2013 Nov 14 & 56610.8 & 17.37 (0.08) & 17.46 (0.10)  & 16.25 (0.06)  & 15.03 (0.05)  & 15.55 (0.04)  & 15.20 (0.05)  \\
2013 Nov 15 & 56611.3 & 17.53 (0.08) & 17.50 (0.08)  & 16.44 (0.07)  & 15.05 (0.05)  & 15.46 (0.04)  & 15.26 (0.05)  \\
2013 Nov 16 & 56612.1 & 17.54 (0.08) & 17.70 (0.08)  & 16.55 (0.07)  & 15.21 (0.05)  & 15.43 (0.04)  & 15.15 (0.05)  \\
2013 Nov 17 & 56613.1 & 17.57 (0.08) & 17.93 (0.12)  & 16.69 (0.07)  & 15.26 (0.05)  & 15.46 (0.04)  & 15.11 (0.05)  \\
2013 Nov 18 & 56614.6 & 17.72 (0.08) & \nodata  & 16.79 (0.07)  & 15.30 (0.05)  & 15.47 (0.04)  & 14.98 (0.05)  \\
2013 Nov 19 & 56615.1 & 17.76 (0.08) & 18.07 (0.09)  & 16.80 (0.07)  & 15.39 (0.05)  & 15.42 (0.04)  & 14.94 (0.04)  \\
2013 Nov 20 & 56616.7 & 17.86 (0.09) & 17.92 (0.09)  & 16.82 (0.07)  & 15.39 (0.05)  & 15.51 (0.04)  & 14.95 (0.05)  \\
2013 Nov 21 & 56617.1 & 17.75 (0.08) & 17.94 (0.09)  & 16.85 (0.07)  & 15.42 (0.05)  & 15.36 (0.04)  & 14.83 (0.04)  \\
2013 Nov 22 & 56618.6 & \nodata & \nodata  & 16.83 (0.06)  & 15.46 (0.04)  & 15.39 (0.04)  & 14.78 (0.06)  \\
2013 Nov 23 & 56619.2 & \nodata & \nodata  & 16.98 (0.06)  & 15.47 (0.04)  & 15.38 (0.04)  & 14.87 (0.08)  \\
2013 Nov 24 & 56620.1 & \nodata & \nodata  & 17.15 (0.07)  & 15.61 (0.04)  & 15.54 (0.04)  & 14.76 (0.07)  \\
2013 Nov 27 & 56623.1 & \nodata & \nodata  & 17.25 (0.08)  & 15.84 (0.05)  & 15.55 (0.04)  & 14.81 (0.04)  \\
2013 Nov 29 & 56625.9 & \nodata & \nodata  & 17.40 (0.08)  & 16.16 (0.05)  & 15.77 (0.05)  & 14.93 (0.06)  \\
2013 Dec 1 & 56627.7 & \nodata & \nodata  & 17.39 (0.07)  & 16.49 (0.05)  & 16.07 (0.04)  & 15.00 (0.05)  \\
2013 Dec 3 & 56629.5 & \nodata & \nodata  & 17.38 (0.07)  & 16.70 (0.05)  & 16.10 (0.04)  & 15.03 (0.06)  \\
2013 Dec 5 & 56631.1 & \nodata & \nodata  & 17.61 (0.08)  & 16.77 (0.06)  & 16.30 (0.05)  & 15.24 (0.06)  \\
2013 Dec 7 & 56633.5 & \nodata & \nodata  & 17.74 (0.10)  & 16.91 (0.07)  & 16.46 (0.06)  & 15.36 (0.06)  \\
2013 Dec 9 & 56635.2 & \nodata & \nodata  & 17.90 (0.10)  & 17.08 (0.07)  & 16.62 (0.06)  & 15.48 (0.07)  
\enddata                   
\tablecomments{Data is presented in the photometric system of \citet{Breeveld2011}.}
\end{deluxetable}

\begin{deluxetable}{lc cccc}
\tabletypesize{\tiny}
\setlength{\tabcolsep}{0.02in}
\tablecaption{CAO Photometry \label{Tab:IdahoPhotom}}
\tablewidth{0pt}
\tablehead{
\colhead{UT Date} &
\colhead{MJD} &
\colhead{$B$ (err)} &
\colhead{$V$ (err)} &
\colhead{$R$ (err)} &
\colhead{$I$ (err)}  \\
& & \colhead{mag} &\colhead{mag} &\colhead{mag} &\colhead{mag}
}
\startdata 
2013 Nov 12 & 56608.5  & \nodata & 15.62 (0.09)  & 15.67 (0.13)  & 15.43 (0.22) \\
2013 Nov 22 & 56618.5  & 15.51  (0.10) & 14.88 (0.08)  & 14.70 (0.05)  & 14.41 (0.05) \\
2013 Nov 23 & 56619.5  & 15.46  (0.09) & 14.80 (0.07)  & 14.55 (0.04)  & 14.33 (0.06) \\
2013 Nov 24 & 56620.5  & 15.53  (0.09) & 14.83 (0.07)  & 14.51 (0.06)  & 14.30 (0.05) \\
2013 Nov 25 & 56621.5  & 15.64  (0.14) & 14.87 (0.06)  & 14.68 (0.06)  & 14.35 (0.08) \\
2013 Nov 26 & 56622.5  & 15.63  (0.09) & 14.82 (0.10)  & 14.66 (0.05)  & 14.33 (0.05) \\
2013 Nov 28 & 56624.5  & 15.83  (0.10) & 14.94 (0.06)  & 14.58 (0.04)  & 14.25 (0.05) \\
2013 Nov 29 & 56625.5  & 15.93  (0.12) & 14.96 (0.10)  & 14.58 (0.05)  & 14.27 (0.06) \\
2013 Dec 5 & 56631.5  & \nodata & 15.41 (0.14)  & 14.80 (0.10)  & 14.42 (0.16) \\
2013 Dec 11 & 56637.5  & \nodata & 15.60 (0.21)  & 14.98 (0.15)  & 14.56 (0.19) \\
2013 Dec 12 & 56638.5  & \nodata  & 15.85 (0.09)  & 15.09 (0.06)  & 14.63 (0.09) \\
2013 Dec 14 & 56640.5  & \nodata  & 15.87 (0.09)  & 15.10 (0.06)  & 14.64 (0.09) 
\enddata     
\tablecomments{Data is presented in the Bessell photometric system.}              
\end{deluxetable}

\begin{deluxetable}{lc cccc}
\tabletypesize{\tiny}
\setlength{\tabcolsep}{0.02in}
\tablecaption{FLWO Photometry \label{Tab:FLWOPhotom}}
\tablewidth{0pt}
\tablehead{
\colhead{UT Date} &
\colhead{MJD} &
\colhead{$B$ (err)} &
\colhead{$V$ (err)} &
\colhead{$r'$ (err)} &
\colhead{$i'$ (err)}  \\
& & \colhead{mag} &\colhead{mag} &\colhead{mag} &\colhead{mag}
}
\startdata
2013 Dec 7 & 56633.4  & 16.70 (0.02)  & 15.54 (0.02)  & 15.15 (0.03)  & 15.06 (0.03)  \\
2013 Dec 7 & 56633.5  & 16.71 (0.03)  & 15.55 (0.03)  & 15.15 (0.05)  & 15.08 (0.04)  \\
2013 Dec 8 & 56634.4  & 16.70 (0.02)  & 15.54 (0.02)  & 15.14 (0.01)  & 15.04 (0.01)  \\
2013 Dec 14 & 56640.3  & 17.16 (0.02)  & 15.93 (0.02)  & 15.47 (0.02)  & 15.35 (0.03)  \\
2013 Dec 23 & 56649.5  & 17.36 (0.03)  & 16.27 (0.01)  & 15.85 (0.01)  & 15.72 (0.01)  \\
2013 Dec 24 & 56650.5  & 17.57 (0.03)  & 16.35 (0.02)  & 15.91 (0.03)  & 15.78 (0.03) \\
2013 Dec 26 & 56652.5  & 17.54 (0.02)  & 16.36 (0.01)  & 15.93 (0.01)  & 15.84 (0.02) \\
2013 Dec 27 & 56653.5  & 17.53 (0.02)  & 16.40 (0.01)  & 15.95 (0.01)  & 15.85 (0.01)  \\
2013 Dec 28 & 56654.5  & 17.62 (0.02)  &  \nodata  &  16.02 (0.03) & 15.91 (0.03) \\
2013 Dec 29 & 56655.5  &  \nodata  &  \nodata  &  \nodata  & 15.92 (0.05)  \\
2013 Dec 30 & 56656.3  & 17.52 (0.02)  & 16.42 (0.01) & 15.96 (0.01)  & 15.84 (0.01)  \\
2014 Jan 1 & 56658.5  & 17.65 (0.02)  & 16.48 (0.02)  &  16.08 (0.02)  &  \nodata  \\
2014 Jan 2 & 56659.5  &  \nodata  & 16.49 (0.01)  & 16.07 (0.01)  & 15.99 (0.01)  \\
2014 Jan 3 & 56660.3  & 17.71 (0.04)  & 16.54 (0.03)  & 16.13 (0.04)  & 15.93 (0.02)  \\
2014 Jan 6 & 56663.5  &  17.71 (0.02)  & 16.58 (0.05)  & 16.19 (0.05)  & 16.10 (0.06)  \\
2014 Jan 9 & 56666.4  & 17.76 (0.03)  & 16.64 (0.02)  &  \nodata  & 16.19 (0.03)  \\
2014 Jan 11 & 56668.3  & 17.76 (0.02)  & 16.65 (0.01)  & 16.29 (0.02)  & 16.23 (0.02)  \\
2014 Jan 12 & 56669.5  & 17.80 (0.03)  & 16.70 (0.03)  & 16.30 (0.03)  & 16.27 (0.04)  \\
2014 Jan 13 & 56670.5  & 17.78 (0.02)  & 16.70 (0.03)  & 16.34 (0.02)  & 16.31 (0.03)  \\
2014 Jan 14 & 56671.4  & 17.83 (0.03)  & 16.73 (0.04)  & 16.38 (0.04)  & 16.32 (0.04)  \\
2014 Jan 15 & 56672.5  &  \nodata  & 16.70 (0.01)  & 16.32 (0.01)  & 16.29 (0.01)  \\
2014 Jan 16 & 56673.5  & 17.88 (0.04)  & 16.78 (0.03)  & 16.44 (0.05)  &  \nodata  \\
2014 Jan 17 & 56674.4  &  \nodata  & 16.69 (0.01)  & 16.27 (0.02)  & 16.35 (0.01)  \\
2014 Jan 18 & 56675.4  & 17.85 (0.03)  & 16.77 (0.02)  & 16.44 (0.03)  & 16.44 (0.03)  \\
2014 Jan 19 & 56676.5  &  \nodata  & 16.82 (0.03)  &  \nodata  & 16.50 (0.04)  \\
2014 Jan 20 & 56677.4  & 17.92 (0.04)  & 16.85 (0.03)  & 16.49 (0.03)  & 16.48 (0.03)  \\
2014 Jan 21 & 56678.4  & 17.89 (0.02)  &  \nodata  & 16.50 (0.02)  & 16.49 (0.02)  \\
2014 Jan 23 & 56680.4  & 17.86 (0.02)  & 16.87 (0.02)  & 16.53 (0.03)  & 16.55 (0.03)  \\
2014 Jan 26 & 56683.5  & 17.92 (0.02)  & 16.91 (0.03)  & 16.59 (0.03)  & 16.61 (0.02)  \\
2014 Jan 28 & 56685.4  & 17.90 (0.02)  & 16.95 (0.02)  & 16.60 (0.03)  & 16.63 (0.03)  \\
2014 Feb 3 & 56691.3  & 17.97 (0.02)  & 17.00 (0.02)  & 16.70 (0.02)  & 16.77 (0.03)  \\
2014 Feb 6 & 56694.4  &  \nodata  & 17.05 (0.03)  &  \nodata  &  \nodata  \\
2014 Feb 9 & 56697.4  & 18.08 (0.02)  & 17.13 (0.02)  & 16.81 (0.02)  & 16.89 (0.02)  \\
2014 Feb 12 & 56700.3  & 18.14 (0.02)  & 17.20 (0.02)  & 16.87 (0.02)  & 16.96 (0.02)  \\
2014 Feb 13 & 56701.4  & 18.16 (0.03)  & 17.21 (0.03)  & 16.89 (0.03)  & 17.01 (0.04)  \\
2014 Feb 14 & 56702.3  & 18.20 (0.03)  & 17.21 (0.03)  & 16.90 (0.02)  & 17.03 (0.03)  \\
2014 Feb 15 & 56703.5  & 18.18 (0.04)  &  \nodata  &  \nodata  &  \nodata  \\
2014 Feb 18 & 56706.4  & 18.20 (0.02)  & 17.28 (0.03)  & 16.97 (0.03)  & 17.09 (0.04)  \\
2014 Feb 20 & 56708.5  & 18.11 (0.02)  & 17.31 (0.02)  & 17.02 (0.02)  & 17.15 (0.03)  \\
2014 Feb 25 & 56713.2  & 18.19 (0.03)  & 17.34 (0.04)  & 17.05 (0.03)  & 17.15 (0.02)  \\
2014 Feb 25 & 56713.3  & 18.30 (0.03)  & 17.42 (0.05)  & 17.07 (0.03)  &  \nodata  \\
2014 Feb 27 & 56715.2  & 18.24 (0.03)  & 17.35 (0.02)  & 17.07 (0.02)  & 17.22 (0.02)  \\
2014 Feb 27 & 56715.5  & 18.28 (0.03)  & 17.39 (0.02)  & 17.11 (0.02)  & 17.25 (0.03)  \\
2014 Mar 5 & 56721.5  &  \nodata  &  \nodata  & 17.08 (0.02)  & 17.24 (0.02)  \\
2014 Mar 6 & 56722.3  & 18.38 (0.03)  & 17.54 (0.04)  & 17.19 (0.04)  & 17.35 (0.03)  \\
2014 Mar 14 & 56730.5  & 18.50 (0.04)  & 17.67 (0.03)  & 17.29 (0.02)  & 17.60 (0.04)  \\
2014 Mar 15 & 56731.4  &  \nodata  & 17.63 (0.03)  &  \nodata  &  \nodata  \\
2014 Mar 17 & 56733.3  & 18.63 (0.04)  & 17.77 (0.03)  & 17.38 (0.03)  & 17.62 (0.03)  \\
2014 Mar 19 & 56735.3  & 18.50 (0.05)  & 17.77 (0.07)  & 17.33 (0.07)  & 17.61 (0.13)  \\
2014 Mar 25 & 56741.3  & 18.68 (0.03)  & 17.88 (0.04)  & 17.50 (0.04)  & 17.71 (0.03)  \\
2014 Mar 30 & 56746.3  & 18.73 (0.10)  & 17.75 (0.05)  &  \nodata  &  \nodata  \\
2014 Apr 2 & 56749.3  & 18.64 (0.03)  & 18.00 (0.02)  & 17.58 (0.02)  & 17.71 (0.02)  \\
2014 Apr 7 & 56754.3  & 18.64 (0.03)  & 18.00 (0.02)  & 17.57 (0.02)  & 17.91 (0.03)  \\
2014 Apr 30 & 56777.2  &  \nodata  &  \nodata  &  \nodata  &  \nodata  \\
2014 May 3 & 56780.2  & 19.46 (0.03)  & 18.70 (0.02)  & 18.12 (0.03)  & 18.39 (0.03)  
\enddata          
\tablecomments{BV data is presented in the Bessell photometric system and ri data is presented in the SDSS photometric system.}         
\end{deluxetable}

\begin{deluxetable}{lc ccc r}
\tabletypesize{\tiny}
\setlength{\tabcolsep}{0.02in}
\tablecaption{MMTCam, LBT, and IMACS Photometry \label{Tab:MMTPhotom}}
\tablewidth{0pt}
\tablehead{
\colhead{UT Date} &
\colhead{MJD} &
\colhead{$r$ (err)} &
\colhead{$i$ (err)} &
\colhead{$z$ (err)} & 
\colhead{Instrument} \\
& & \colhead{mag} &\colhead{mag} &\colhead{mag} &
}
\startdata
2013 Nov 19 & 56615.5 & 15.01 (0.03) & 14.98 (0.04) & 15.04 (0.05) & MMTCam \\
2014 Apr 1 & 56748.1 & 17.59 (0.05) & 17.85 (0.02) & \nodata & MMTCam \\
2014 Apr 4 & 56751.3 & 17.69 (0.03) & 17.91 (0.02) & 17.04 (0.01) & MMTCam \\
2014 May 21 & 56798.2 & 18.41 (0.02) & 18.71 (0.03) & \nodata & MMTCam \\
2014 June 7 & 56815.1 & 18.71 (0.04) & 18.92 (0.01) & \nodata & MMTCam \\
2014 Oct 22 & 56952.1 & 20.48 (0.05) & 20.96 (0.02) & \nodata & LBT \\
2014 Nov 20 & 56981.5 & 21.18 (0.03) & 21.62 (0.03) & \nodata & MMTCam \\
2014 Dec 18 & 57009.3 & 21.36 (0.02) & \nodata & \nodata & IMACS \\
2015 Jan 15 & 57037.3 & 21.65 (0.03) & \nodata & \nodata & IMACS \\
2015 Feb 10 & 57063.4 & 22.13 (0.06) & 22.58 (0.04) & \nodata & MMTCam \\
2015 Apr 16 & 57128.1 & $>$ 22.3 & $>$ 22.4 & \nodata & MMTCam
\enddata     
\tablecomments{Data is presented in the SDSS photometric system.}              
\end{deluxetable}

\begin{deluxetable}{lc cr}
\tabletypesize{\tiny}
\setlength{\tabcolsep}{0.02in}
\tablecaption{VLA Observations \label{Tab:VLA}}
\tablewidth{0pt}
\tablehead{
\colhead{UT Date} &
\colhead{MJD} &
\colhead{Frequency} &
\colhead{$F_{\nu}$\tablenotemark{a}} \\
& & \colhead{GHz} & \colhead{$\mu$Jy}
}
\startdata
2013 Nov 16 & 56612.8 & 4.8 & $<$45.6\\
\nodata & \nodata & 7.1 & $<$42.0 \\
2013 Nov 26 & 56622.4 & 4.8 & $<$36.0 \\
\nodata & \nodata & 7.1 & $<$30.9 \\
2014 Jan 4 & 56661.3 & 4.8 & $<$36.0 \\
\nodata & \nodata & 7.1 & $<$29.3
\enddata                   
\tablenotetext{a}{All quoted flux limits are 3$\sigma$.}
\end{deluxetable}

\begin{deluxetable}{lccc}
\tabletypesize{\scriptsize}
\tablecaption{Optical and Infrared Spectroscopy \label{tab:spectra}}
\tablewidth{0pt}
\tablehead{
\colhead{UT Date} &
\colhead{MJD} &
\colhead{Telescope} &
\colhead{Instrument\tablenotemark{a}}
 }
\startdata
2013 Nov 9 & 56605 & MMT & BC \\
2013 Nov 10 & 56606 & MMT & BC\tablenotemark{b} \\
2013 Nov 10 & 56606 & MMT & BC \\
2013 Nov 11 & 56607 & MMT & BC \\
2013 Nov 20 & 56616 & Magellan-Baade & FIRE \\
2013 Nov 21 & 56617 & MMT & Hectospec \\
2013 Nov 29 & 56625 & MMT & Hectospec \\
2013 Nov 30 & 56626 & Magellan-Baade & FIRE \\
2013 Dec 6 & 56632 & Tillinghast 60-in & FAST \\
2013 Dec 7 & 56633 & Tillinghast 60-in & FAST \\
2013 Dec 9 & 56634 & Magellan-Baade & FIRE \\
2013 Dec 10 & 56636 & Tillinghast 60-in & FAST \\
2013 Dec 13 & 56639 & MDM 2.4-m & OSMOS \\
2013 Dec 24 & 56650 & MMT & BC \\
2013 Dec 27 & 56653 & MMT & BC \\
2013 Dec 28 & 56654 & MMT & BC \\
2013 Dec 31 & 56657 & Magellan-Clay & LDSS-3 \\
2014 Jan 1 & 56658 & Magellan-Baade & FIRE \\
2014 Jan 3 & 56660 & Tillinghast 60-in & FAST \\
2014 Jan 6 & 56663 & Tillinghast 60-in & FAST \\
2014 Jan 7 & 56664 & Tillinghast 60-in & FAST \\
2014 Jan 9 & 56666 & Tillinghast 60-in & FAST \\
2014 Jan 9 & 56666 & Magellan-Baade & FIRE \\
2014 Jan 25 & 56682 & Shane 3-m & Kast \\
2014 Jan 28 & 56685 & Tillinghast 60-in & FAST \\
2014 Jan 30 & 56687  & Tillinghast 60-in & FAST \\
2014 Feb 2 & 56690 & Tillinghast 60-in & FAST \\
2014 Feb 2 & 56690 & Magellan-Baade & FIRE \\
2014 Feb 3 & 56691 & Magellan-Baade & IMACS \\
2014 Feb 26 & 56714 & Tillinghast 60-in & FAST \\
2014 Feb 27 & 56715 & Magellan-Baade & FIRE \\
2014 Mar 4 & 56720 & Tillinghast 60-in & FAST \\
2014 Mar 6 & 56722 & MMT & BC \\
2014 Mar 7 & 56723 & Tillinghast 60-in & FAST \\
2014 Mar 25 & 56741 & Magellan-Baade & FIRE \\
2014 Apr 28 & 56775 & MMT & BC \\
2014 Oct 23 & 56953 & LBT & MODS \\
2015 Jan 15 & 57037 & Magellan-Baade & IMACS 
\enddata
\tablenotetext{a}{Instrument References: FAST spectrograph \citep{Fabricant1998} on the FLWO 60-inch Tillinghast telescope; Blue Channel (BC) spectrograph \citep{Schmidt1989} on the 6.5 MMT; hectospec multi-fiber spectrograph \citep{Fabricant2005} on the MMT; Low Dispersion Survey Spectrograph-3 (LDSS-3; \citealt{Allington-Smith1994}) on Magellan-Clay, the Inamori-Magellan Areal Camera \& Spectrograph (IMACS; \citealt{Dressler2006}) on Magellan-Baade, OSMOS on the MDM 2.4-m \citep{Martini2011}, the Kast spectrograph \citep{Miller93} on Shane 3-m at Lick Observatory, and the Multi-Object Double Spectrograph (MODS; \citealt{pab+10}) mounted on the 2 $\times$ 8.4-m LBT; Folded-port InfraRed Echellette spectrograph (FIRE; \citealt{Simcoe2013}) on Magellan-Baade}
\tablenotetext{b}{Observation with the 1200 lines/mm grating.}
\end{deluxetable}

\begin{deluxetable}{lccccc}
\tabletypesize{\scriptsize}
\tablecaption{Basic Photometric Properties \label{tab:PhotomProps}}
\tablewidth{0pt}
\tablehead{
\colhead{Band} &
\colhead{T$_{\rm{max}}$} &
\colhead{m$_{\rm{obs,max}}$} &
\colhead{M$_{\rm{abs,max}}$} &
\colhead{$\Delta$m$_{15}$} &
\colhead{Neb.\ Decline Rate\tablenotemark{a}} \\
& \colhead{(MJD)} & \colhead{(AB mag)} & \colhead{(AB mag)} & \colhead{(mag)} & \colhead{(mag day$^{-1}$)}
 }
\startdata
$w2$ & 56608.7 (1.1) & 18.51 (0.06) & $-$13.36 (0.05) & $>$0.65 & \nodata \\
$m2$ & 56607.7 (1.0) & 18.40 (0.05) & $-$13.47 (0.05) & $>$0.70 & \nodata \\
$w1$ & 56608.8 (1.0) & 17.18 (0.04) & $-$14.69 (0.04) & 1.23 (0.08) & \nodata \\
$u$ & 56609.4 (1.1) & 15.56 (0.04) & $-$16.31 (0.04) & 1.16 (0.05) & \nodata \\
$b$  & 56615.7 (1.5) & 14.93 (0.03) & $-$16.94 (0.03) & 1.11 (0.05) & 0.0119 (0.0004) \\
$v$ & 56618.6 (1.5) & 14.52 (0.05) & $-$17.35 (0.05) & 0.74 (0.11) &  0.0156 (0.0003)\\
$R$ & 56621.1 (1.5)  & 14.55 (0.03) & $-$17.32 (0.03) & 0.44 (0.05) & 0.0155 (0.0003) \\
$I$  & 56623.1 (1.5)  & 14.54 (0.03) & $-$17.33 (0.03) & 0.38 (0.06) &  0.0198 (0.0004)
\enddata
\tablecomments{Values for peak magnitudes, time of maximum, and $\Delta$m$_{15}$ were measured from low order polynomials to photometry from a single source for each band.  For w2$-$through$-$v bands, the polynomial was fit to the data from \emph{Swift} UVOT, while for R$-$band and I$-$band the polynomial was fit to CAO data. The peak magnitudes presented here have been shifted to the AB photometric zeropoint as described in Section~\ref{sec:LCcomb}.}
\tablenotetext{a}{Measured from a linear fit to the FLWO data between $+$60 and $+$120 days.}
\end{deluxetable}

\end{document}